\documentclass[oneside,english]{elsart}
\usepackage[T1]{fontenc}
\usepackage[latin1]{inputenc}
\usepackage{subfigure}
\usepackage{amsmath}
\usepackage{graphicx}
\usepackage{amssymb}
\usepackage[authoryear]{natbib}

\makeatletter

\providecommand{\tabularnewline}{\\}

\usepackage{babel}
\makeatother

\begin{document}
\begin{frontmatter}

\title{Gene regulation in continuous cultures: A unified theory for bacteria
and yeasts}

\author{Jason T. Noel}

\address{Department of Chemical Engineering, University of Florida, Gainesville,
FL~32611-6005.}

\author{Atul Narang}

\address{Department of Chemical Engineering, University of Florida, Gainesville,
FL~32611-6005.}

\ead{narang@che.ufl.edu}

\thanks{Corresponding author. Tel: + 1-352-392-0028; fax: + 1-352-392-9513}

\begin{keyword}
\noindent Mathematical model, mixed substrate growth, substitutable
substrates, chemostat, gene expression.
\end{keyword}
\begin{abstract}
During batch growth on mixtures of two growth-limiting substrates,
microbes consume the substrates either sequentially or simultaneously.
These growth patterns are manifested in all types of bacteria and
yeasts. The ubiquity of these growth patterns suggests that they are
driven by a universal mechanism common to all microbial species. In
previous work, we showed that a minimal model accounting only for
enzyme induction and dilution explains the phenotypes observed in
batch cultures of various wild-type and mutant/recombinant cells.
Here, we examine the extension of the minimal model to continuous
cultures. We show that: (1)~Several enzymatic trends, usually attributed
to specific regulatory mechanisms such as catabolite repression, are
completely accounted for by dilution. (2)~The bifurcation diagram
of the minimal model for continuous cultures, which classifies the
substrate consumption pattern at any given dilution rate and feed
concentrations, provides a a precise explanation for the empirically
observed correlation between the growth patterns in batch and continuous
cultures. (3)~Numerical simulations of the model are in excellent
agreement with the data. The model captures the variation of the steady
state substrate concentrations, cell densities, and enzyme levels
during the single- and mixed-substrate growth of bacteria and yeasts
at various dilution rates and feed concentrations. (4)~This variation
is well approximated by simple analytical expressions that furnish
physical insights into the steady states of continuous cultures. Since
the minimal model describes the behavior of the cells in the absence
of any regulatory mechanisms, it provides a framework for rigorously
quantitating the effect of these mechanisms. We illustrate this by
analyzing several data sets from the literature.
\end{abstract}
\end{frontmatter}

\section{Introduction}

The mechanisms of gene regulation are of fundamental importance in
biology. They play a crucial role in development by determining the
fate of isogenic embroyonic cells, and there is growing belief that
the diversity of biological organisms reflects the variation of their
regulatory mechanisms~\citep{Carroll,Ptashne2}.

Many of the key principles of gene regulation, such as positive, negative,
and allosteric control, were discovered by studying the growth of
bacteria and yeasts in batch cultures containing one or more growth-limiting
substrates. A model system that played a particularly important role
is the growth of \emph{Escherichia coli} on lactose and a mixture
of lactose and glucose~\citep{Muller-Hill}.

It turns out that the enzymes catalyzing the transport and peripheral
catabolism of lactose, such as lactose permease and $\beta$-galactosidase,
are synthesized or \emph{induced} only if lactose is present in the
environment. The molecular mechanism of induction was discovered by
Monod and coworkers~\citep{jacob61}. They showed that the genes
encoding the peripheral enzymes for lactose are contiguous and transcribed
sequentially, an arrangement referred to as the \emph{lac} operon.
In the absence of lactose, the \emph{lac} operon is not transcribed
because the \emph{lac} repressor is bound to a specific site on the
\emph{lac} operon called the \emph{operator}. This prevents RNA polymerase
from attaching to the operon and initiating transcription. In the
presence of lactose, transcription of \emph{lac} is triggered because
allolactose, a product of $\beta$-galactosidase, binds to the repressor,
and renders it incapable of binding to the operator.

When \emph{Escherichia coli }is grown on a mixture of glucose and
lactose, \emph{lac} transcription, and hence, the consumption of lactose,
is somehow suppressed until glucose is exhausted. During this period
of preferential growth on glucose, the peripheral enzymes for lactose
are diluted to very small levels. The consumption of lactose begins
upon exhaustion of glucose, but only after a certain lag during which
the lactose enzymes are built up to sufficiently high levels. Thus,
the cells exhibit two exponential growth phases separated by an intermediate
lag. Monod referred to this phenomenon as \emph{diauxic} or {}``double''
growth~\citep{monod1}.

Two major molecular mechanisms have been proposed to explain the repression
of \emph{lac} transcription in the presence of glucose:

\begin{enumerate}
\item \emph{Inducer exclusion}~\citep{Postma1993}: In the presence of
glucose, enzyme IIA$^{{\rm glc}}$, a peripheral enzyme for glucose,
is dephosphorylated. The dephosphorylated~IIA$^{{\rm glc}}$ inhibits
lactose uptake by binding to lactose permease. This reduces the intracellular
concentration of allolactose, and hence, the transcription rate of
the \emph{lac} operon.\\
Genetic evidence suggests that phosphorylated~IIA$^{{\rm glc}}$
activates adenylate cyclase, which catalyzes the synthesis of cyclic
AMP (cAMP). Since glucose uptake results in dephosphorylation of IIA$^{{\rm glc}}$,
one expects the cAMP level to decrease in the presence of glucose.
This forms the basis of the second mechanism of \emph{lac} repression.
\item \emph{cAMP activation}~\citep{Ptashne2}: It has been observed that
the recruitment of RNA polymerase to promoter is inefficient unless
a protein called catabolite activator protein (CAP) is bound to a
specific \emph{CAP site} on the promoter. Furthermore, CAP has a low
affinity for the CAP site, but when bound to cAMP, its affinity increases
dramatically. The inhibition of \emph{lac} transcription by glucose
is then explained as follows.\\
In the presence of lactose alone (i.e., no glucose), the cAMP level
is high. Hence, CAP becomes cAMP-bound, attaches to the CAP site,
and promotes transcription by recruiting RNA polymerase. When glucose
is added to the culture, the cAMP level decreases by the mechanism
described above. Consequently, CAP, being cAMP-free, fails to bind
to the CAP site, and \emph{lac} transcription is abolished.
\end{enumerate}
Mechanistic mathematical models of the glucose-lactose diauxie appeal
to both these mechanisms~\citep{Kremling2001,Santillan2007,vandedem75,Wong1997}

Over the last few decades, microbial physiologists have accumulated
a vast amount of data showing that diauxic growth is ubiquitous~\citep[reviewed in][]{egli95,harder82,kovarova98}.
It occurs in diverse microbial species and numerous pairs of \emph{substitutable}
substrates (satisfying identical nutrient requirements). In many of
these systems, the above mechanisms play no role. For instance, cAMP
is not detectable in gram-positive bacteria~\citep{Mach1984}, and
has little effect on the transcription of peripheral enzymes in pseudomonads~\citep{Collier1996}
and yeasts~\citep{Eraso1984}. It has been proposed that in these
cases, the mechanisms of repression are analogous to those of the
\emph{lac} operon, but involve different transcription factors, such
as CcpA in gram-positive bacteria~\citep{Chauvaux1996}, Crc in pseudomonads~\citep{Morales2004},
and MigA in yeasts~\citep{Johnston1999}.

Paradoxically, there is growing evidence that cAMP activation and
inducer exclusion are not sufficient for explaining \emph{lac} repression.
Not long after the discovery of cAMP activation~\citep{Perlman1968,Ullmann1968},
Ullmann presented several lines of evidence showing that this mechanism
played a relatively modest role in diauxic growth~\citep{Ullmann1974}.
But the most compelling evidence was obtained by Aiba and coworkers,
who showed that the cAMP levels are essentially the same during the
first and second growth phases; furthermore, glucose-mediated \emph{lac}
repression persists in the presence of high (5mM) exogenous cAMP levels,
and in mutants of \emph{E. coli} in which the ability of cAMP to influence
\emph{lac} transcription is completely abolished~\citep{Inada96,Kimata97}.
In Section~\ref{sub:Regulation}, we shall show that the positive
correlation between \emph{lac} expression and intracellular cAMP levels~\citep{Epstein1975},
which forms the foundation of the cAMP activation mechanism, does
not reflect a causal relationship: The observed variation of \emph{lac}
expression is almost entirely due to dilution (rather than cAMP activation).

The persistence of \emph{lac} repression in cAMP-independent cells
has led to the hypothesis that inducer exclusion is the sole cause
of repression~\citep{Kimata97}. However, this mechanism exerts a
relatively mild effect. The lactose uptake rate is inhibited only
$\sim$50\% in the presence of glucose~\citep{McGinnis1969}.

Thus, transcriptional repression, by itself, cannot explain the several
hundred-fold repression of \emph{lac} during the first exponential
growth phase of diauxic growth. In earlier work, we have shown that
complete repression is predicted by a minimal model accounting for
only induction and growth~\citep{narang98b,Narang2006a}.

Yet another phenomenon that warrants an explanation is the empirically
observed correlation between the substrate consumption pattern and
the specific growth on the individual substrates. For instance, in
the case of diauxic growth, it has been observed that:

\begin{quote}
In most cases, although not invariably, the presence of a substrate
permitting a higher growth rate prevents the utilization of a second,
`poorer' substrate in batch culture~\citep{harder82}.
\end{quote}
It turns out that sequential consumption of substrates is not the
sole growth pattern in batch cultures. Monod observed that in several
cases of mixed-substrate growth, there was no diauxic lag~\citep{monod1,monod47},
and subsequent studies have shown that both substrates are often consumed
simultaneously~\citep{egli95}. The occurrence of simultaneous substrate
consumption also appears to correlate with the specific growth rates
on the individual substrates. Based on a comprehensive review of the
literature, Egli notes that:

\begin{quote}
Especially combinations of substrates that support medium or low maximum
specific growth rates are utilized simultaneously~\citep{egli95}.
\end{quote}
Recently, we have shown that the minimal model provides a natural
explanation for the foregoing correlations, which hinges upon the
fact that the enzyme dilution rate is proportional to the specific
growth rate~\citep{Narang2007a}. Roughly, substrates that support
high growth rates lead to such high enzyme dilution rates that the
enzymes of the {}``less preferred'' substrates are diluted to near-zero
levels. On the other hand, substrates that support low growth rates
fail to efficiently dilute the enzymes of the other substrates. A
more precise statement of this argument is given in Section~\ref{sec:MixedSubstrateGrowth}.

Now, all the experimental results described above were obtained in
batch cultures. However, since the invention of the chemostat, microbial
physiologists have acquired extensive data on microbial growth in
continuous cultures. Detailed analyses of this data have revealed
well-defined patterns or motifs that occur in diverse microbial species
growing on various substrates~\citep{egli95,harder82,kovarova98}.
The goal of this work is to show that the minimal model also accounts
for the patterns observed in continuous cultures. We begin by giving
a brief preview of these patterns in single- and mixed-substrate cultures.

\begin{figure}
\noindent \begin{centering}
\subfigure[]{\includegraphics[width=2.6in]{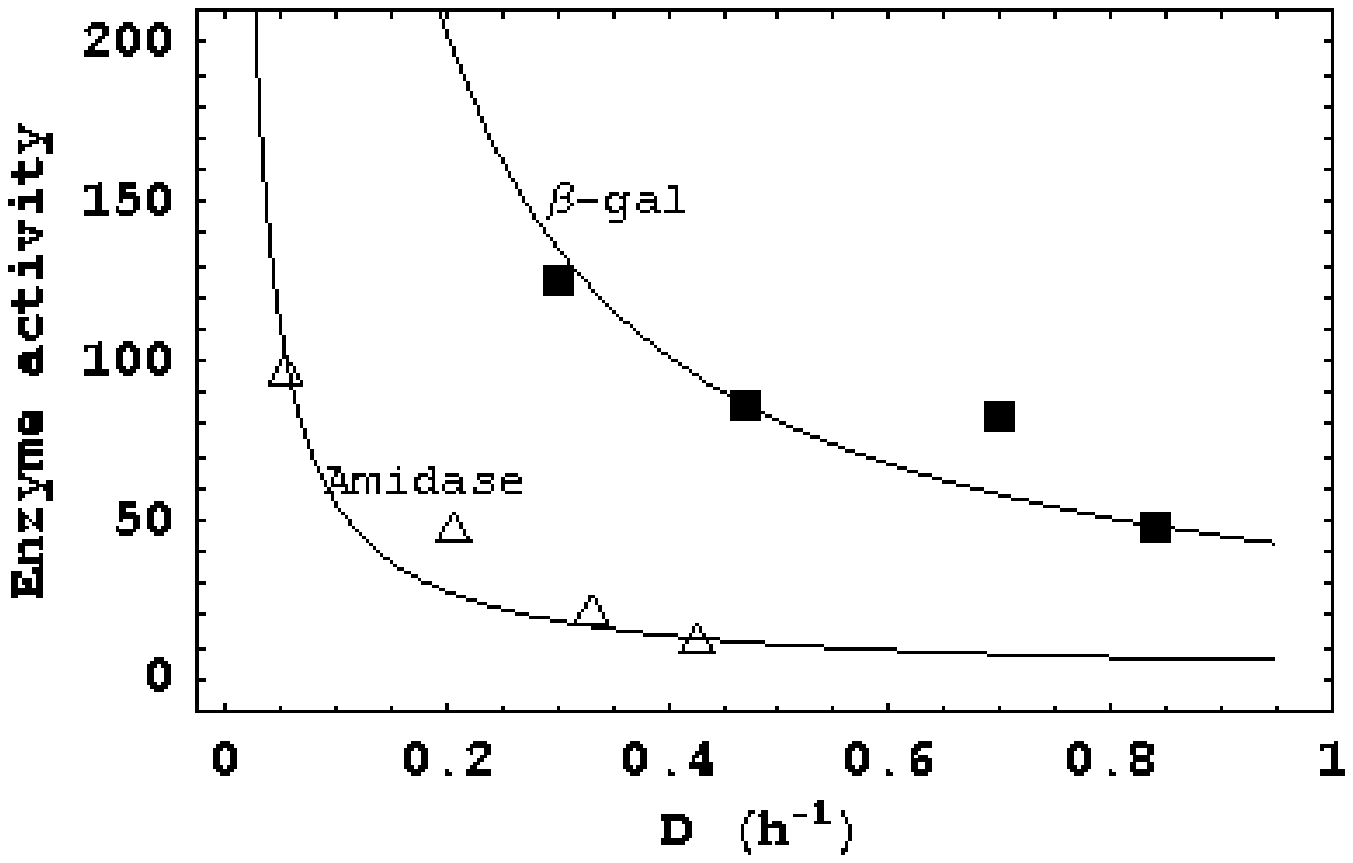}}\hspace*{0.1in}\subfigure[]{\includegraphics[width=2.6in]{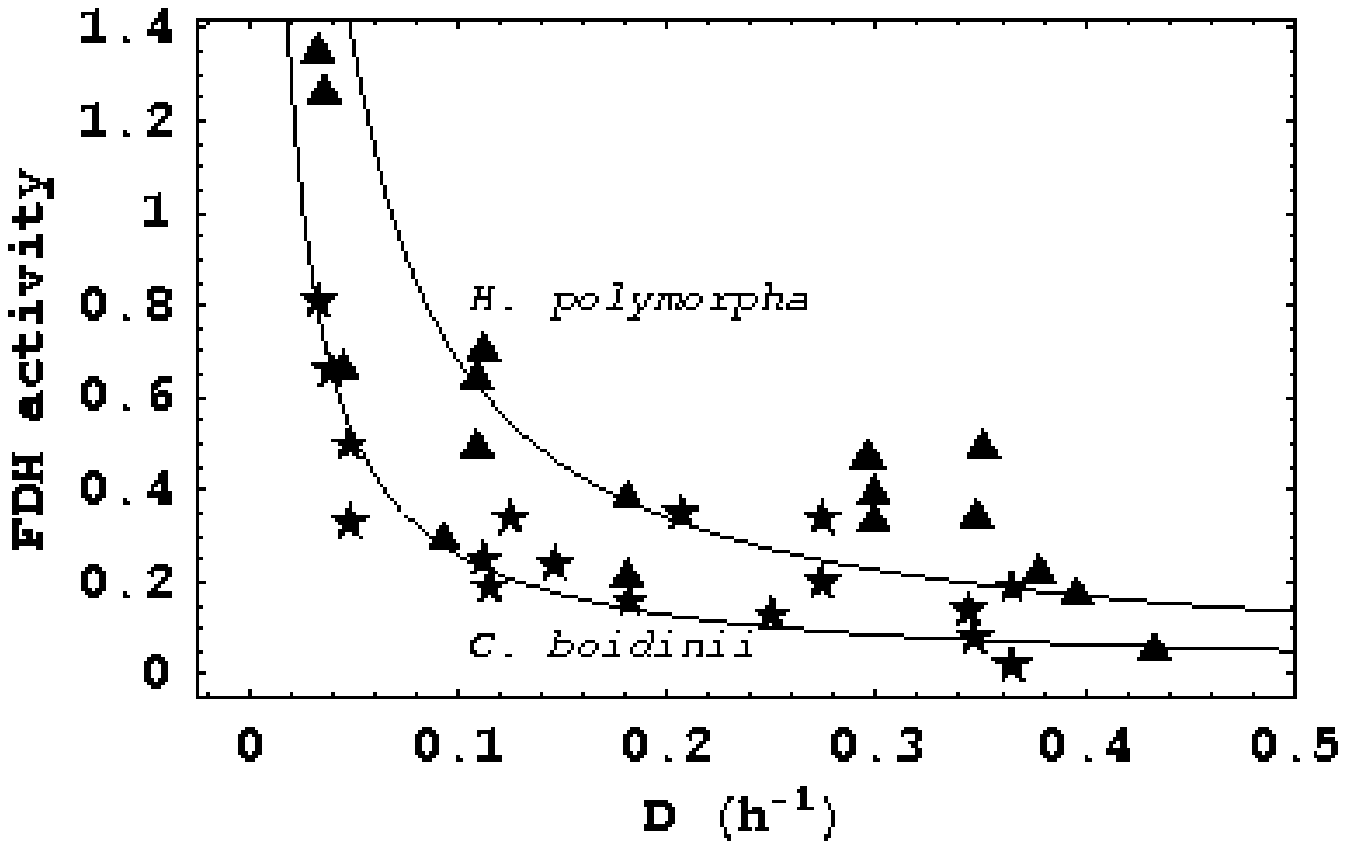}}
\par\end{centering}

\caption{\label{fig:NonInducibleEnzymes}If the induction rate of the peripheral
enzyme is independent of the inducer level, its activity decreases
with the dilution rate. (a)~The activities of $\beta$-galactosidase~\citep[Fig.~1]{silver69}
and amidase~\citep[Fig.~2]{clarke68} during growth of the constitutive
mutants, \emph{E. coli }B6b2 and \emph{P. aeruginosa} C11, on lactose
and acetamide, respectively. (b)~The activity of formaldehyde dehydrogenase
(FDH) during glucose-limited growth of \emph{H. polymorpha} and \emph{C.
boidinii}~\citep[Fig.~1c,d]{egli80}. The curves in (a) and (b) shows
the fits to eqs.~\eqref{eq:eConstitutiveChemostat1} and \eqref{eq:eConstitutiveChemostat2},
respectively.}

\end{figure}

\begin{figure}
\noindent \begin{centering}
\subfigure[]{\includegraphics[width=2.6in]{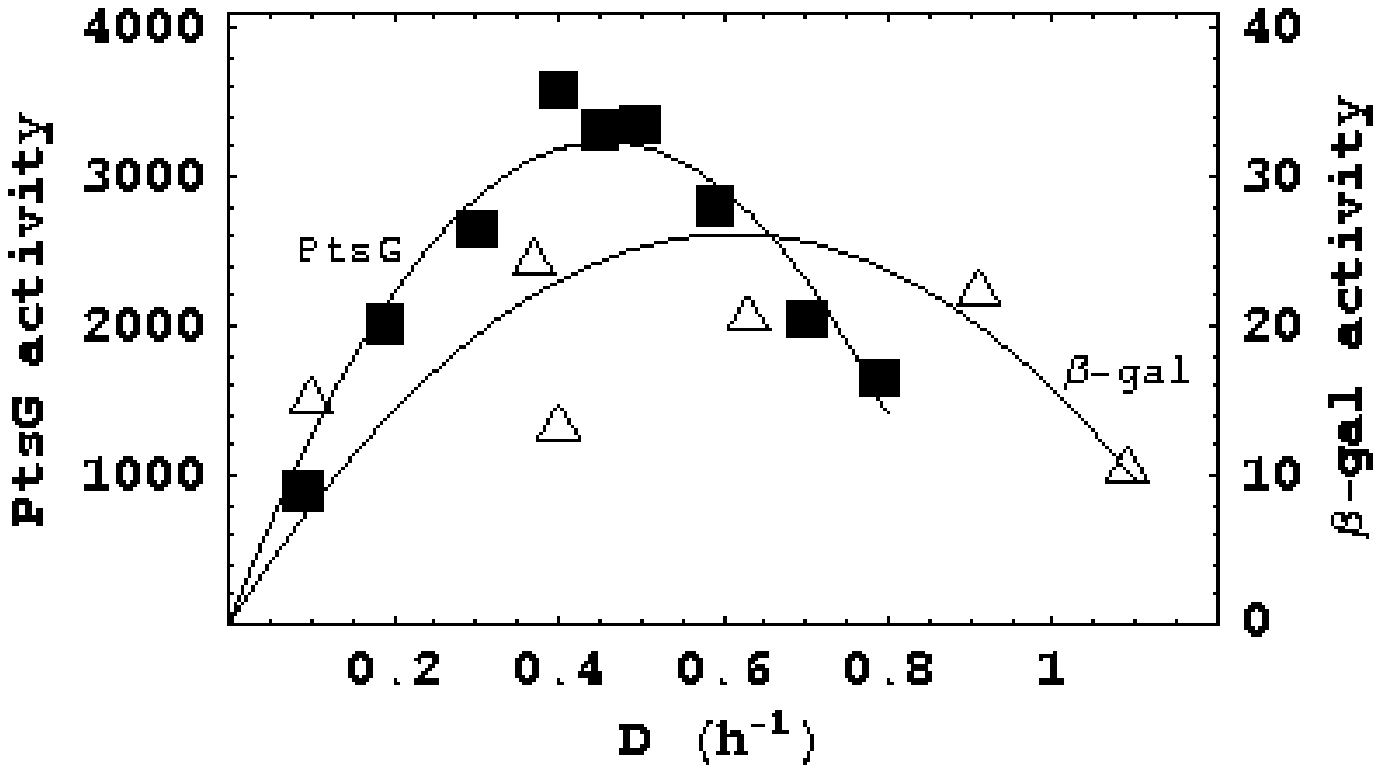}}\hspace*{0.1in}\subfigure[]{\includegraphics[width=2.6in]{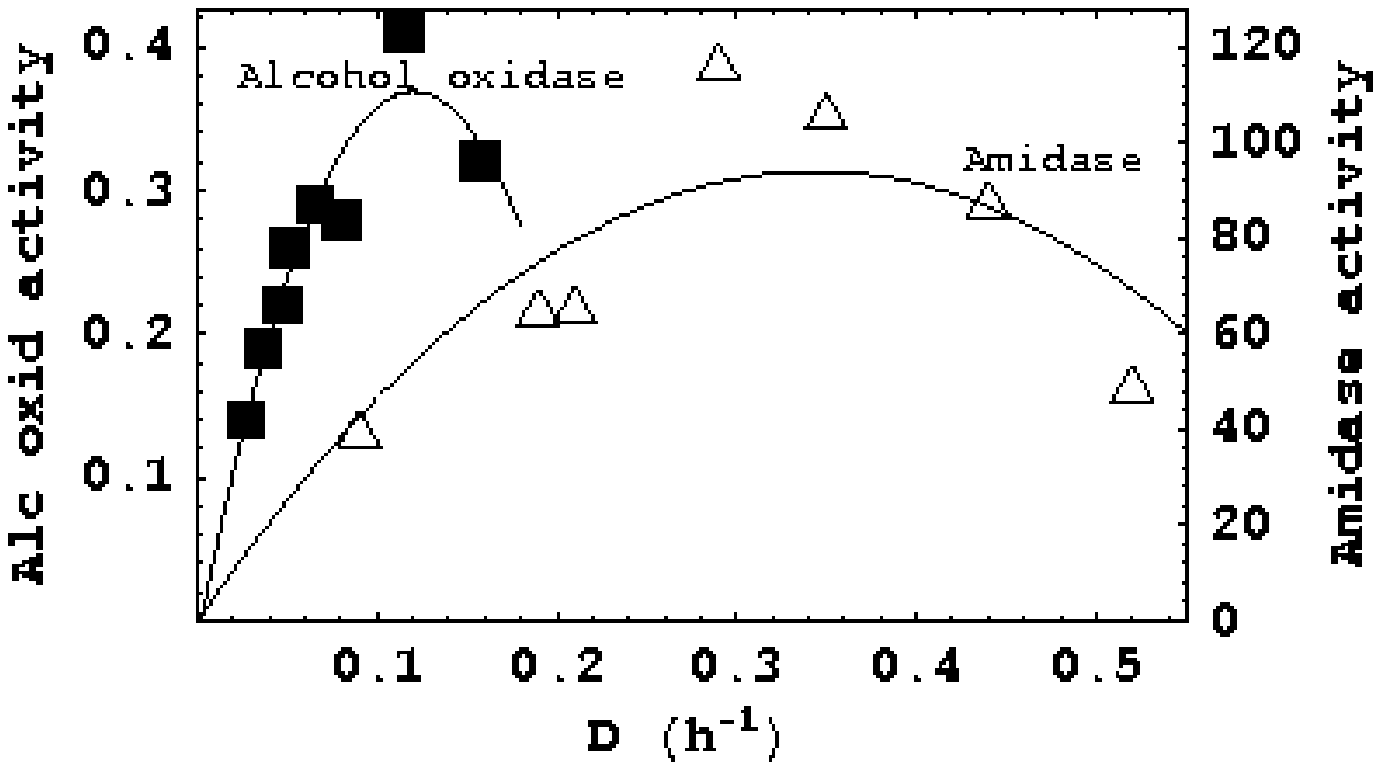}}
\par\end{centering}

\caption{\label{fig:InducibleEnzymes}The activity of inducible enzymes passes
through a maximum at an intermediate dilution rate. (a)~Activities
of PtsG~\citep[Fig.~1]{Seeto2004} and $\beta$-galactosidase~\citep[Fig.~4a]{silver69}
during glucose-limited growth of \emph{E. coli }K12 and lactose-limited
growth of \emph{E. coli }B6, respectively. (b)~Activities of alcohol
oxidase~\citep[Fig.~1]{Egli1983a} and amidase~\citep[Fig.~1]{clarke68}
during methanol-limited growth of \emph{H. polymorpha} and acetamide-limited
growth of \emph{P. aeroginosa}, respectively.}

\end{figure}

In chemostats limited by a single substrate, steady growth can be
maintained at all dilution rates up to the \emph{critical} dilution
rate at which the cells wash out. At subcritical dilution rates, the
peripheral enzyme activities of various cell types and growth-limiting
substrates invariably show one of the following two trends~\citep{harder82}:

\begin{enumerate}
\item If the peripheral enzyme is fully constitutive, its activity decreases
monotonically with the dilution rate (Fig.~\ref{fig:NonInducibleEnzymes}a).
The very same trend is observed even if the peripheral enzyme is inducible,
but the nutrient medium lacks substrates that can induce the synthesis
of the enzyme (Fig.~\ref{fig:NonInducibleEnzymes}b and Fig.~\ref{fig:RoleOfcAMP}d).
\item If the peripheral enzyme is inducible and the medium contains the
inducing substrate, the enzyme activity passes through a maximum at
an intermediate dilution rate (Fig.~\ref{fig:InducibleEnzymes}).
\end{enumerate}
Clarke and coworkers were the first to observe these trends during
acetamide-limited growth of wild-type and constitutive mutants of
\emph{P. aeruginosa}~\citep{clarke68}. They hypothesized that in
wild-type cells, the amidase activity exhibits a maximum (Fig.~\ref{fig:InducibleEnzymes}b)
due to the balance between induction and catabolite repression. Specifically,

\begin{quotation}
at low dilution rates, catabolite repression is minimal and the rate
of amidase synthesis is dependent mainly on the rate at which acetamide
is presented to the bacteria. Thus, with increase in dilution rate
the amidase specific activity under steady state conditions increases.
However, above $D=0.30$~h$^{-1}$ the growth rate has increased
to the point where metabolic intermediates are being formed at a sufficiently
high rate to cause significant catabolite repression. At higher dilution
rates catabolite repression becomes dominant and at $D=0.6$~h$^{-1}$
the enzyme concentration has decreased considerably~\citep[p.~232]{clarke68}.
\end{quotation}
In constitutive mutants, the enzyme activity decreases monotonically
(Fig.~\ref{fig:NonInducibleEnzymes}a) because {}``they lack completely
the part where, in the curves for the wild-type strain, \ldots{}
induction is dominant.'' These hypotheses have subsequently been
invoked to rationalize the similar trends found in the wild-type cells
and constitutive mutants of many other microbial systems~\citep[reviewed in][]{dean72,matin78,toda81}.
We shall show below that the decline of the enzyme activity (at all
$D$ in constitutive mutants, and at high $D$ in wild-type cells)
is entirely due to dilution, rather than catabolite repression.

\begin{figure}
\noindent \begin{centering}
\includegraphics[width=2.5in,height=2in]{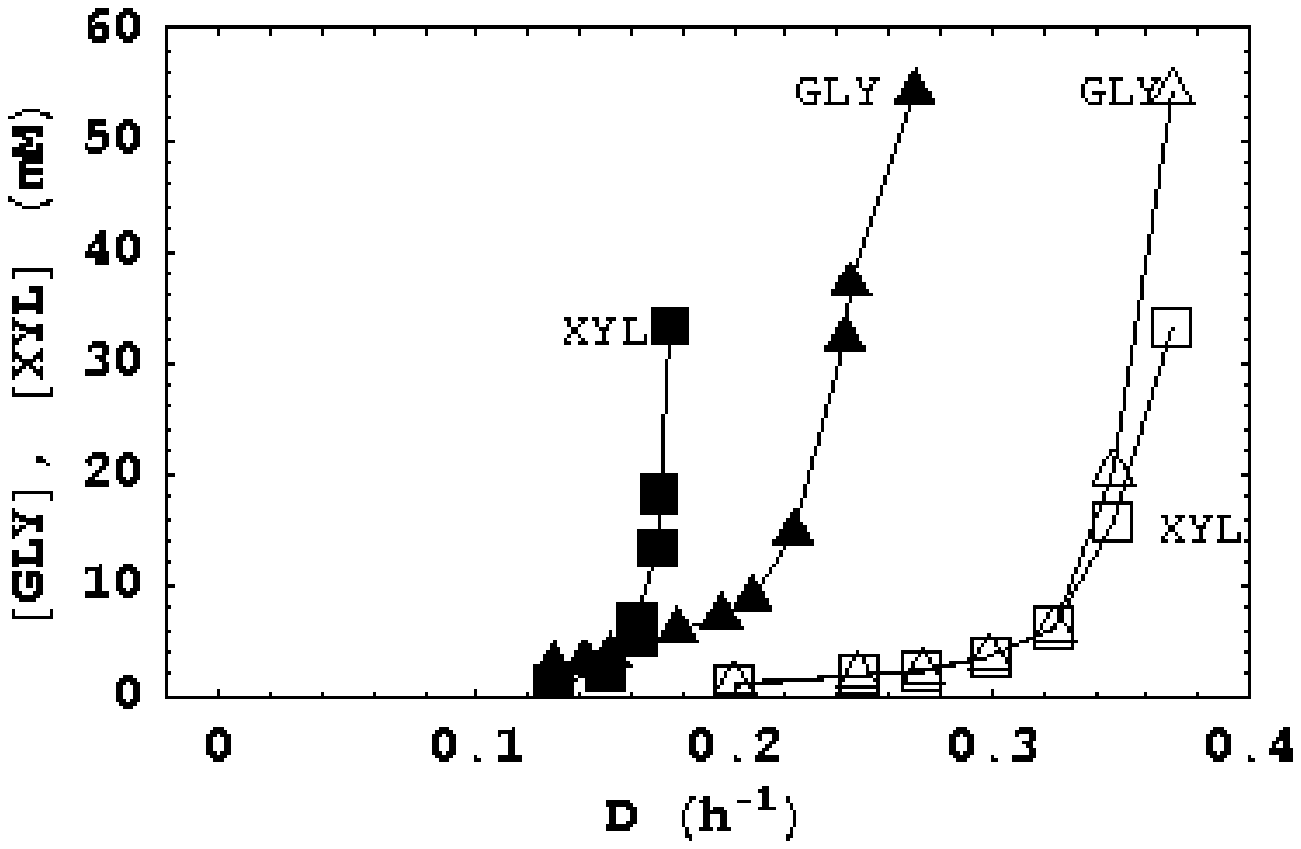}\hspace*{0.1in}\includegraphics[width=2.5in,height=2in]{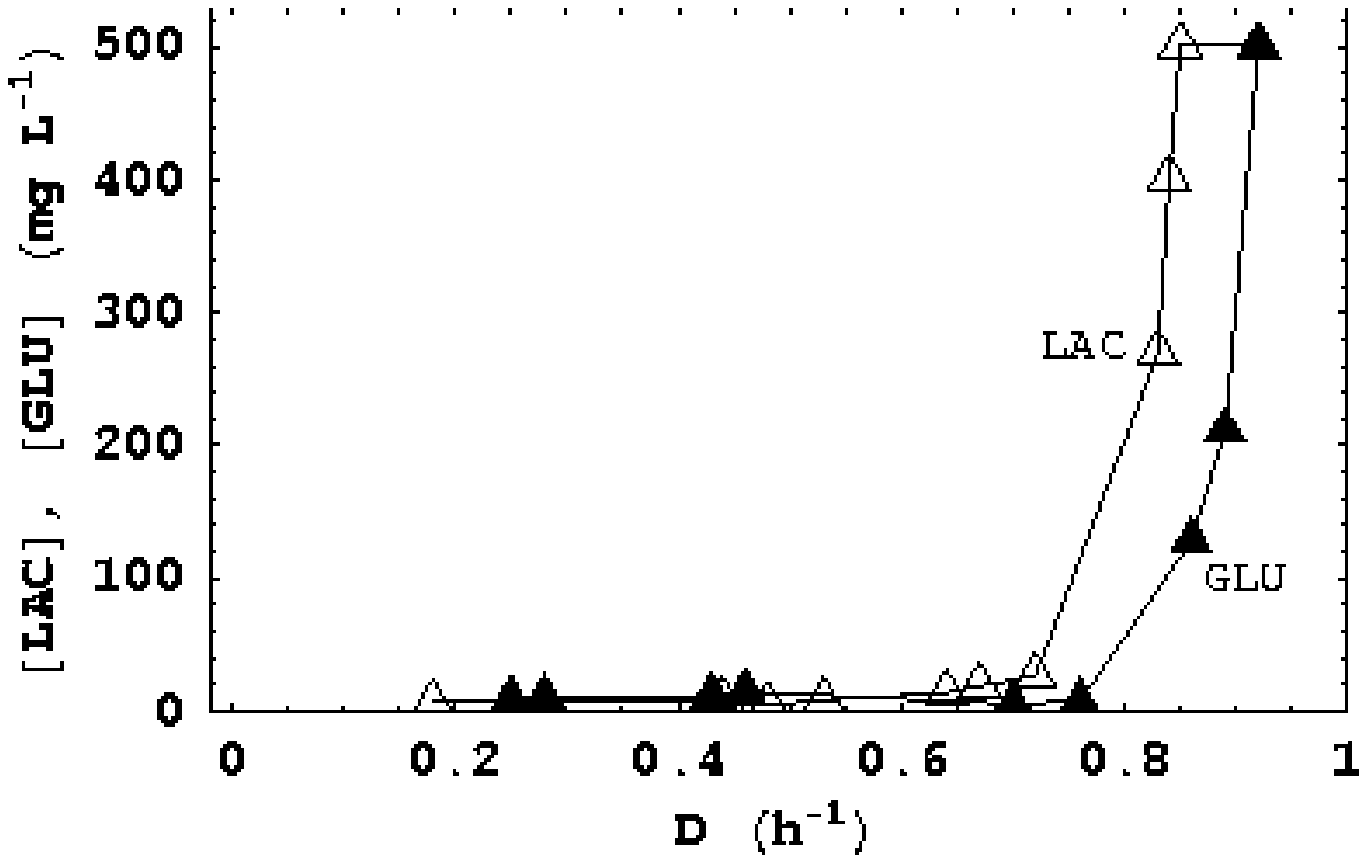}
\par\end{centering}

\caption{\label{fig:MSgrowthPatterns}Simultaneous and preferential growth
patterns in continuous cultures. (a)~During growth of \emph{H. polymorpha}
on a mixture of xylose and glycerol, both substrates are consumed
at all dilution rates up to washout~\citep[calculated from Fig.~3 of][]{brinkmann92}.
Closed and open symbols show the substrate concentrations during single-
and mixed-substrate growth, respectively. (b)~During growth of \emph{E.
coli} B on a mixture of glucose and lactose, consumption of lactose
ceases at a dilution rate below the (washout) dilution rate at which
consumption of glucose ceases~\citep[Fig.~3]{silver69}.}

\end{figure}

\begin{figure}
\noindent \begin{centering}
\subfigure[]{\includegraphics[width=2.6in]{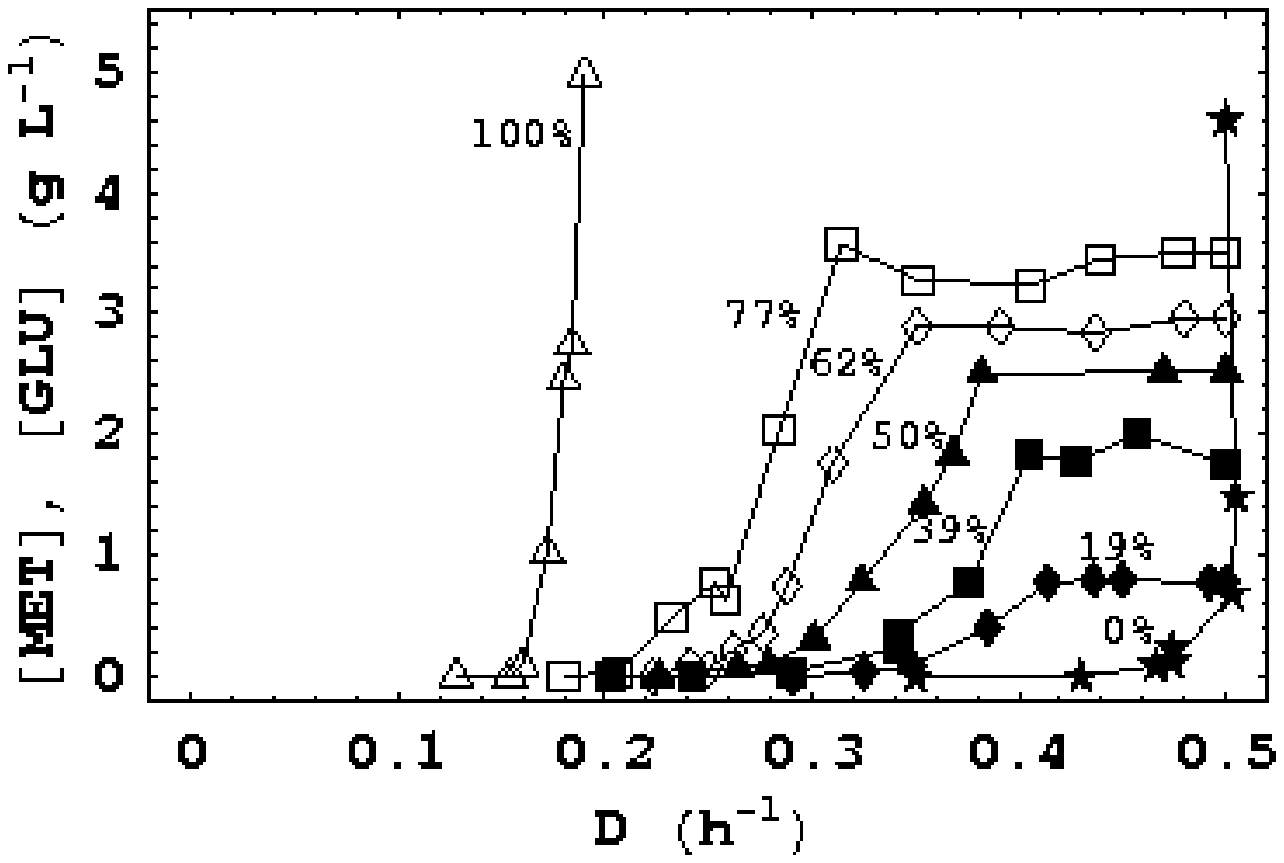}}\hspace*{0.1in}\subfigure[]{\includegraphics[width=2.6in]{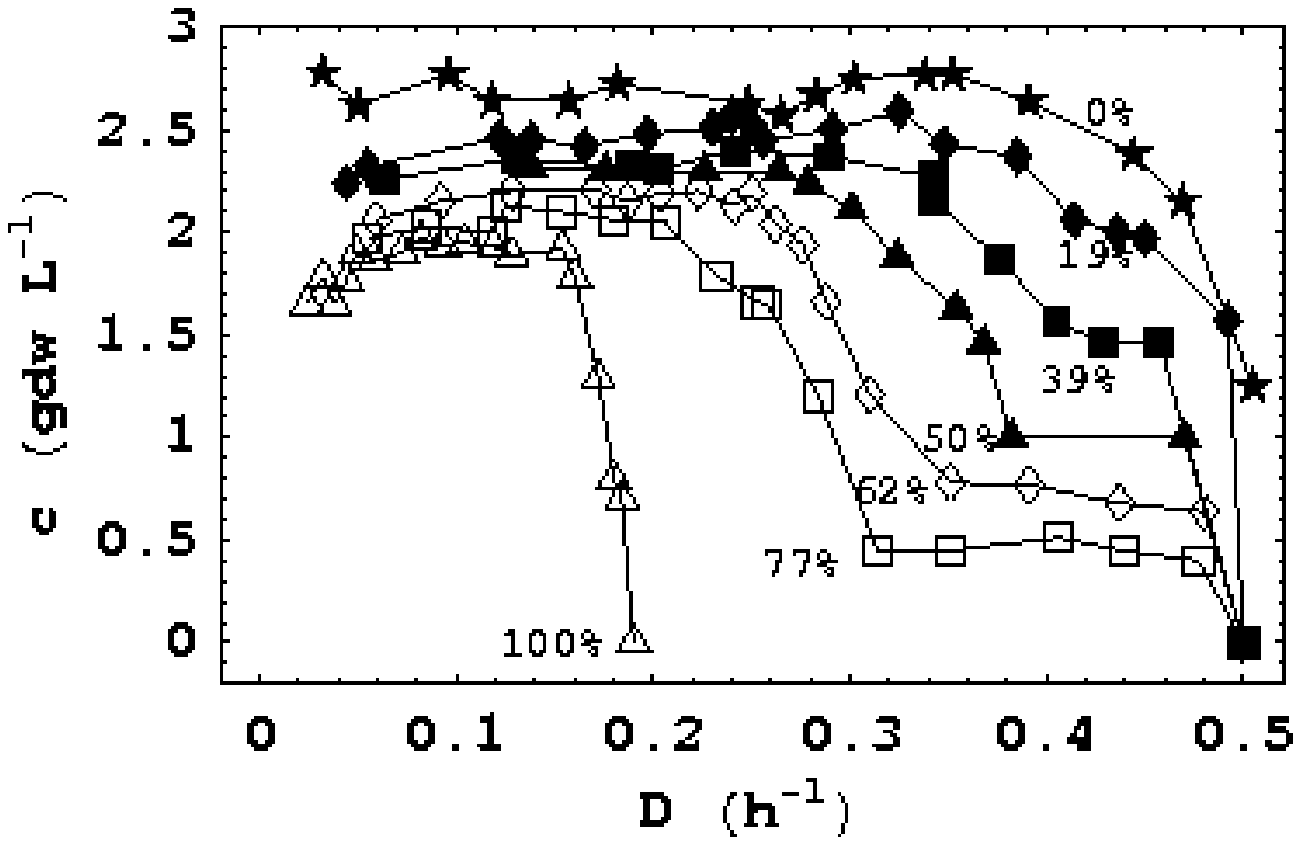}}
\par\end{centering}

\caption{\label{f:GlucoseMethanol}Variation of the substrate concentrations
and cell density during mixed-substrate growth of \emph{H. polymorpha}
on a mixture of glucose and methanol~\citep[Fig.~2]{egli86b}. The
total feed concentration of glucose and methanol was 5 g~L$^{-1}$
in all the experiments. The percentages show the mass percent of methanol
in the feed. The critical dilution rates on pure methanol (curve labeled
100\%) and pure glucose (curve labeled 0\%) are $\sim$0.2 and $\sim$0.5~h$^{-1}$,
respectively.}

\end{figure}

In chemostats limited by pairs of carbon sources, two types of steady
state profiles have been observed. Importantly, these profiles correlate
with, and in fact, can be predicted by, the substrate consumption
pattern observed in substrate-excess batch cultures. Indeed, pairs
of substrates consumed simultaneously in substrate-excess batch cultures
are consumed simultaneously in continuous cultures at all dilution
rates up to washout~(Fig.~\ref{fig:MSgrowthPatterns}a). In contrast,
pairs of substrates that show diauxic growth in substrate-excess batch
cultures are consumed simultaneously in continuous cultures only if
the dilution rate is sufficiently small. For instance, during growth
of \emph{Escherichia coli} B on a mixture of glucose and lactose,
consumption of the {}``less preferred'' substrate, lactose, declines
sharply at an intermediate dilution rate, beyond which only glucose
is consumed (Fig.~\ref{fig:MSgrowthPatterns}b). This growth pattern
has been observed in several other systems~\citep[reviewed in][]{harder76,harder82},
but the most comprehensive data was obtained in studies of the methylotrophic
yeasts, \emph{Hansenula polymorpha} and \emph{Candida boidinii}, with
mixtures of methanol + glucose~\citep{egli80,egli82a,egli82b,egli86b}.%
\footnote{We shall constantly appeal to this extensive data on the growth of
methylotrophic yeasts. A detailed exposition of the metabolism and
gene regulation in these organisms can be found in a recent review~\citep{Hartner2006}.%
} These studies, which were performed with various feed concentrations
of glucose and methanol, revealed several well-defined patterns:

\begin{enumerate}
\item The sharp decline in methanol consumption at an intermediate dilution
rate is analogous to the phenomenon of diauxic growth in batch cultures
inasmuch as it is triggered by a precipitous drop in the activities
of the peripheral enzymes for methanol, the {}``less preferred''
substrate. The \emph{transition} dilution rate was empirically defined
by Egli as the dilution rate at which the concentration of the {}``less
preferred substrate'' achieves a sufficiently high value, e.g., half
of the feed concentration~\citep[Fig.~1]{egli86b}.
\item The transition dilution rate is always higher than the critical dilution
rate on methanol (Fig.~\ref{f:GlucoseMethanol}a). This was referred
to as the \emph{enhanced growth rate effect,} since methanol was consumed
at dilution rates significantly higher than the critical dilution
rate in cultures fed with pure methanol.
\item The transition dilution rate varies with the feed composition ---
the larger the fraction of methanol in the feed, the smaller the transition
dilution rate (Fig.~\ref{f:GlucoseMethanol}a).
\end{enumerate}
These and several other general trends, discussed later in this work,
have been observed in a wide variety of bacteria and yeasts~\citep[reviewed in][]{egli95,kovarova98}.

The occurrence of the very same growth patterns in such diverse organisms
suggests that they are driven by a universal mechanism common to all
species. Thus, we are led to consider the minimal model, which accounts
for only those processes --- induction and growth --- that occur in
all microbes. In earlier work, computational studies of a variant
of the minimal model showed that it captures all the growth patterns
discussed above~\citep{narang98a}. Here, we perform a rigorous bifurcation
analysis of the model to obtain a complete classification of the growth
patterns at any given dilution rate and feed concentrations. The analysis
reveals new features, such as threshold effects, that were missed
in our earlier work. It also provides simple explanations for the
following questions:

\begin{enumerate}
\item During single-substrate growth of wild-type cells, why does the activity
of the peripheral enzymes pass through a maximum?
\item During growth on mixtures of substrates, what are conditions under
which (a)~both substrates are consumed at all dilution rates up to
washout (Fig.~\ref{fig:MSgrowthPatterns}a), and (b)~consumption
of one of the substrates ceases at an intermediate dilution rate (Fig.~\ref{fig:MSgrowthPatterns}b)?
\item Why does the transition dilution rate exist? Why is it always larger
than the critical dilution rate of the {}``less preferred'' substrate,
and why does it vary with the feed concentrations (Fig.~\ref{f:GlucoseMethanol}a)?
\end{enumerate}
Finally, we show that under the conditions typically used in experiments,
the physiological (intracellular) steady states are completely determined
by only two parameters, namely, the dilution rate and the mass fraction
of the substrates in the feed (rather than the feed concentrations
\emph{per se}). In fact, we derive explicit expressions for the steady
state substrate concentrations, cell density, and enzyme levels, which
provide physical insight into the variation of these quantities with
the dilution rate and feed concentrations.

\section{Theory}

\begin{figure}
\begin{centering}
\includegraphics[width=8cm,height=6cm]{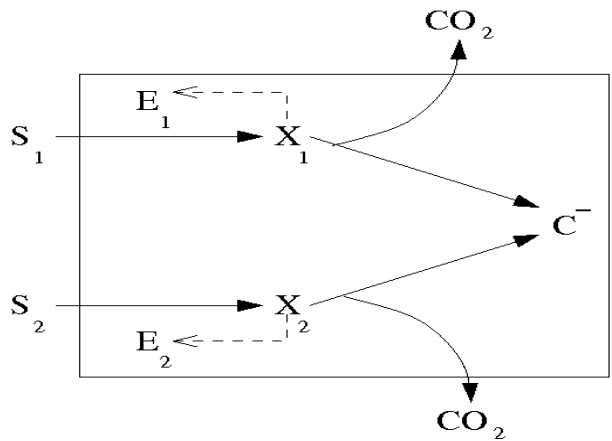}
\par\end{centering}

\caption{\label{f:Scheme}Kinetic scheme of the minimal model~\citep{narang98b}.}

\end{figure}

Fig.~\ref{f:Scheme} shows the kinetic scheme of the minimal model.
Here, $S_{i}$ denotes the $i^{{\rm th}}$ exogenous substrate, $E_{i}$
denotes the {}``lumped'' peripheral enzymes for $S_{i}$, $X_{i}$
denotes internalized $S_{i}$, and $C^{-}$ denotes all intracellular
components except $E_{i}$ and $X_{i}$ (thus, it includes precursors,
free amino acids, and macromolecules).

We assume that:

\begin{enumerate}
\item The concentrations of the intracellular components, denoted $e_{i}$,
$x_{i}$, and $c^{-}$, are based on the dry weight of the cells (g
per g dry weight of cells, i.e., g~gdw$^{-1}$). The concentrations
of the exogenous substrate and cells, denoted $s_{i}$ and $c$, are
based on the volume of the reactor (g~L$^{-1}$ and gdw~L$^{-1}$,
respectively). The rates of all the processes are based on the dry
weight of the cells (g~gdw$^{-1}$~h$^{-1}$). We shall use the
term \emph{specific rate} to emphasize this point.\\
The choice of these units implies that if the concentration of any
intracellular component, $Z$, is $z$ g~gdw$^{-1}$, then the evolution
of $z$ in a continuous culture operating at dilution rate, $D$,
is given by\[
\frac{dz}{dt}=r_{z}^{+}-r_{z}^{-}-\left(D+\frac{1}{c}\frac{dc}{dt}\right)z,\]
where $r_{z}^{+}$ and $r_{z}^{-}$ denote the specific rates of synthesis
and degradation of $Z$, and $Dz$ is the specific rate of efflux
of $Z$ from the chemostat. It is shown below that the sum, $Dz+(1/c)(dc/dt)z$,
is precisely the rate of dilution of $Z$ due to growth.
\item The transport and peripheral catabolism of $S_{i}$ is catalyzed by
a unique system of peripheral enzymes, $E_{i}$. The specific uptake
rate of $S_{i}$, denoted $r_{s,i}$, follows the modified Michaelis-Menten
kinetics \[
r_{s,i}\equiv V_{s,i}e_{i}\frac{s_{i}}{K_{s,i}+s_{i}},\]
where $V_{s,i}$ is a fixed constant, i.e., inducer exclusion is negligible.
\item Part of the internalized substrate, denoted $X_{i}$, is converted
to $C^{-}$. The remainder is oxidized to ${\rm CO_{2}}$ in order
to generate energy.

\begin{enumerate}
\item The conversion of $X_{i}$ to $C^{-}$ and ${\rm CO_{2}}$ follows
first-order kinetics, i.e., \[
r_{x,i}\equiv k_{x,i}x_{i}.\]

\item The fraction of $X_{i}$ converted to $C^{-}$, denoted $Y_{i}$,
is constant. It is shown below that $Y_{i}$ is essentially identical
to the yield of biomass on $S_{i}$. This assumption is therefore
tantamount to assuming that the yield of biomass on a substrate is
unaffected by the presence of another substrate in the medium.
\end{enumerate}
\item The internalized substrate also induces the synthesis of $E_{i}$,
which is assumed to controlled entirely by the initiation of transcription
(i.e., mechanisms such as attenuation and proteolysis are neglected).

\begin{enumerate}
\item The specific synthesis rate of $E_{i}$ follows the hyperbolic Yagil
\& Yagil kinetics~\citep{yagil71}, \begin{equation}
r_{e,i}^{+}\equiv V_{e,i}\frac{1+x_{i}}{1+\alpha_{i}+x_{i}},\label{eq:InductionRate}\end{equation}
where $V_{e,i}$ is a fixed constant, i.e., catabolite repression
is negligible. In the particular case of negatively controlled operons
(modulated by repressors), $V_{e,i}$ and $\alpha_{i}$ can be expressed
in terms of parameters associated with kinetics of RNAP-promoter and
repressor-operator binding, respectively. In other cases, \eqref{eq:InductionRate}
must be viewed as a phenomenological description of the induction
kinetics.
\item Enzyme degradation is a first-order process, i.e., the specific rate
of enzyme degradation is \[
r_{e,i}^{-}=k_{e,i}e_{i}.\]

\item The enzymes are synthesized at the expense of $C^{-}$, and their
degradation produces $C^{-}$.
\end{enumerate}
\end{enumerate}
Given these assumptions, the mass balances yield the equations\begin{align}
\frac{ds_{i}}{dt} & =D\left(s_{f,i}-s_{i}\right)-r_{s,i}c,\label{eq:sO}\\
\frac{dx_{i}}{dt} & =r_{s,i}-r_{x,i}-\left(D+\frac{1}{c}\frac{dc}{dt}\right)x_{i},\label{eq:xO}\\
\frac{de_{i}}{dt} & =r_{e,i}^{+}-r_{e,i}^{-}-\left(D+\frac{1}{c}\frac{dc}{dt}\right)e_{i},\label{eq:eO}\\
\frac{dc^{-}}{dt} & =\left(Y_{1}r_{x,1}+Y_{2}r_{x,2}\right)-\left(r_{e,1}^{+}-r_{e,1}^{-}\right)-\left(r_{e,2}^{+}-r_{e,2}^{-}\right)-\left(D+\frac{1}{c}\frac{dc}{dt}\right)c^{-},\label{eq:cMO}\end{align}
where $s_{f,i}$ denotes the concentration of $S_{i}$ in the feed
to the chemostat.

Following Fredrickson, we observe that eqs.~\eqref{eq:xO}--\eqref{eq:cMO}
implicitly define the specific growth rate and the evolution of the
cell density~\citep{fredrickson76}. Indeed, since $x_{1}+x_{2}+e_{1}+e_{2}+c^{-}=1$,
addition of~\eqref{eq:xO}--\eqref{eq:cMO} yields\begin{equation}
0=r_{g}-\left(D+\frac{1}{c}\frac{dc}{dt}\right)\Leftrightarrow\frac{dc}{dt}=\left(r_{g}-D\right)c,\label{eq:cO}\end{equation}
where\begin{equation}
r_{g}\equiv\sum_{i=1}^{2}r_{s,i}-\sum_{i=1}^{2}(1-Y_{i})r_{x,i}\label{eq:rG}\end{equation}
is the specific growth rate. As expected, it is the net rate of substrate
accumulation in the cells. It follows from \eqref{eq:cO} that the
last term in eqs.~\eqref{eq:xO}--\eqref{eq:cMO} is precisely the
dilution rate of the corresponding physiological (intracellular) variables.

The equations can be simplified further because $k_{x,i}^{-1}$, the
turnover time for $X_{i}$, is small (on the order of seconds to minutes).
Thus, $x_{i}$ rapidly attains quasisteady state, resulting in the
equations\begin{align}
\frac{ds_{i}}{dt} & =D\left(s_{f,i}-s_{i}\right)-r_{s,i}c,\label{eq:ContS}\\
\frac{de_{i}}{dt} & =r_{e,i}^{+}-\left(r_{g}+k_{e,i}\right)e_{i},\label{eq:ContE}\\
\frac{dc}{dt} & =\left(r_{g}-D\right)c,\label{eq:ContC}\\
x_{i} & \approx\frac{V_{s,i}e_{i}s_{i}/\left(K_{s,i}+s_{i}\right)}{k_{x,i}},\label{eq:ContX}\\
r_{g} & \approx Y_{1}r_{s,1}+Y_{2}r_{s,2},\label{eq:ContRg}\end{align}
where \eqref{eq:ContX}--\eqref{eq:ContRg} follow from the quasisteady
state relation, $0\approx r_{s,i}-r_{x,i}$.

It is worth noting that:

\begin{enumerate}
\item The parameter, $Y_{i}$, which was defined as the fraction of $X_{i}$
converted to $C^{-}$, is essentially the yield of biomass on $S_{i}$.
Indeed, substituting the relation, $r_{x,i}\approx r_{s,i}$ in \eqref{eq:rG}
shows that after quasisteady state has been attained, $Y_{i}$ approximates
the fraction of $S_{i}$ that accumulates in the cell.
\item Although we have assumed that $Y_{i}$ is constant, the validity of
\eqref{eq:ContRg} does not rest upon this assumption. It is true
even if the yields are variable (i.e., depend on the physiological
state).
\item Enzyme induction is subject to positive feedback. This becomes evident
if \eqref{eq:ContX} is substituted in \eqref{eq:InductionRate}:
The quasisteady state induction rate, $r_{e,i}^{+}$, is an increasing
function of $e_{i}$. The positive feedback reflects the cyclic structure
of enzyme synthesis that is inherent in Fig.~\ref{f:Scheme}: $E_{i}$
promotes the synthesis of $X_{i}$, which, in turn, stimulates the
induction of $E_{i}$.
\end{enumerate}
We shall appeal to these facts later.

The equations for batch growth are obtained from the above equations
by letting $D=0$. In what follows, we shall be particularly concerned
with the initial evolution of the enzymes in experiments performed
with sufficiently \emph{small} inocula. Under this condition, the
substrate consumption for the first few hours is so small that the
substrate concentrations remain essentially constant at their initial
levels, $s_{0,i}$. In the face of this quasiconstant environment,
the enzyme activities move toward a quasisteady state corresponding
to exponential (balanced) growth. This motion is given by the equations\begin{align}
\frac{de_{i}}{dt} & =r_{e,i}^{+}-\left(r_{g}+k_{e,i}\right)e_{i},\label{eq:BatchReducedE}\\
x_{i} & \approx\frac{V_{s,i}e_{i}s_{0,i}/\left(K_{s,i}+s_{0,i}\right)}{k_{x,i}},\label{eq:BatchReducedX}\\
r_{g} & \approx Y_{1}V_{s,1}e_{1}\frac{s_{0,1}}{K_{s,1}+s_{0,1}}+Y_{2}V_{s,2}e_{2}\frac{s_{0,2}}{K_{s,2}+s_{0,2}},\label{eq:BatchReducedC}\end{align}
obtained by letting $s_{i}\approx s_{0,i}$ in eqs.~\eqref{eq:ContX}--\eqref{eq:ContRg}.
The steady state(s) of these equations represent the quasisteady state
enzyme activities attained during the first exponential growth phase.

We note finally that the cells sense their environment through the
ratio,\[
\sigma_{i}\equiv\frac{s_{i}}{K_{s,i}+s_{i}},\]
which characterizes the extent to which the permease is saturated
with $S_{i}$. Not surprisingly, we shall find later on that the physiological
steady states depend on the ratio, $\sigma_{i}$, rather than the
substrate concentrations \emph{per se}. It is, therefore, convenient
to define the ratios\[
\sigma_{0,i}\equiv\frac{s_{0,i}}{K_{s,i}+s_{0,i}},\;\sigma_{f,i}\equiv\frac{s_{f,i}}{K_{s,i}+s_{f,i}}.\]
For a given cell type, $\sigma_{i}$, $\sigma_{0,i}$, $\sigma_{f,i}$
are increasing functions of $s_{i}$, $s_{i,0}$, $s_{f,i}$, respectively,
and can be treated as surrogates for the corresponding substrate concentrations.
In what follows, we shall frequently refer to $\sigma_{i}$, $\sigma_{0,i}$,
and $\sigma_{f,i}$ as the substrate concentration, initial substrate
concentration, and feed concentration, respectively.

\section{Results}

\subsection{\label{sub:Regulation}The role of regulation}

Before analyzing the foregoing equations for batch and continuous
cultures, we pause to consider the role of regulatory mechanisms.
This is important because the literature places great emphasis on
the regulatory mechanisms, whereas the model neglects them completely
($V_{s,i}$ and $V_{e,i}$ are assumed to be constants). It is therefore
relevant to ask: To what extent do the regulatory mechanisms control
gene expression? To address this question, we shall focus on the well-characterized
\emph{lac} operon in \emph{E. coli}. The analysis shows that cAMP
activation and inducer exclusion cannot explain the strong repression
observed in experiments.

\subsubsection{cAMP activation}

\begin{figure}
\noindent \begin{centering}
\subfigure[]{\includegraphics[width=2.6in]{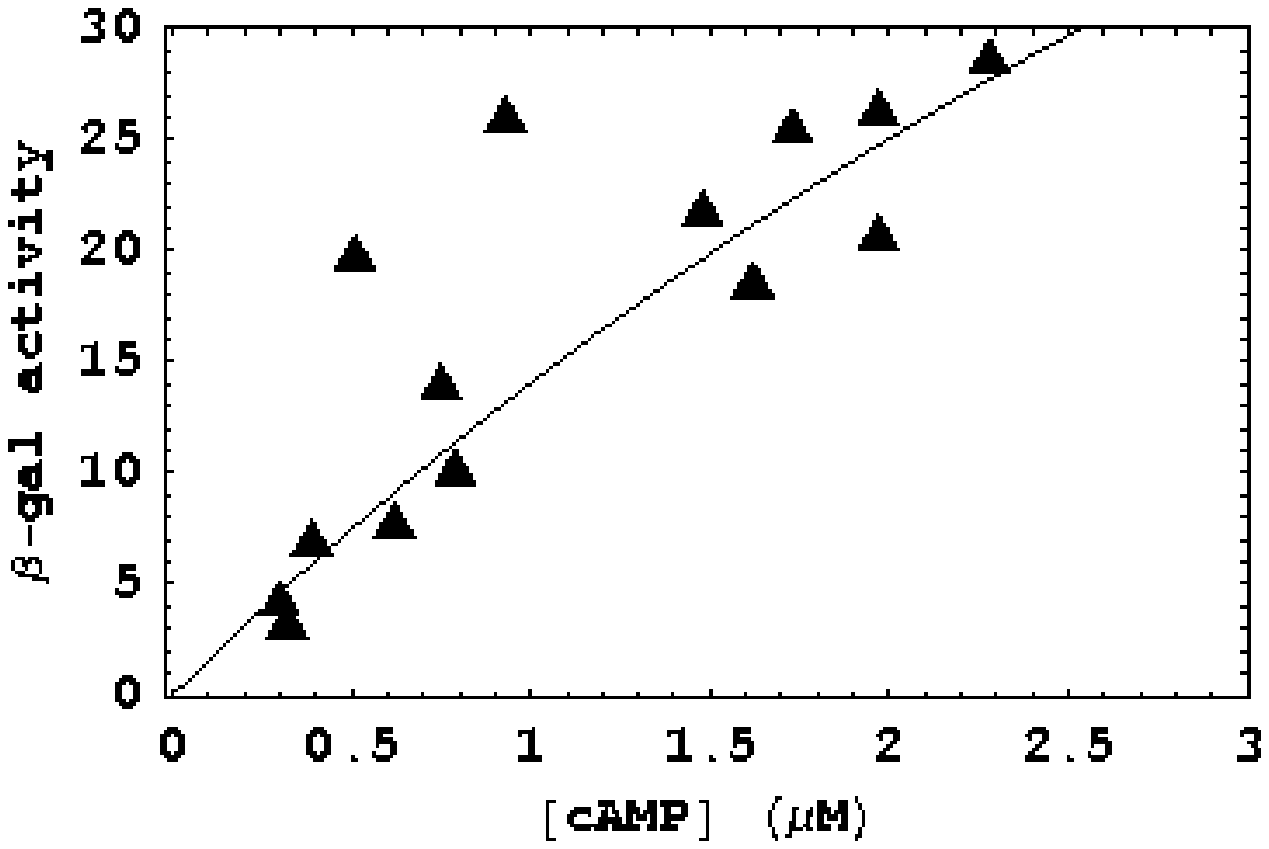}}\hspace*{0.1in}\subfigure[]{\includegraphics[width=2.6in]{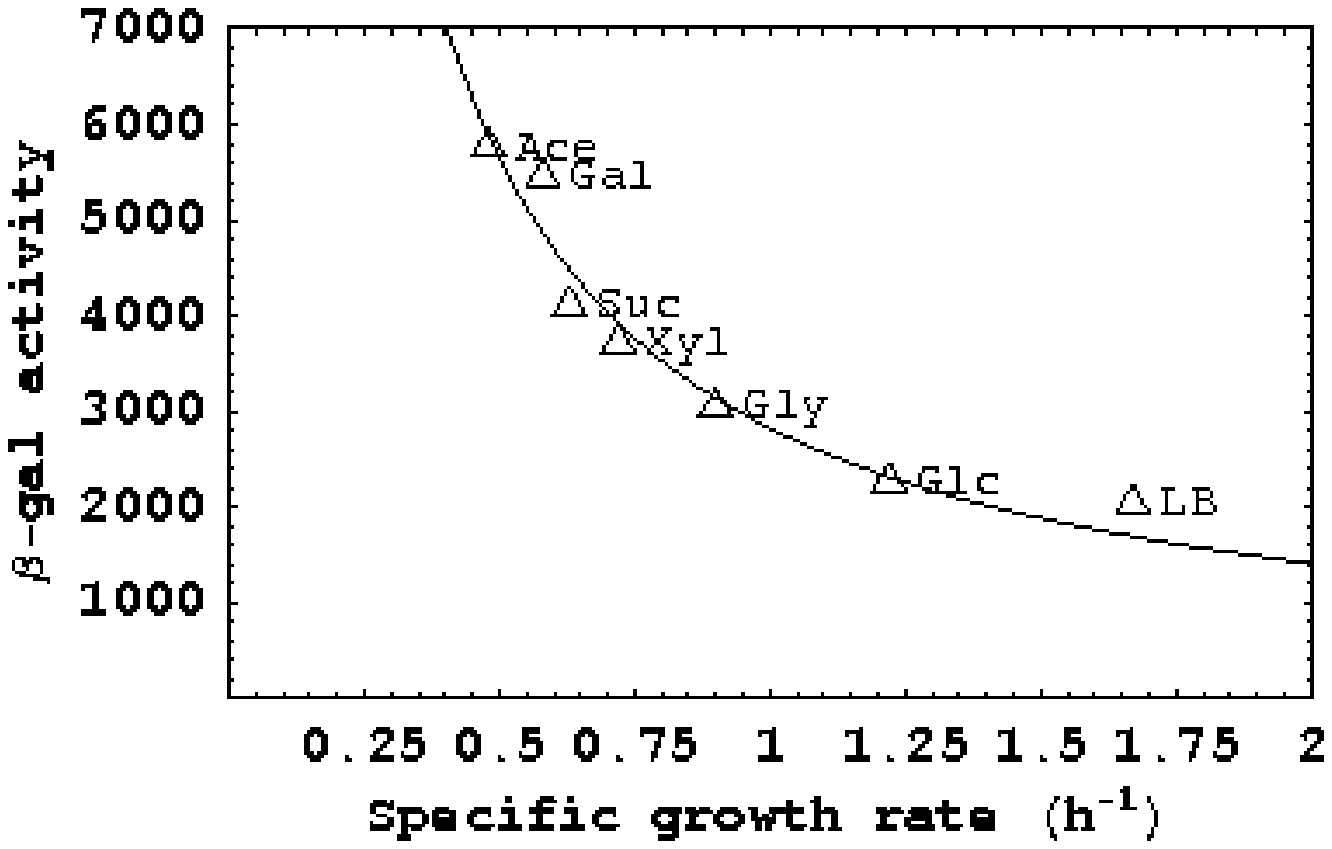}}
\par\end{centering}

\noindent \begin{centering}
\subfigure[]{\includegraphics[width=2.6in]{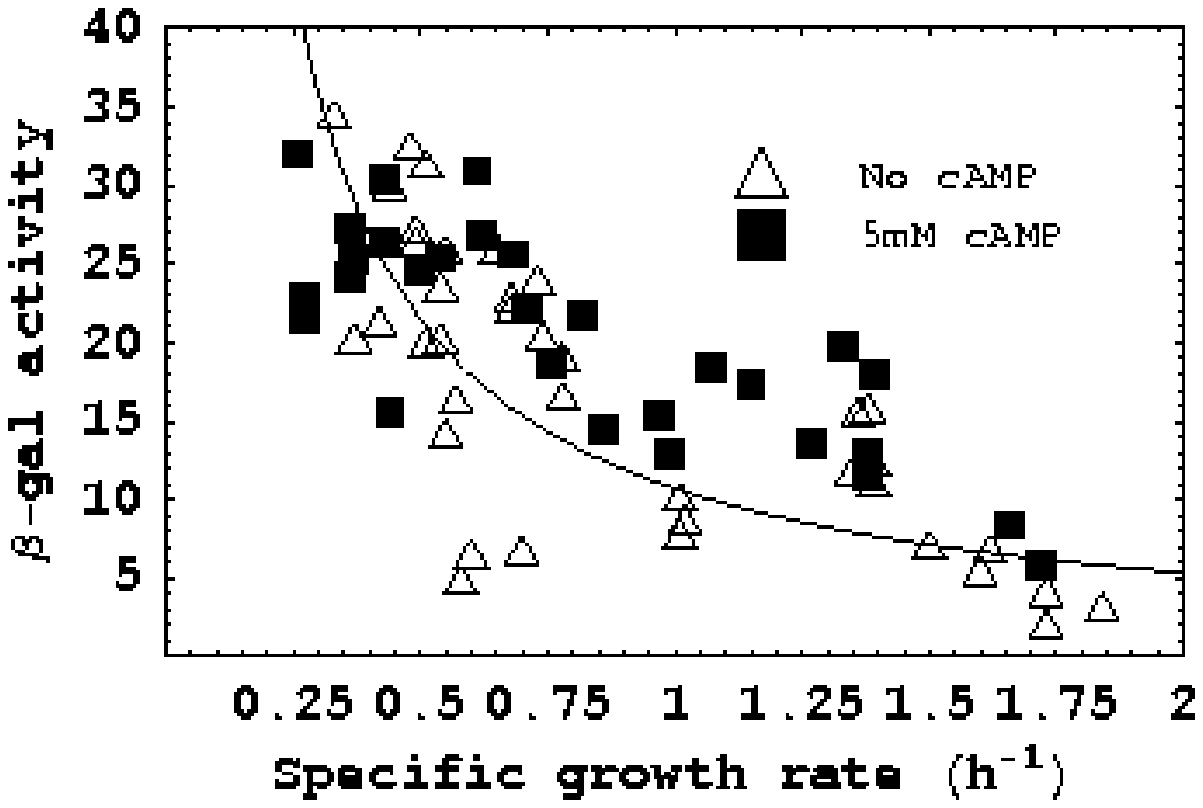}}\hspace*{0.1in}\subfigure[]{\includegraphics[width=2.6in]{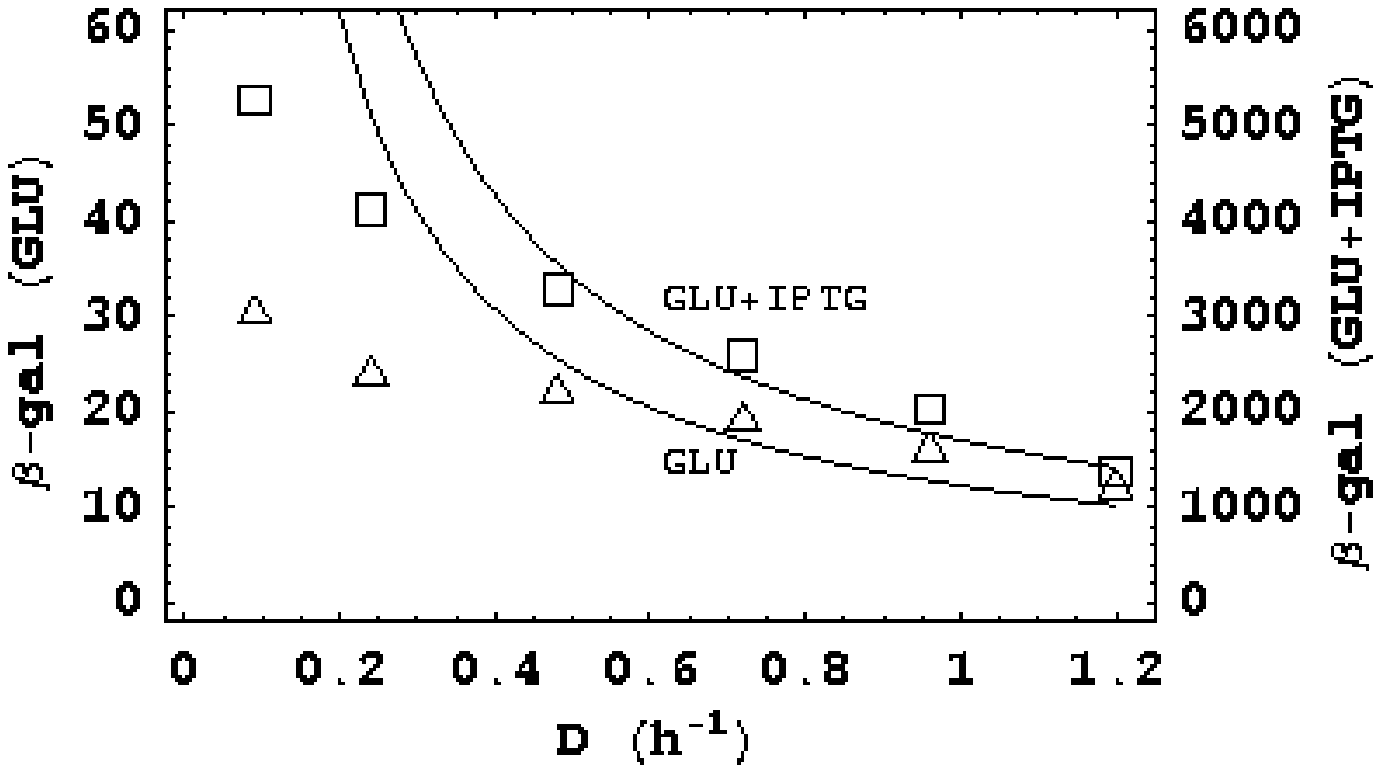}}
\par\end{centering}

\caption{\label{fig:RoleOfcAMP}The relative roles of dilution and cAMP activation
in \emph{lac} induction. (a)~Variation of the $\beta$-galactosidase
activity with intracellular cAMP level during exponential growth of
the \emph{lac}-constitutive strain \emph{E. coli} X2950 on various
carbon sources~\citep[Fig.~2]{Epstein1975}. (b,c)~During batch
growth of (b) \emph{E. coli} MC4100 $\lambda$CPT100~\citep[Fig.~2]{Kuo2003}
and (c) \emph{E. coli} NC3~\citep[Fig.~1]{Wanner1978} on various
carbon sources, the steady state $\beta$-galactosidase activity is
inversely proportional to the specific growth rate on the carbon source.
Moreover, the $\beta$-galactosidase activity increases at most 2-fold
if 5mM cAMP is added to the medium ($\blacksquare$). (d)~The inverse
relationship is satisfied at $D\gtrsim0.4$~h$^{-1}$ in continuous
cultures of \emph{E. coli} K12 W3110 grown on glucose and glucose
+ 1mM IPTG~\citep[Fig.~1]{Kuo2003}. The curves in (b)--(c) and (d)
show the fits to \eqref{eq:eConstitutiveBatch} and \eqref{eq:eConstitutiveChemostat1},
respectively.}

\end{figure}

Interest in cAMP as a \emph{lac} regulator began when it was discovered
that the \emph{lac} induction rate increased $\sim$2-fold upon addition
of cAMP to glucose-limited cultures~\citep{Perlman1968,Ullmann1968}.
However, this remained a hypothesis until Epstein et al showed that
the \emph{lac} induction rate increased with the intracellular cAMP
levels (Fig.~\ref{fig:RoleOfcAMP}a). They obtained this relationship
by measuring the $\beta$-galactosidase activities and intracellular
cAMP levels during exponential growth of the \emph{lac}-constitutive
strain \emph{E. coli} X2950 on a wide variety of carbon sources. The
use of a constitutive strain served to eliminate inducer exclusion:
In such cells, the repressor-operator binding is impaired so severely
that $\alpha_{i}\approx0$, and the induction rate is unaffected by
the inducer level ($r_{e,i}^{+}\approx V_{e,i}$). The diverse carbon
sources provided a way of varying the intracellular cAMP levels: In
general, the lower the specific growth rate on a substrate, the higher
the intracellular cAMP level~\citep{Buettner1973}.

Although Fig.~\ref{fig:RoleOfcAMP}a shows a correlation between
the cAMP level and the \emph{lac} induction rate, it does not prove
that there is a causal relationship between them. Indeed, since protein
degradation is negligible in substrate-excess batch cultures~\citep{Mandelstam1958},
\eqref{eq:BatchReducedE} yields\begin{equation}
e_{i}\approx\frac{r_{e,i}^{+}}{r_{g}}\approx\frac{V_{e,i}}{r_{g}}.\label{eq:eConstitutiveBatch}\end{equation}
Thus, the steady state enzyme level depends on the ratio of the specific
induction and growth rates. Since the substrates that yield high intracellular
cAMP levels also support low specific growth rates, it is conceivable
that Fig.~\ref{fig:RoleOfcAMP}a reflects the variation of the specific
growth rate, $r_{g}$, rather than the specific induction rate, $V_{e,i}$.

Eq.~\eqref{eq:eConstitutiveBatch} provides a simple criterion for
checking if the specific induction rate varied in the experiments:
Evidently, $V_{e,i}$ varies if and only if $e_{i}$ is \emph{not}
inversely proportional to $r_{g}$. Unfortunately, we cannot check
the data in Fig.~\ref{fig:RoleOfcAMP}a with this criterion, since
the authors did not report the specific growth rates of the carbon
sources. However, the very same experiments have been performed by
others~\citep{Kuo2003,Wanner1978}, and this data shows that the
$\beta$-galactosidase activity is inversely proportional to $r_{g}$
(Figs.~\ref{fig:RoleOfcAMP}b,c). It follows that in Figs.~\ref{fig:RoleOfcAMP}b
and \ref{fig:RoleOfcAMP}c, the $\beta$-galactosidase activity varies
almost entirely due to dilution. The influence of cAMP on \emph{lac}
transcription is very modest, a fact that is further confirmed by
the very marginal (<2-fold) improvement of the $\beta$-galactosidase
activity in the presence of 5mM cAMP (Figs.~\ref{fig:RoleOfcAMP}c).
The same is likely to be true of the data in Fig.~\ref{fig:RoleOfcAMP}a.
Thus, the data from \citealp{Epstein1975}, which is extensively cited
to support the existence of cAMP-mediated \emph{lac} control, forces
upon us just the opposite conclusion: cAMP has virtually no effect
on the \emph{lac} induction rate.

The data from continuous cultures also implies the absence of significant
cAMP activation. Indeed, \eqref{eq:ContE} implies that when \emph{lac}-constitutive
mutants of \emph{E. coli} are grown in continuous cultures, the steady
state $\beta$-galactosidase activity at sufficiently high dilution
rates is given by the relation\begin{equation}
e_{i}\approx\frac{V_{e,i}}{D}.\label{eq:eConstitutiveChemostat1}\end{equation}
This relationship also holds for wild-type cells if the medium contains
saturating concentrations of the gratuitous inducer, IPTG. In both
cases, cAMP control is significant only if $e_{i}$ is not inversely
proportional to $D$. But experiments performed with \emph{lac}-constitutive
mutants (Fig.~\ref{fig:NonInducibleEnzymes}a) and wild-type cells
exposed to saturating IPTG levels (Fig.~\ref{fig:RoleOfcAMP}d) show
that $\beta$-galactosidase activity is inversely proportional to
$D$ for sufficiently large $D$.%
\footnote{At small $D$, the enzyme activity is significantly \emph{lower} than
that predicted by \eqref{eq:eConstitutiveChemostat1}. Evidently,
this discrepancy cannot be due to cAMP activation. It cannot be resolved
by enzyme degradation either: Although the data can be fitted to the
equation, $e_{i}=V_{e,i}/(D+k_{e,i})$, the values of $k_{e,i}$ required
for these fits are physiologically implausible ($\sim$0.5~h$^{-1}$).
Based on the detailed study of the \emph{ptsG} operon at various dilution
rates, the discrepancy is probably due to enhanced expression of the
starvation sigma factor, RpoS, at low dilution rates~\citep[Fig.~3]{Seeto2004}%
}

Catabolite repression also appears to be insignificant in cell types
other than \emph{E. coli}. The amidase activity is inversely proportional
to $D$ in the constitutive mutant \emph{P. aeroginosa} C11 (Fig.~\ref{fig:NonInducibleEnzymes}a).
Moreover, this declining trend is not due to catabolite repression,
as postulated by Clarke and coworkers --- it reflects the effect of
dilution. Likewise, the activity of alcohol oxidase is inversely proportional
to $D$ when \emph{H. polymorpha} and \emph{C. boidinii} are grown
in glucose-limited chemostats. In this case, \eqref{eq:ContE} implies
that at sufficiently large dilution rates\begin{equation}
e_{i}=\frac{V_{e,i}}{1+\alpha_{i}}\frac{1}{D},\label{eq:eConstitutiveChemostat2}\end{equation}
which is formally similar to \eqref{eq:eConstitutiveChemostat1}.

\subsubsection{Inducer exclusion}

The data shows that inducer exclusion certainly exists. When glucose
or glucose-6-phosphate are added to a culture growing exponentially
on lactose, the specific uptake rate of lactose declines $\sim$2-fold~\citep[Figs.~3,5]{McGinnis1969}.
Likewise, in the presence of glucose, the intracellular inducer concentration,
as measured by the accumulation of radioactively labeled TMG, also
decreases $\sim$2-fold~\citep[Fig.~1]{Winkler1967}. But this modest
decrease cannot account for the dramatic repression of the \emph{lac}
operon during diauxic growth: The addition of glucose-6-phosphate
to a culture growing on lactose decreases the activity of $\beta$-galactosidase
$\sim$300-fold after only 4 generations of growth~\citep[Table~1]{Hogema1998a}.

We conclude that cAMP activation and inducer exclusion do influence
the \emph{lac} induction rate, but their effect is not commensurate
with the experimentally observed repression. We show below that the
minimal model predicts complete transcriptional repression, despite
the absence of any regulatory mechanism.

\subsection{\label{sec:SingleSubstrateGrowth}Single-substrate growth}

In wild-type cells, the repressor-operator binding is so tight that
$\alpha_{i}$ is large compared to 1. This is true even in the case
of glucose ($\alpha_{i}$=5--10), which is often treated as a substrate
consumed by constitutive enzymes~\citep[Table~3]{Notley-McRobb2000b}.
Under these conditions, the basal induction rate is small compared
to the fully induced induction rate. Hence, at all but the smallest
inducer levels, the specific induction rate can be approximated by
the expression\begin{equation}
V_{e,i}\frac{x_{i}}{K_{e,i}+x_{i}},\; K_{e,i}\equiv\frac{\alpha_{i}}{K_{x,i}},\label{eq:ApproxInductionRate}\end{equation}
and the quasisteady state induction rate, obtained by substituting
\eqref{eq:ContX} in \eqref{eq:ApproxInductionRate}, is\[
V_{e,i}\frac{e_{i}\sigma_{i}}{\bar{K}_{e,i}+e_{i}\sigma_{i}},\;\bar{K}_{e,i}\equiv\frac{K_{e,i}k_{x,i}}{V_{s,i}}.\]
It is useful to make this approximation because the equations become
amenable to rigorous analysis and yield simple physical insights.
In the rest of the paper, we shall confine our attention to these
idealized \emph{perfectly inducible} systems.

We begin by considering the growth of cultures limited by a single
substrate, say, $S_{1}$. Under this condition, $s_{2}=e_{2}=0$,
and the steady state values of the remaining variable satisfy the
equations\begin{align}
0 & =\frac{ds_{1}}{dt}=D(s_{f,1}-s_{1})-V_{s,1}e_{1}\sigma_{1}c,\label{eq:SS_s1}\\
0 & =\frac{de_{1}}{dt}=V_{e,1}\frac{e_{1}\sigma_{1}}{\bar{K}_{e,1}+e_{1}\sigma_{1}}-\left(Y_{1}V_{s,1}e_{1}\sigma_{1}+k_{e,1}\right)e_{1},\label{eq:SS_e1}\\
0 & =\frac{dc}{dt}=\left(Y_{1}V_{s,1}e_{1}\sigma_{1}-D\right)c,\label{eq:SS_c}\\
x_{1} & \approx\frac{V_{s,1}e_{1}\sigma_{1}}{k_{x,1}}.\label{eq:SS_x1}\end{align}
These equations have three types of steady states: $e_{1}>0$, $e_{2}=0$,
$c>0$, denoted $E_{101}$; $e_{1}>0$, $e_{2}=0$, $c=0$, denoted
$E_{100}$; and $e_{1}=0$, $e_{2}=0$, $c=0$, denoted $E_{000}$.
The first two steady states correspond to the persistence and washout
steady states of the classical Monod model~\citep{herbert56}. The
third steady state, $E_{000}$, is a consequence of perfect inducibility
--- it exists precisely because the specific induction rate is zero
in the absence of the inducer.

\begin{figure}
\noindent \begin{centering}
\includegraphics[width=3in]{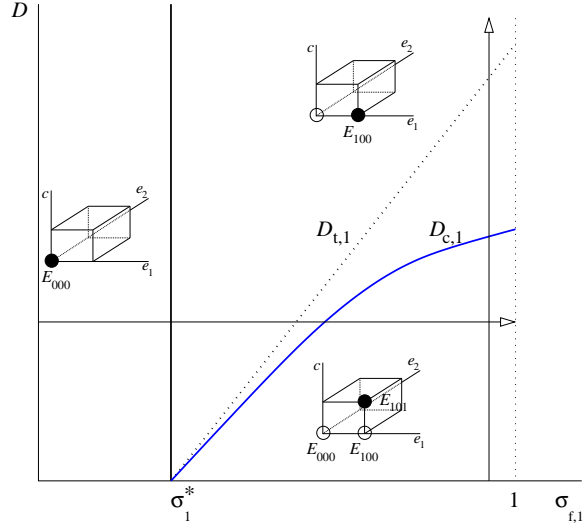}
\par\end{centering}

\caption{\label{fig:BDss}Bifurcation diagram for single-substrate growth on
$S_{1}$. The blue curve denotes the curve of the critical dilution
rate, $D_{c,1}$, and $\sigma_{1}^{*}$ denotes the threshold feed
concentration. The 3-D insets depict the existence and stability of
the steady states at any given $\sigma_{f,1}$ and $D$. Stable and
unstable steady states are represented by full and open circles, respectively.
The dashed line shows the transition dilution rate, $D_{t,1}$, which
plays a crucial role in mixed-substrate growth (Section~\ref{sec:MixedSubstrateGrowth}).
The figure is not drawn to scale: Since $k_{e,1}\sim0.05$~h$^{-1}$
and $V_{e,1}/\bar{K}_{e,1}\sim1$~h$^{-1}$, the threshold, $\sigma_{1}^{*}\equiv k_{e,1}/(V_{e,1}/\bar{K}_{e,1})$,
is on the order of 0.05.}

\end{figure}

Fig.~\ref{fig:BDss} shows the bifurcation diagram for eqs.~\eqref{eq:SS_s1}--\eqref{eq:SS_x1}.
It can be inferred from the following facts derived in Appendix~\ref{a:StabilityContinuous}.

\begin{enumerate}
\item $E_{000}$ always exists, but it is stable if and only if\[
\frac{V_{e,1}\sigma_{f,1}}{\bar{K}_{e,1}}<k_{e,1}\Leftrightarrow\sigma_{f,1}<\sigma_{1}^{*}\equiv\frac{k_{e,1}}{V_{e,1}/\bar{K}_{e,1}}.\]
Thus, growth cannot be sustained in a chemostat if the feed concentration
is below the threshold level, $\sigma_{1}^{*}$. Under this condition
\[
r_{e,1}^{+}=V_{e,1}\frac{e_{1}\sigma_{1}}{\bar{K}_{e,1}+e_{1}\sigma_{1}}\le V_{e,1}\frac{e_{1}\sigma_{f,1}}{\bar{K}_{e,1}+e_{1}\sigma_{f,1}}\le k_{e,1}e_{1},\]
i.e., the induction rate of $E_{1}$ never exceeds the degradation
rate. Hence, as $t\rightarrow\infty$, $e_{1}$ approaches zero, and
$c,\sigma_{1}$ tend toward $0,\sigma_{f,1}$, respectively.
\item $E_{100}$ exists if and only if $\sigma_{f,1}>\sigma_{1}^{*}$, in
which case it is given by the relations, $c=0$, $\sigma_{1}=\sigma_{f,1}$,
and \[
e_{1}=\frac{-\left(\bar{K}_{e,1}+\frac{k_{e,1}}{Y_{1}V_{s,1}}\right)+\sqrt{\left(\bar{K}_{e,1}+\frac{k_{e,1}}{Y_{1}V_{s,1}}\right)^{2}+\frac{4V_{e,1}}{Y_{1}V_{s,1}}\left(\sigma_{f,1}-\sigma_{1}^{*}\right)}}{2\sigma_{f,1}}.\]
It is stable if and only if $D$ exceeds the critical dilution rate
\begin{equation}
D_{c,1}\equiv Y_{1}V_{s,1}e_{1}\sigma_{f,1}.\label{eq:Dc1}\end{equation}
Evidently, $D_{c,1}$ is an increasing function of $\sigma_{f,1}$
that is zero when $\sigma_{f,1}=\sigma_{1}^{*}$. We show below that
$D_{c,1}$ is the maximum specific growth rate that can be sustained
in the chemostat ($r_{g}<D_{c,1})$.\\
At first sight, this steady state poses a paradox: There are no cells
in the chemostat, and yet, these non-existent cells have positive
enzyme levels. The paradox disappears once it is recognized that $E_{100}$
reflects the long-run behavior of a small inoculum introduced into
a sterile chemostat operating at $\sigma_{f,1}>\sigma_{1}^{*}$ and
$D>D_{c,1}$. As $t\rightarrow\infty$, the specific growth rate of
the cells approaches the maximum level, $D_{c,1}$. However, since
$r_{g}<D_{c,1}$, \[
\frac{dc}{dt}=\left(r_{g}-D\right)c<\left(D_{c,1}-D\right)c<0,\]
so that $c\rightarrow0$, $\sigma_{1}\rightarrow\sigma_{f,1}$, and
$e_{1}$ approaches the positive value given above. Thus, in contrast
to $E_{000}$, the cells fail to accumulate despite sustained induction
of the enzymes because the dilution rate is higher than the maximum
specific growth rate that can be attained.
\item $E_{101}$ exists if and only if $\sigma_{f,1}>\sigma_{1}^{*}$ and
$D<D_{c,1}$, in which case it is given by the relations\begin{align}
x_{1} & =\frac{D}{Y_{1}k_{x,1}},\label{eq:E101_x1}\\
e_{1} & =\frac{V_{e,1}}{D+k_{e,1}}\frac{D}{D+Y_{1}V_{s,1}\bar{K}_{e,1}},\label{eq:E101_e1}\\
\sigma_{1} & =\frac{D}{Y_{1}V_{s,1}e_{1}}=\frac{\left(D+k_{e,1}\right)\left(D+Y_{1}V_{s,1}\bar{K}_{e,1}\right)}{Y_{1}V_{s,1}V_{e,1}},\label{eq:E101_s1}\\
c & =Y_{1}(s_{f,1}-s_{1}),\label{eq:E101_c}\end{align}
all of which follow from the fact that the specific substrate uptake
rate is proportional to $D$, i.e., \begin{equation}
r_{s,1}=\frac{D}{Y_{1}}.\label{eq:E101_rS1}\end{equation}
Furthermore, $E_{101}$ is stable whenever it exists. Evidently, this
is the only steady state that allows positive cell densities to be
sustained.\\
The critical dilution rate, $D_{c,1}$, is the maximum specific growth
rate that can be maintained at steady state. To see this, observe
that at $E_{101}$, the specific growth rate equals the dilution rate.
Thus, \eqref{eq:E101_s1} says that specific growth rate increases
with $\sigma_{1}$, and is maximal when $\sigma_{1}=\sigma_{f,1}$.
Letting $\sigma_{1}=\sigma_{f,1}$ in \eqref{eq:E101_s1} and solving
for $D$ yields \eqref{eq:Dc1}.
\end{enumerate}
Fig.~\eqref{fig:BDss} shows that at any given dilution rate and
feed concentration, exactly one of the three steady states is stable.
It is therefore straightforward to deduce the variation of the steady
states along any path in the $\sigma_{f,1},D$-plane. The experiments
are typically performed by varying either $\sigma_{f,1}$ or $D$,
while the other one is held constant. We shall confine our attention
to these two cases.

\subsubsection{Variation of steady states with $\sigma_{f,1}$ at fixed $D$}

\begin{figure}
\noindent \begin{centering}
\subfigure[]{\includegraphics[width=2.6in]{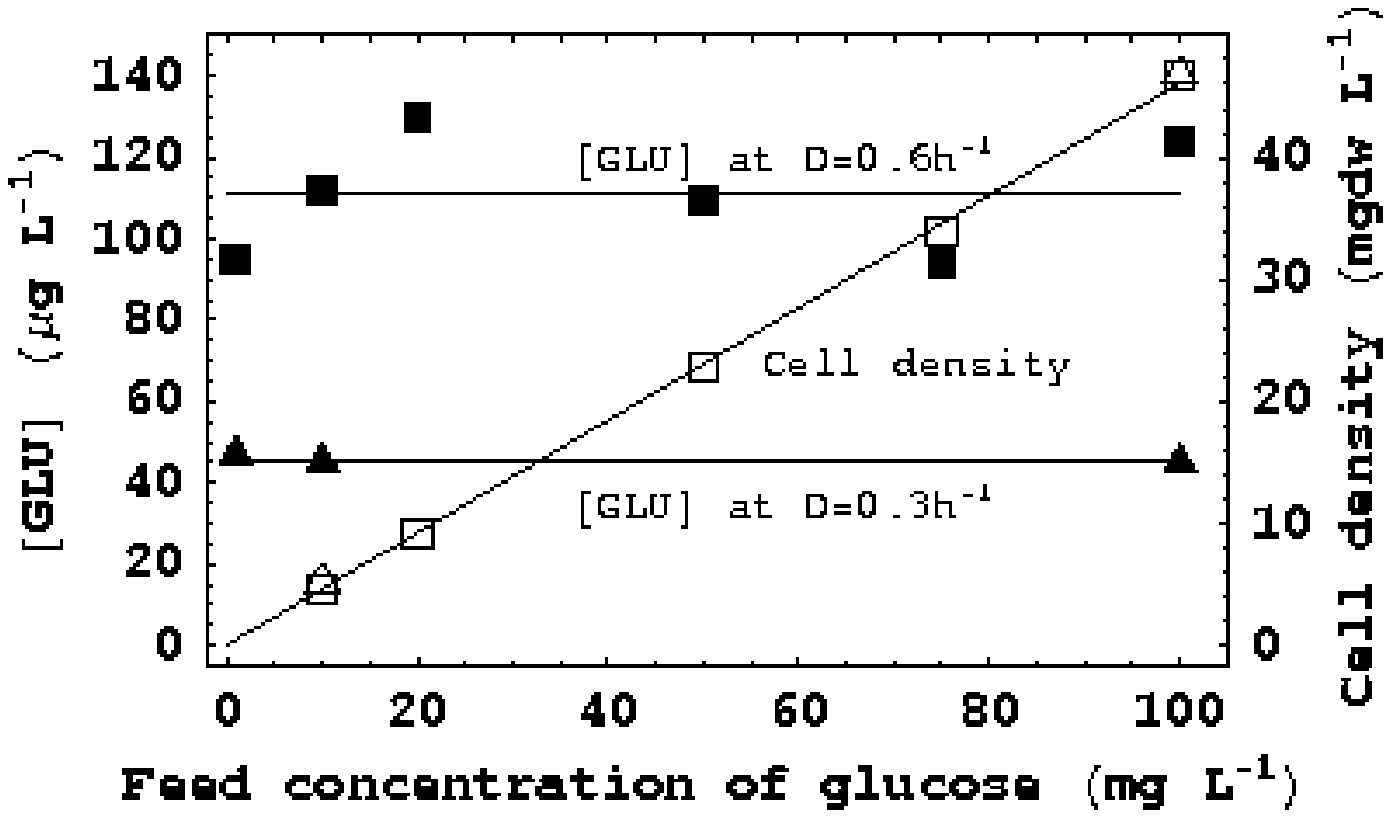}}\hspace*{0.2in}\subfigure[]{\includegraphics[width=2.6in]{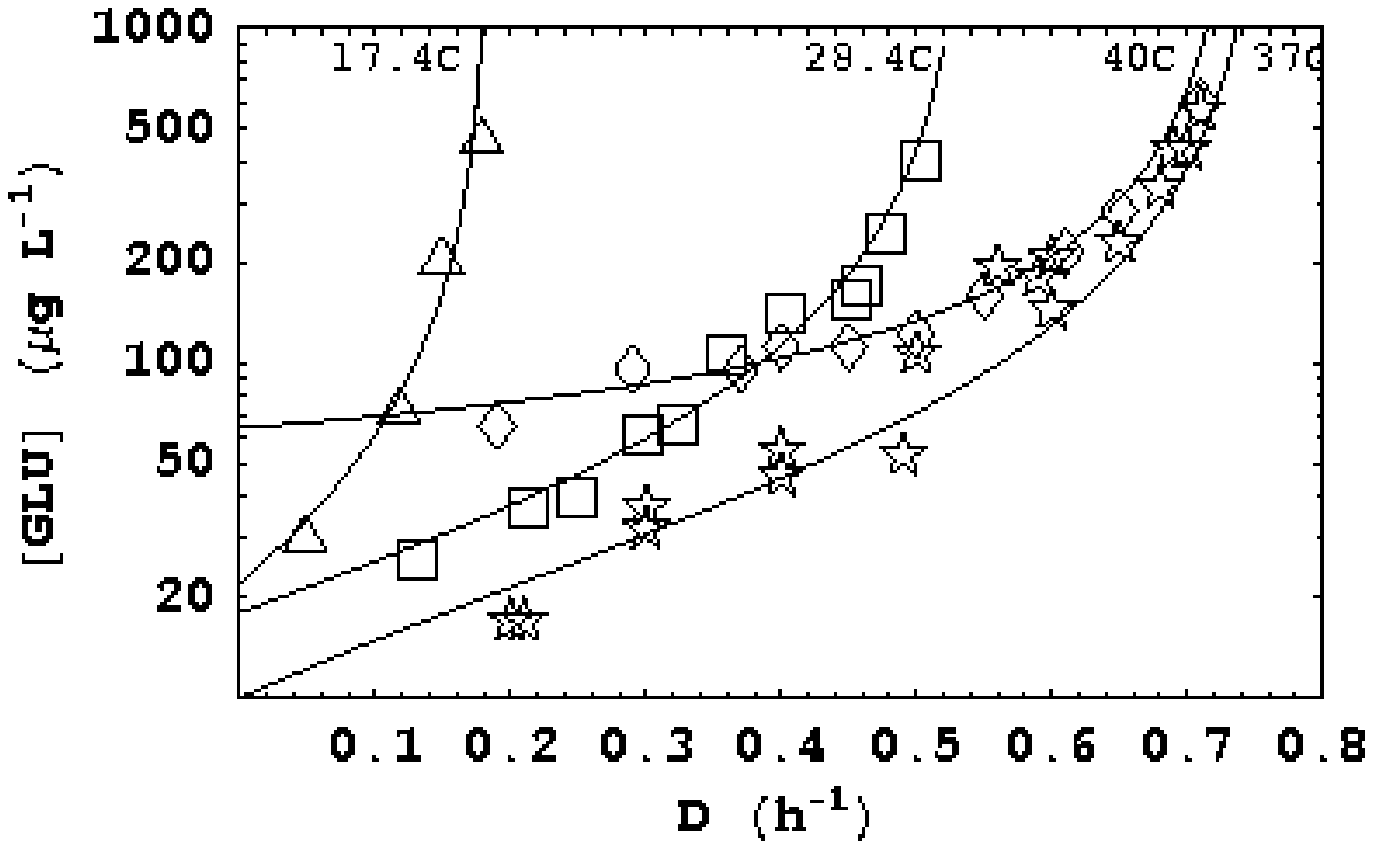}}
\par\end{centering}

\caption{\label{fig:SSdata}Growth of \emph{E. coli} ML308 in a glucose-limited
chemostat. (a)~When the feed concentration of glucose is increased
at a fixed dilution rate, the residual glucose concentration (closed
symbols) is constant, and the cell density (open symbols) increases
linearly~\citep[Appendix~B]{lendenmann94}. The data for $D=0.3$
h$^{-1}$ and $D=0.6$ h$^{-1}$ is represented by the symbols, $\triangle,\blacktriangle$
and $\square,\blacksquare$, respectively. (a)~When the dilution
rate is increased at a fixed feed concentration (100 mg~L$^{-1}$),
the residual substrate concentration increases monotonically with
$D$, and approaches a threshold at small dilution rates~\citep[Fig.~2]{Kovarova1996}.
The data was obtained at 4 temperatures.}

\end{figure}

If the feed concentration is increased at a sufficiently small fixed
$D$ (horizontal arrow in Fig.~\ref{fig:BDss}), steady growth is
feasible only if the feed concentration lies in the region to the
right of the blue curve, where $E_{101}$ is stable. The variation
of the steady states in this region is given by \eqref{eq:E101_x1}--\eqref{eq:E101_c},
which imply that as the feed concentration increases, $x_{1},e_{1},\sigma_{1}$
remain constant, and $c$ increases linearly. Thus, any increase in
the feed concentration is compensated by a proportional increase in
the cell density --- all other variables remain unchanged. We shall
appeal to this fact later.

No data is available for the inducer or enzyme levels, but experiments
performed with glucose-limited cultures of \emph{E. coli} ML308 show
that at dilution rates of $0.3$ and 0.6~h$^{-1}$, the residual
substrate concentration is indeed independent of the feed concentration,
and the cell density does increase linearly with the feed concentration~(Fig.~\ref{fig:SSdata}a).
No threshold was observed in these experiments because the feed concentrations
were relatively high ($\ge1$ mg~L$^{-1}$). As we show below, the
threshold concentration for glucose is $\sim$10--50 $\mu$g~L$^{-1}$.

\subsubsection{Variation of steady states with $D$ at fixed $\sigma_{f,1}$}

If the dilution rate is increased at any fixed $\sigma_{f,1}<\sigma_{1}^{*}$,
there is no growth regardless of the dilution rate at which the chemostat
is operated. If $\sigma_{f,1}>\sigma_{1}^{*}$, the variation of the
steady states with $D$ is qualitatively identical to that of the
Monod model: The persistence steady state, $E_{101}$, is stable for
$0<D<D_{c,1}$, and the washout steady state, $E_{100}$, is stable
for $D\ge D_{c,1}$ (vertical arrow in Fig.~\ref{fig:BDss}). We
shall confine our attention to $E_{101}$, since $E_{100}$ is independent
of $D$.

\paragraph*{Variation of inducer and enzyme levels}

It follows from \eqref{eq:E101_x1}--\eqref{eq:E101_e1} that as the
dilution rate increases over the range, $(0,D_{c,1})$, the inducer
level increases linearly, whereas the enzyme level passes through
a maximum at $D=\sqrt{\left(Y_{1}V_{s,1}\bar{K}_{e,1}\right)k_{e,1}}.$
The latter is consistent with the data for systems with inducible
enzymes (Fig.~\ref{fig:InducibleEnzymes}).

In contrast to the hypothesis of Clarke et al., the model explains
the non-monotonic profile of the enzyme level as the outcome of a
balance between induction and dilution/degradation (as opposed to
catabolite repression). To see this, observe that the steady state
enzyme level is given by the ratio of the induction and dilution/degradation
rates, i.e.,\[
e_{1}=\frac{r_{e,1}^{+}}{D+k_{e,1}}.\]
Furthermore, since $x_{1}$ increases with $D$, so does $r_{e,1}^{+}$,
but it saturates at sufficiently large $D$. At low values of $D$,
the enzyme level increases with $D$ because the induction rate increases
with $D$. At high $D$, the steady state enzyme level inversely proportional
to $D$ because the induction rate saturates.

The validity of the above explanation rests upon the hypothesis that
the induction rate saturates at large $D$. This hypothesis is supported
by the data for the \emph{lac} and \emph{ptsG} operons. When batch
cultures of \emph{E. coli} growing exponentially on saturating concentrations
of lactose (which are physiologically identical to cultures growing
in a lactose-limited chemostat near the critical dilution rate) are
exposed to 0.4~mM IPTG, there is almost no change in the $\beta$-galactosidase
activity~\citep[Table 1]{Hogema1998a}. Similarly, the PtsG activity
of wild-type and \emph{ptsG}-constitutive \emph{E. coli} is the same
in glucose-excess batch cultures and continuous cultures operating
at high dilution rates~\citep[Fig.~3]{Seeto2004}. Induction is therefore
already saturated during growth of \emph{E. coli} on lactose or glucose,
provided the dilution rate is sufficiently large. It remains to be
seen if this hypothesis is also valid for the other systems. However,
at large dilution rates, the enzyme activity is more or less inversely
proportional to $D$ for all the systems shown Fig.~\ref{fig:InducibleEnzymes},
a result expected only if the induction rate saturates at large $D$.
Thus, the decline of the enzyme activity in both constitutive mutants
and wild-type cells is due to dilution, rather than catabolite repression.

\paragraph*{Variation of the substrate concentration and cell density}

According to the Monod model, the residual substrate concentration
is proportional to the dilution rate, i.e., $\sigma_{1}=D/r_{g,1}^{\textnormal{max}}$,
where $r_{g,1}^{\textnormal{max}}$ denotes the maximum specific growth
rate on $S_{1}$~\citep{herbert56}. In contrast, eq.~\eqref{eq:E101_s1}
yields\[
\sigma_{1}=\sigma_{1}^{*}+\left(\frac{k_{e,1}+Y_{1}V_{s,1}\bar{K}_{e,1}}{Y_{1}V_{s,1}V_{e,1}}\right)D+\left(\frac{1}{Y_{1}V_{s,1}V_{e,1}}\right)D^{2},\]
which implies that at small dilution rates, $\sigma_{1}$ approaches
the threshold level, $\sigma_{1}^{*}$, and at large dilution rates,
$\sigma_{1}$ is proportional to $D^{2}$. Both these properties follow
from the relation, $r_{s,1}\equiv V_{s,1}e_{1}\sigma_{1}=D/Y_{1}$,
and the steady state profiles of the peripheral enzyme activity. Indeed,
at low dilution rates, $\sigma_{1}$ is constant because the enzyme
level increases linearly. Likewise, at high dilution rates, $\sigma_{1}\sim D^{2}$
because $e_{1}\sim1/D$.

The experiments provide clear evidence of threshold concentrations.
Kovarova et al.~have shown that when \emph{E. coli} ML308 is grown
in a glucose-limited chemostat, the residual glucose concentration
approaches threshold levels of 20--40~$\mu$g/L at small dilution
rates (Fig.~\ref{fig:SSdata}b). The authors did not identify the
specific mechanism underlying the threshold. It may exist, for example,
simply because the maintenance requirements preclude growth. However,
we shall show below that in some cases, the threshold exists precisely
because the enzymes are not induced at sufficiently low values of
the substrate concentration.

\subsection{\label{sec:MixedSubstrateGrowth}Mixed-substrate growth}

We have shown above that in chemostats limited by a single substrate,
enzyme synthesis is abolished if the feed concentration is below the
threshold level, and this occurs because the induction rate of the
enzyme cannot match the degradation rate. Our next goal is to show
that in mixed-substrate cultures, enzyme synthesis is often abolished
because the induction rate of the enzyme cannot match its \emph{dilution}
rate. To this end, we shall begin with the case of mixed-substrate
batch cultures, where this phenomenon is particularly transparent.
This analysis of batch cultures will also enable us to establish the
formal similarity between the classification of the growth patterns
in batch and continuous cultures.

\subsubsection{Batch cultures}

\paragraph{Mathematical classification of the growth patterns}

The growth of mixed-substrate batch cultures is empirically classified
based on the substrate consumption pattern (preferential or simultaneous)
during the first exponential growth phase. The mathematical correlate
of the foregoing empirical classification is provided by the bifurcation
diagram for the equations\begin{align}
\frac{de_{1}}{dt} & =V_{e,1}\frac{e_{1}\sigma_{0,1}}{\bar{K}_{e,1}+e_{1}\sigma_{0,1}}-\left(Y_{1}V_{s,1}e_{1}\sigma_{0,1}+Y_{2}V_{s,2}e_{2}\sigma_{0,2}+k_{e,1}\right)e_{1},\label{eq:BatchReduced1}\\
\frac{de_{2}}{dt} & =V_{e,2}\frac{e_{2}\sigma_{0,2}}{\bar{K}_{e,2}+e_{2}\sigma_{0,2}}-\left(Y_{1}V_{s,1}e_{1}\sigma_{0,1}+Y_{2}V_{s,2}e_{2}\sigma_{0,2}+k_{e,2}\right)e_{2},\label{eq:BatchReduced2}\end{align}
which describe the motion of the enzymes toward the quasisteady state
corresponding to the first exponential growth phase.

We gain physical insight into the foregoing equations by observing
that they are formally similar to the Lotka-Volterra model for two
competing species\begin{align}
\frac{dN_{1}}{dt} & =a_{1}N_{1}-\left(b_{11}N_{1}+b_{12}N_{2}\right)N_{1},\label{eq:LV1}\\
\frac{dN_{2}}{dt} & =a_{2}N_{2}-\left(b_{21}N_{1}+b_{22}N_{2}\right)N_{2},\label{eq:LV2}\end{align}
where $N_{i}$ is the population density of the $i$-th species; $a_{i}N_{i}$
is the (unrestricted) growth rate of the $i$-th species in the absence
of any competion; and $b_{ii}N_{i}^{2}$, $b_{ij}N_{i}N_{j}$ represent
the growth rate reduction due to intra- and inter-specific competition,
respectively. Evidently, the net induction rate, $V_{e,i}e_{i}\sigma_{0,i}/(\bar{K}_{e,i}+e_{i}\sigma_{0,i})-k_{e,i}e_{i}$,
and the two dilution terms in \eqref{eq:BatchReduced1}--\eqref{eq:BatchReduced2}
are the correlates of the unrestricted growth rate and the two competition
terms in the Lotka-Volterra model. Thus, the interaction between the
enzymes is competitive (the enzymes inhibit each other via the dilution
terms), despite the absence of any transcriptional repression (the
induction rate of $E_{i}$ is unaffected by activity of $E_{j},j\ne i$).

Now, the Lotka-Volterra equations are known to yield coexistence or
extinction steady states, depending on the values of the parameters,
$b_{ij}$~\citep{murray}. The formal similarity of eqs.~\eqref{eq:BatchReduced1}--\eqref{eq:BatchReduced2}
and \eqref{eq:LV1}--\eqref{eq:LV2} suggests that the enzymes will
exhibit coexistence and extinction steady states, depending on the
values of the physiological parameters and the initial substrate concentrations.
We show below that this is indeed the case. Moreover, the coexistence
and extinction steady states are the correlates of simultaneous and
preferential growth, respectively.

\begin{figure}
\noindent \begin{centering}
\includegraphics[width=4in]{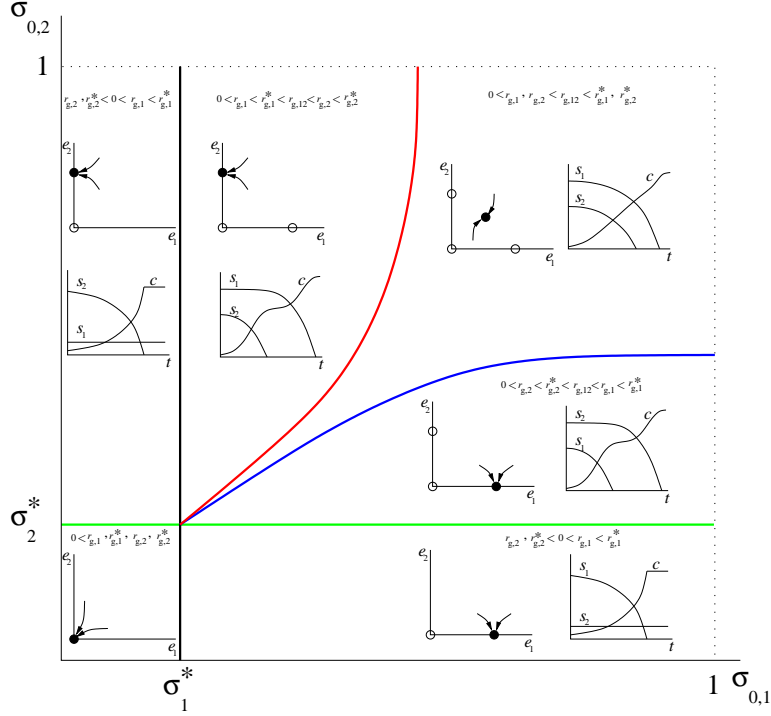}
\par\end{centering}

\caption{\label{f:BDbatch}Bifurcation diagram classifying the growth patterns
of mixed-substrate batch cultures at various initial substrate concentrations,
$\sigma_{0,1}$ and $\sigma_{0,2}$. The green and black lines represent
the threshold concentrations for induction of $E_{1}$ and $E_{2}$,
respectively. The blue and red curves represent the locus of initial
substrate concentrations defined by the equations, $r_{g,1}=r_{g,2}^{*}$
and $r_{g,2}=r_{g,1}^{*}$, respectively. The insets in each of the
six regions show the dynamics at the corresponding initial substrate
concentrations. The first inset shows the motion of the enzymes during
the first exponential growth phase; the second one shows the resultant
evolution of the substrate concentrations and cell density.}

\end{figure}

Eqs.~\eqref{eq:BatchReduced1}--\eqref{eq:BatchReduced2} have 4
types of steady states: $e_{1}=0$, $e_{2}=0$; $e_{1}>0$, $e_{2}=0$;
$e_{1}=0$, $e_{2}>0$; and $e_{1}>0$, $e_{2}>0$, which are denoted
$E_{00}$, $E_{10}$, $E_{01}$, and $E_{11}$, respectively. Each
of these steady states has a simple biological interpretation. If
the enzyme levels reach $E_{00}$ (resp., $E_{11}$), none (resp.,
both) of the two substrates are consumed during the first exponential
growth phase. If the enzyme levels reach $E_{10}$ (resp., $E_{01}$),
only $S_{1}$ (resp., $S_{2}$) is consumed during the first exponential
growth phase. The bifurcation diagram shows the existence and stability
of these steady states (and hence, the growth pattern) at any given
$\sigma_{0,1}$ and $\sigma_{0,2}$ (Fig.~\ref{f:BDbatch}). It can
be inferred from the following facts derived in Appendix~\ref{sec:AnalysisBatch}.

\begin{enumerate}
\item $E_{00}$ always exists (for all $\sigma_{0,1}$ and $\sigma_{0,2}$).
It is stable if and only if\[
r_{g,i}^{*}\equiv\frac{V_{e,i}\sigma_{0,i}}{\bar{K}_{e,i}}-k_{e,i}<0\Leftrightarrow\sigma_{0,i}<\sigma_{i}^{*},\; i=1,2\]
i.e., the point, $\left(\sigma_{0,1},\sigma_{0,2}\right)$, lies below
the green line and to the left of the black line in  Fig.~\ref{f:BDbatch}.
\item $E_{10}$ exists if and only if $\sigma_{0,1}>\sigma_{1}^{*}$, in
which case it is unique, and given by \[
e_{1}=\frac{-\left(\bar{K}_{e,1}+\frac{k_{e,1}}{Y_{1}V_{s,1}}\right)+\sqrt{\left(\bar{K}_{e,1}+\frac{k_{e,1}}{Y_{1}V_{s,1}}\right)^{2}+\frac{4\bar{K}_{e,1}r_{g,1}^{*}}{Y_{1}V_{s,1}}}}{2\sigma_{0,1}},\; e_{2}=0.\]
At this steady state, the cells consume only $S_{1}$, and grow exponentially
at the specific growth rate\[
r_{g,1}\equiv Y_{1}V_{s,1}e_{1}\sigma_{0,1}.\]
It is stable if and only if\begin{equation}
r_{g,2}^{*}<r_{g,1},\label{eq:StabilityE10}\end{equation}
i.e., $\left(\sigma_{0,1},\sigma_{0,2}\right)$ lies below the blue
curve in Fig.~\ref{f:BDbatch}, defined by the equation\begin{equation}
r_{g,2}^{*}=r_{g,1}.\label{eq:BDlowerCurve}\end{equation}
Thus, $E_{10}$ is stable under two distinct sets of conditions. Either
$\sigma_{2,0}$ is below the threshold level, $\sigma_{2}^{*}$, in
which case synthesis of $E_{2}$ cannot be sustained because its induction
rate cannot match its degradation rate. Alternatively, $\sigma_{0,2}$
is above the threshold level, but sustained synthesis of $E_{2}$
remains infeasible because the dilution of $E_{2}$ due to growth
on $S_{1}$ is too large. Thus, $r_{g,2}^{*}$ can be viewed as the
highest specific growth rate tolerated by the induction machinery
for $E_{2}$: Synthesis of $E_{2}$ is abolished whenever $r_{g,1}$
exceeds $r_{g,2}^{*}$. We shall refer to $r_{g,2}^{*}$ as the specific
growth rate for \emph{extinction} of $E_{2}$.\\
For any given $\sigma_{0,1}>\sigma_{1}^{*}$, the exponential growth
rate, $r_{g,1}$, can be higher or lower than $r_{g,2}^{*}$ (depending
on the value of $\sigma_{0,2}$), but it cannot exceed the corresponding
extinction growth rate, $r_{g,1}^{*}$. This reflects the biologically
plausible fact that $e_{1}$ cannot be zero during exponential growth
on $S_{1}$.
\item $E_{01}$ exists if and only if $\sigma_{0,2}>\sigma_{2}^{*}$, in
which case it is (uniquely) given by the expressions\[
e_{1}=0,\; e_{2}=\frac{-\left(\bar{K}_{e,2}+\frac{k_{e,2}}{Y_{2}V_{s,2}}\right)+\sqrt{\left(\bar{K}_{e,2}+\frac{k_{e,2}}{Y_{2}V_{s,2}}\right)^{2}+\frac{4\bar{K}_{e,2}r_{g,2}^{*}}{Y_{2}V_{s,2}}}}{2\sigma_{0,2}}.\]
At this steady state, the cells consume only $S_{2}$, and grow exponentially
at the specific growth rate\[
r_{g,2}\equiv Y_{2}V_{s,2}e_{2}\sigma_{0,2}.\]
It is stable if and only if\begin{equation}
r_{g,1}^{*}<r_{g,2},\label{eq:StabilityE01}\end{equation}
i.e., $\left(\sigma_{0,1},\sigma_{0,2}\right)$ lies to the left of
the red curve in Fig.~\ref{f:BDbatch}, defined by the equation\begin{equation}
r_{g,1}^{*}=r_{g,2}.\label{eq:BDupperCurve}\end{equation}
Furthermore, $r_{g,2}\le r_{g,2}^{*}$ with equality being attained
precisely when $\sigma_{2}=\sigma_{2}^{*}$, in which case $r_{g,2}=r_{g,2}^{*}=0$.
\item $E_{11}$ exists and is unique if and only if $E_{10}$ and $E_{01}$
are unstable, a condition satisfied precisely when\[
r_{g,2}^{*}>r_{g,1},\; r_{g,1}^{*}>r_{g,2},\]
i.e., $\left(\sigma_{0,1},\sigma_{0,2}\right)$ lie between the blue
and red curves in Fig.~\ref{f:BDbatch}. At this steady state, both
enzymes {}``coexist'' because neither substrate supports a specific
growth rate that is large enough to annihilate the enzymes for the
other substrate. Hence, both substrates are consumed, and the cells
grow exponentially at the specific growth rate\[
r_{g,12}\equiv Y_{1}V_{s,1}e_{1}\sigma_{0,1}+Y_{2}V_{s,2}e_{2}\sigma_{0,2},\]
where $e_{1},e_{2}$ are the unique positive solutions of \eqref{eq:BatchReduced1}--\eqref{eq:BatchReduced2}.
\item The blue curve and red curves are given by the equations\begin{align}
r_{g,1}^{*} & =\left(1+\frac{k_{e,1}}{Y_{1}V_{s,1}\bar{K}_{e,1}}\right)r_{g,2}^{*}+\frac{1}{Y_{1}V_{s,1}\bar{K}_{e,1}}\left(r_{g,2}^{*}\right)^{2},\label{eq:DefnBlueCurve}\\
r_{g,2}^{*} & =\left(1+\frac{k_{e,2}}{Y_{2}V_{s,2}\bar{K}_{e,2}}\right)r_{g,1}^{*}+\frac{1}{Y_{2}V_{s,2}\bar{K}_{e,2}}\left(r_{g,2}^{*}\right)^{2},\label{eq:DefnRedCurve}\end{align}
respectively. Both equations define graphs of increasing functions
passing through $(\sigma_{1}^{*},\sigma_{2}^{*})$, but the blue curve
always lies below the red curve.
\end{enumerate}
Thus, the bifurcation diagram consists of 6 distinct regions such
that exactly one of the four steady states is stable in each region.

The bifurcation diagram implies that there are six possible growth
patterns, depending on the initial substrate concentrations. Three
of these growth patterns occur if $\sigma_{0,1}<\sigma_{1}^{*}$ or/and
$\sigma_{0,1}<\sigma_{2}^{*}$. If both $\sigma_{0,1}$ and $\sigma_{0,2}$
are below their respective threshold concentrations, neither substrate
is consumed. If only one of them is below its threshold level, say,
$\sigma_{0,2}<\sigma_{2}^{*}$, only $S_{1}$ is consumed during the
first exponential growth phase. However, growth is not diauxic because
$S_{2}$ is never consumed. If $\sigma_{0,1}>\sigma_{1}^{*}$ and
$\sigma_{0,2}>\sigma_{2}^{*}$, the model predicts the existence of
three additional growth patterns.

\begin{enumerate}
\item If $\sigma_{0,1}$, $\sigma_{0,2}$ lie in the region between the
blue and green curves, $e_{2}$ approaches zero during the first exponential
growth phase. Importantly, since $\sigma_{0,2}$ is higher than the
threshold level, $E_{2}$ is synthesized upon exhaustion of $S_{1}$
at the end of the first exponential growth phase. Hence, growth is
diauxic with preferential consumption of $S_{1}$.
\item If $\sigma_{0,1}$, $\sigma_{0,2}$ lie in the region between the
red and black curves, $e_{1}$ approaches zero during the first exponential
growth phase, i.e., there is diauxic growth with preferential consumption
of $S_{2}$.
\item If $\sigma_{0,1}$, $\sigma_{0,2}$ lie between the red and blue curves,
$e_{1}$ and $e_{2}$ attain positive values during the first exponential
growth phase. Consequently, both substrates are consumed until one
of them is exhausted, and there is no diauxic lag before the remaining
substrate is consumed.
\end{enumerate}
We show below that the foregoing classification is consistent with
the data.

\paragraph{Growth patterns at saturating substrate concentrations}

\begin{figure}
\noindent \begin{centering}
\includegraphics[width=2.6in]{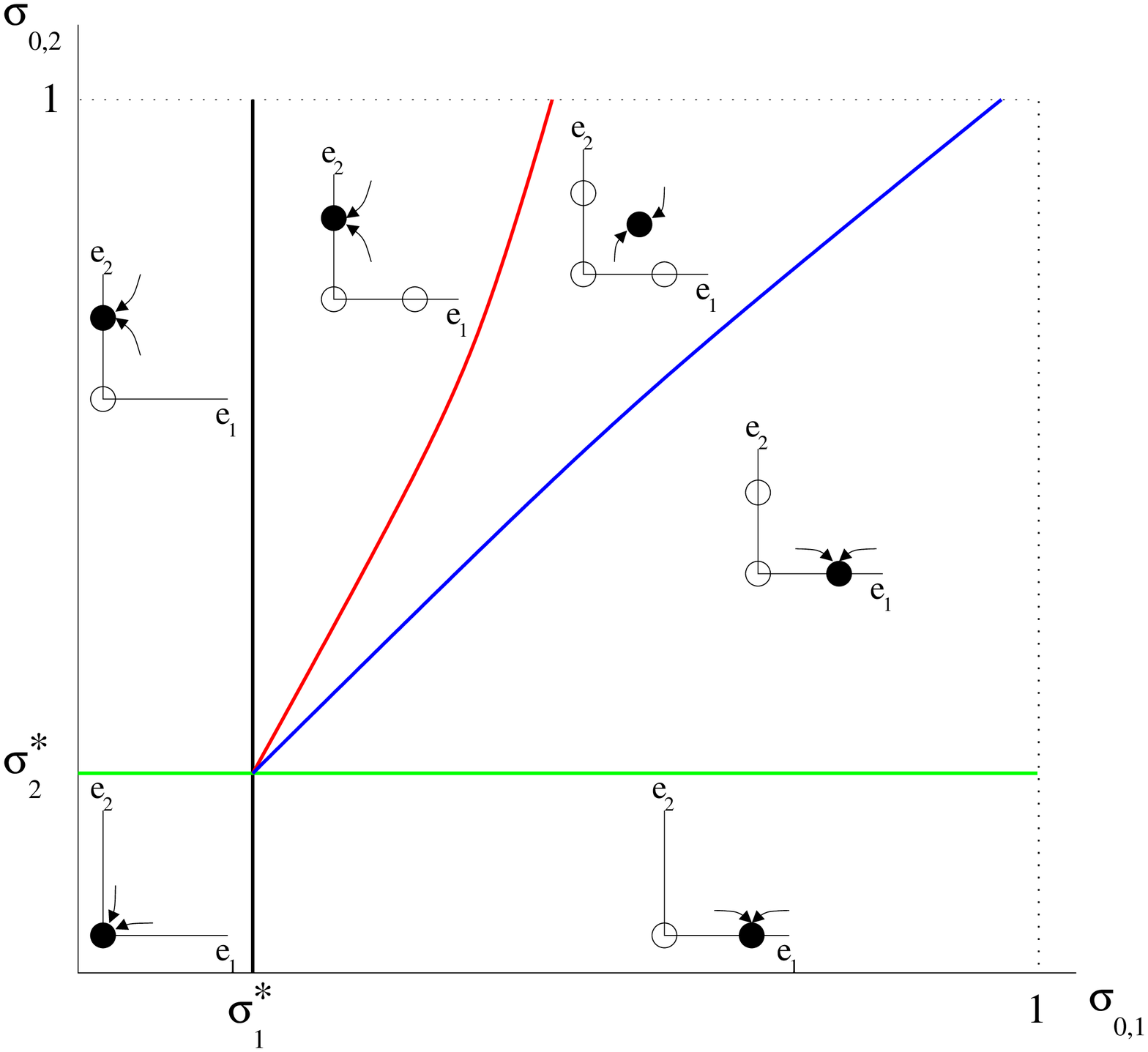}\hspace*{0.1in}\includegraphics[width=2.6in]{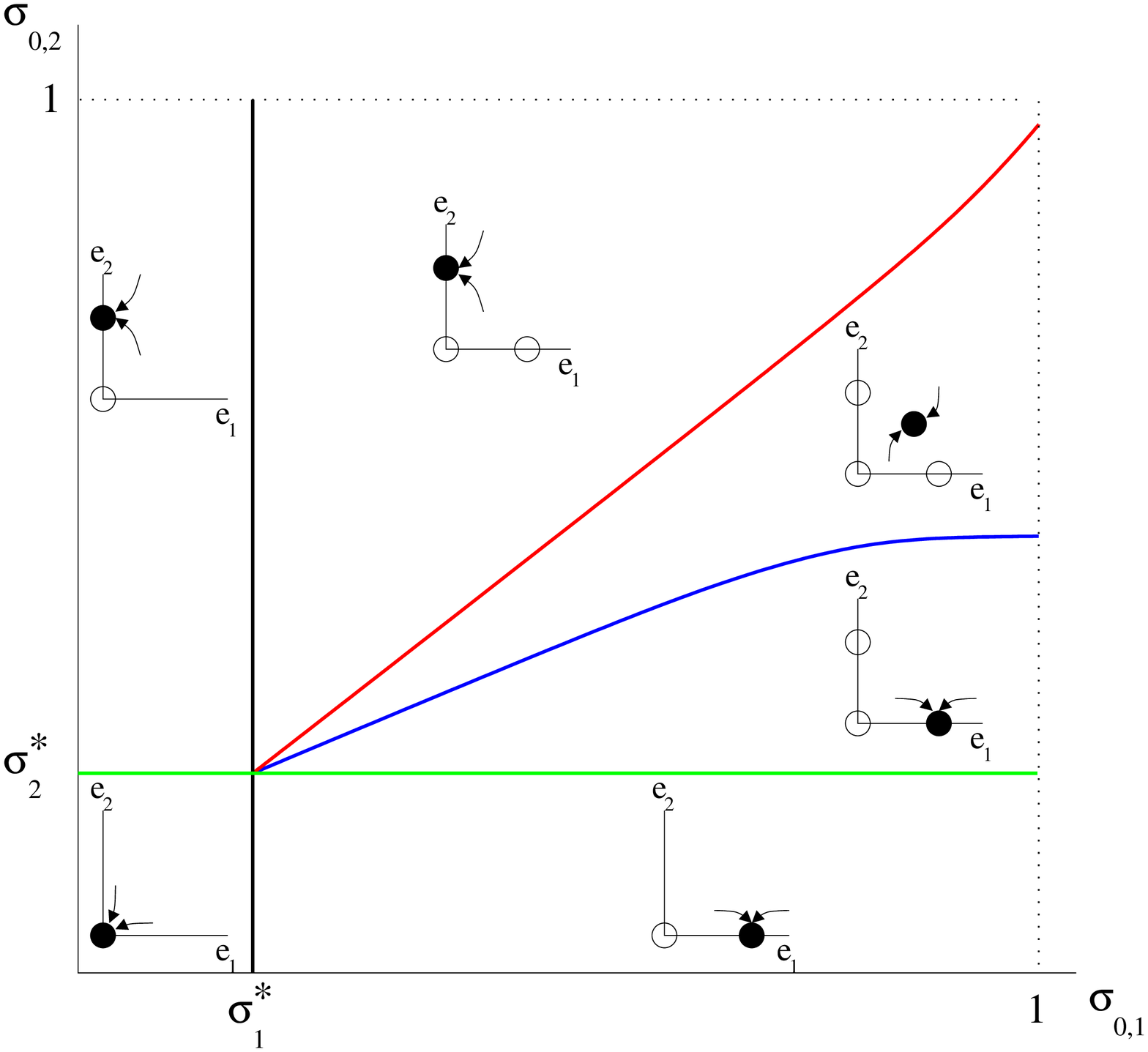}
\par\end{centering}

\caption{\label{fig:BDbatch2}Bifurcation diagrams corresponding to diauxic
growth under substrate-excess conditions ($\sigma_{0,1},\sigma_{0,2}\approx1$).
(a)~Preferential consumption of $S_{1}$. (b)~Preferential consumption
of $S_{2}$.}

\end{figure}

Physiological experiments, which are generally performed with saturating
substrate concentrations ($\sigma_{0,1},\sigma_{0,2}\approx1$), show
that depending on the cell type, the two substrates are consumed sequentially
or simultaneously. The behavior of the model is consistent with this
observation. To see this, observe that depending on the values of
the physiological parameters (which, in effect, determine the cell
type), there are three possible arrangements of the red and blue curves,
each of which yields a distinct growth pattern at saturating substrate
concentrations. Indeed, $S_{1}$ is consumed preferentially if $(1,1)$
lies below the blue curve (Fig.~\ref{fig:BDbatch2}a). This occurs
precisely when the physiological parameters satisfy the condition
\begin{align}
\frac{V_{e,1}}{\bar{K}_{e,1}} & >\frac{V_{e,2}}{\bar{K}_{e,2}}+\frac{1}{Y_{1}V_{s,1}\bar{K}_{e,1}}\left(\frac{V_{e,2}}{\bar{K}_{e,2}}\right)^{2},\label{eq:S1consumedCondition}\end{align}
obtained from \eqref{eq:DefnBlueCurve} by neglecting enzyme degradation
($k_{e,i}=0$). Likewise, $S_{2}$ is consumed preferentially if $(1,1)$
lies above the red curve (Fig.~\ref{fig:BDbatch2}b), i.e., the physiological
parameters satisfy the condition\begin{align}
\frac{V_{e,2}}{\bar{K}_{e,2}} & >\frac{V_{e,1}}{\bar{K}_{e,1}}+\frac{1}{Y_{2}V_{s,2}\bar{K}_{e,12}}\left(\frac{V_{e,1}}{\bar{K}_{e,1}}\right)^{2},\label{eq:S2consumedCondition}\end{align}
obtained from \eqref{eq:DefnRedCurve} upon setting $k_{e,i}=0$.
Finally, the substrates are consumed simultaneously if (1,1) lies
above the blue curve and below the red curve (Fig.~\ref{f:BDbatch}),
i.e., both the foregoing conditions are violated.

The biological meaning of these conditions was discussed in earlier
work~\citep{Narang2007a}. Indeed, one can check that \eqref{eq:S1consumedCondition}
and \eqref{eq:S2consumedCondition} are equivalent to the conditions\begin{align*}
\alpha & <\alpha_{*}\equiv\frac{\kappa_{2}\left(-\kappa_{1}+\sqrt{\kappa_{1}^{2}+4}\right)}{2},\\
\alpha & >\alpha^{*}\equiv\frac{2}{\kappa_{1}\left(-\kappa_{2}+\sqrt{\kappa_{1}^{2}+4}\right)},\end{align*}
respectively, where $\alpha\equiv\sqrt{Y_{2}V_{s,2}V_{e,2}}/\sqrt{Y_{1}V_{s,1}V_{e,1}}$
is a measure of the maximum specific growth rate on $S_{2}$ relative
to that on $S_{1}$, and $\kappa_{i}\equiv\bar{K}_{e,i}/\sqrt{V_{e,i}/(Y_{i}V_{s,i})}$
is a measure of saturation constant for induction of $E_{i}$. Thus,
the substrates are consumed preferentially whenever the maximum specific
growth rate on one of the substrates is sufficiently large compared
to that on the other substrate. Just how large this ratio must be
depends on the saturation constants, $\kappa_{i}$. Enzymes with small
saturation constants are quasi-constitutive --- their synthesis cannot
be abolished even if the other substrate supports a large specific
growth rate. In contrast, synthesis of enzymes with large saturation
constants can be abolished even if the other substrate supports a
comparable specific growth rate.

\paragraph{Transitions triggered by changes in substrate concentrations}

\begin{figure}[t]
\begin{centering}
\subfigure[]{\includegraphics[width=2.6in]{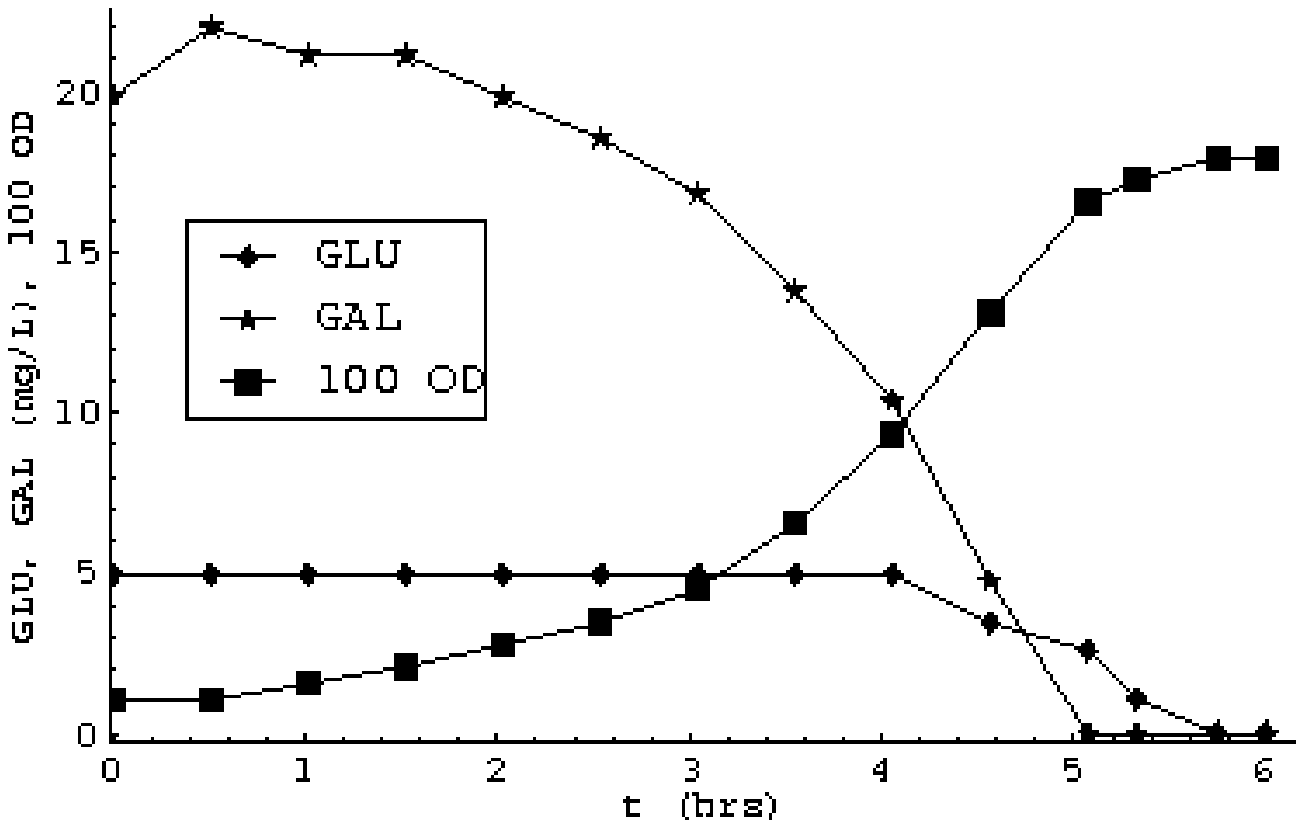}}\hspace*{0.1in}\subfigure[]{\includegraphics[width=2.6in]{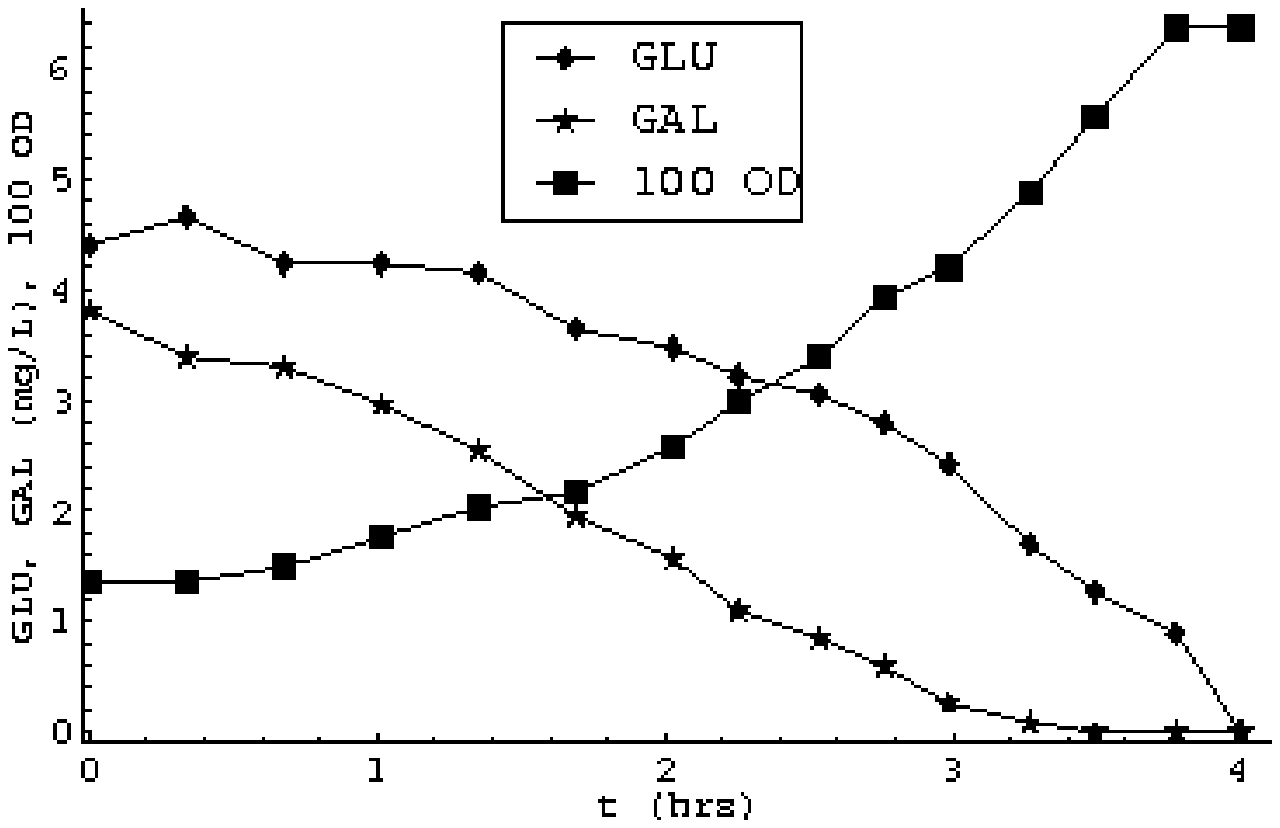}}
\par\end{centering}

\begin{centering}
\subfigure[]{\includegraphics[width=2.6in]{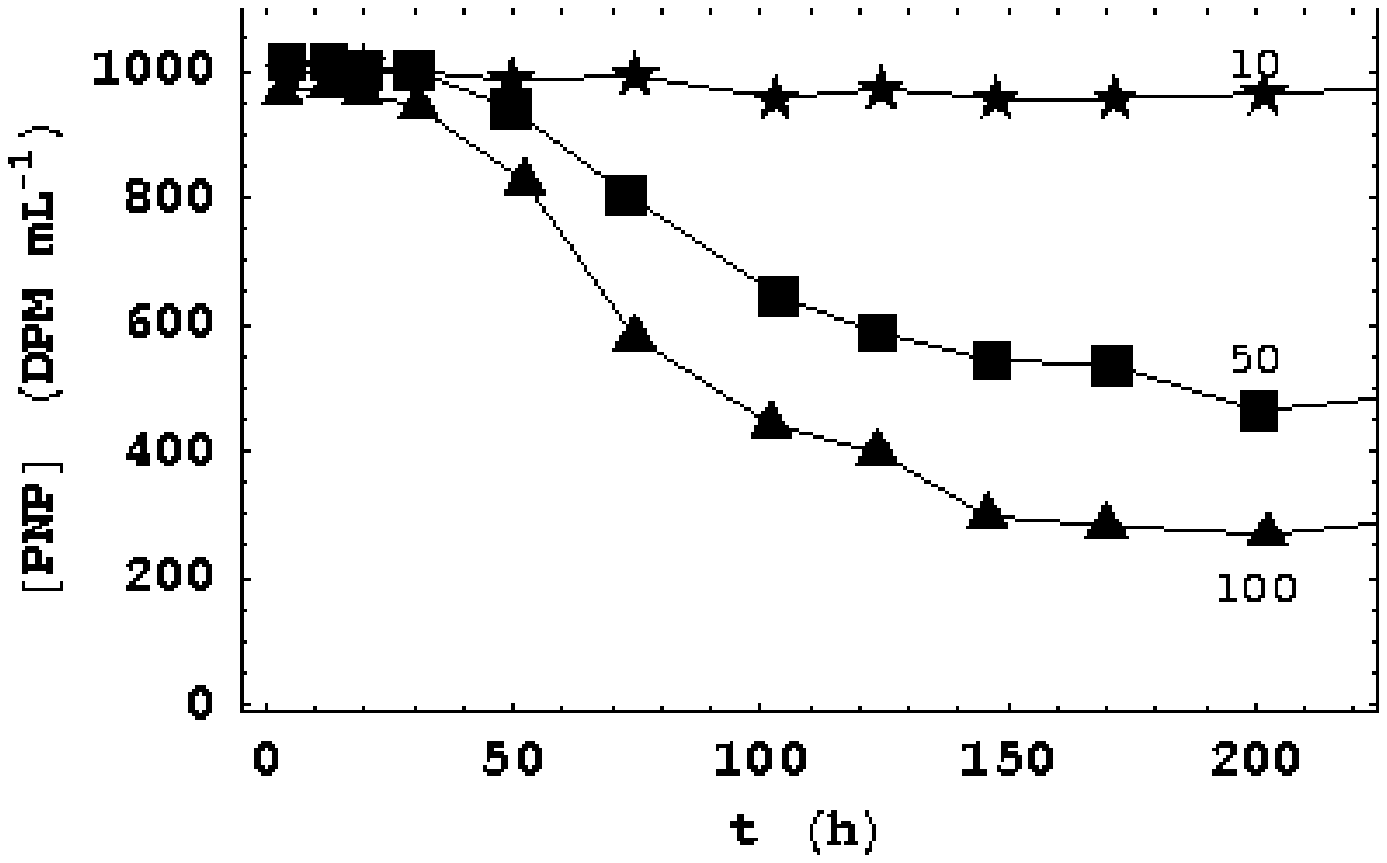}}\hspace*{0.1in}\subfigure[]{\includegraphics[width=2.6in]{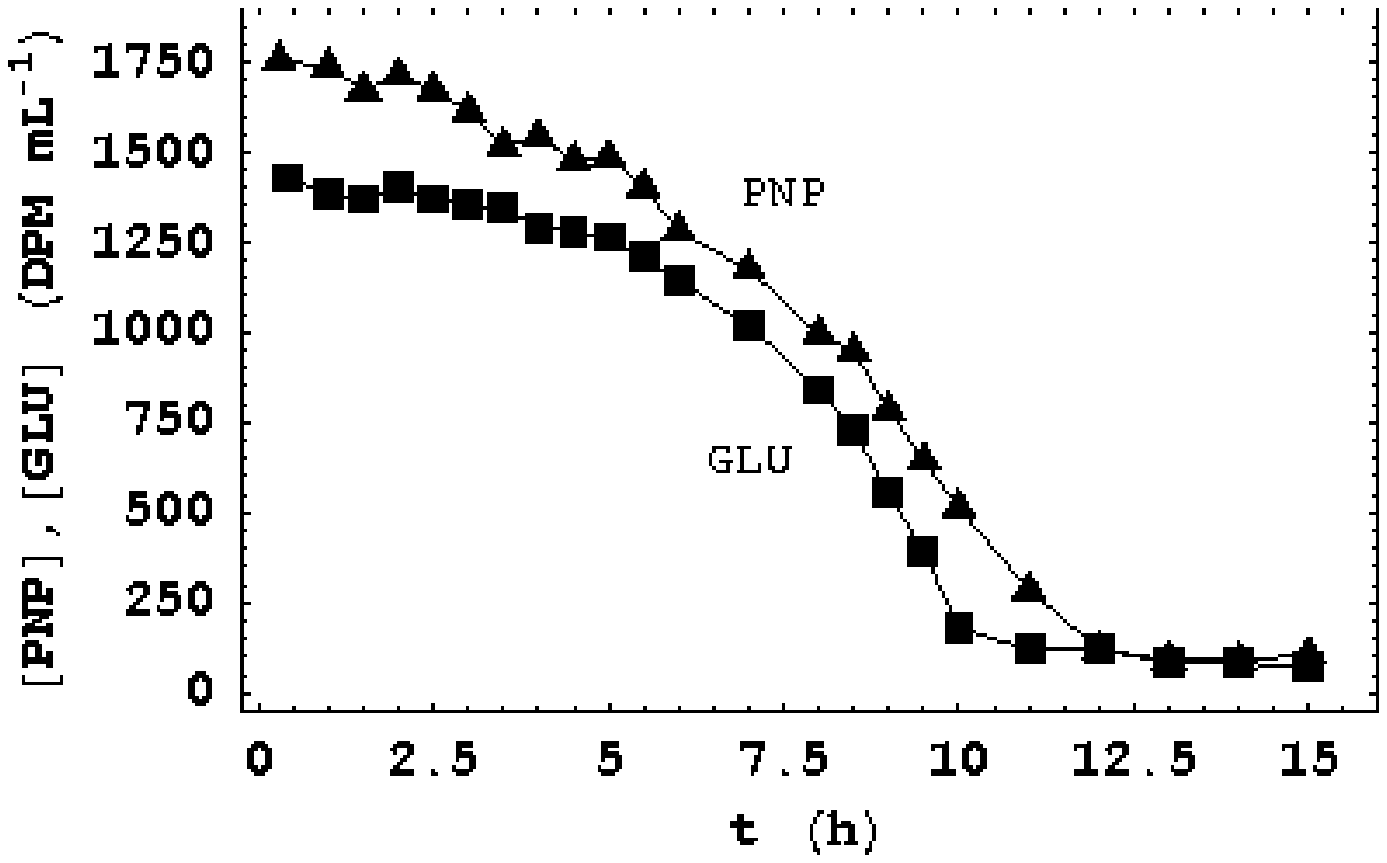}}
\par\end{centering}

\caption{\label{fig:BatchTransitions}Transitions of the growth patterns triggered
by changes in the initial substrate concentrations. \textbf{Upper
panel:} Batch growth of \emph{Escherichia coli} ML308 on mixtures
of glucose and galactose. (a)~At high initial glucose concentrations
(20 mg~L$^{-1}$), glucose is consumed before galactose~\citep{lendenmann94}.
(b)~At low initial glucose concentrations (5~mg~L$^{-1}$), glucose
and galactose are consumed simultaneously~\citep[Fig.~1]{egli93}.
\textbf{Lower panel:} Batch growth of \emph{Pseudomonas} on PNP and
PNP + glucose~\citep[Figs.~1,4]{Schmidt1987}. DPM denotes disintegrations
per min of $^{14}$C-labelled PNP and glucose. (c)~No PNP is consumed
if its initial concentration is 10 $\mu$g~L$^{-1}$. Significant
consumption occurs at initial concentrations of 50 and 100 $\mu$g~L$^{-1}$.
(d)~PNP and glucose are consumed simultaneously if their initial
concentrations are high (3 and 10 mg~L$^{-1}$, respectively).}

\end{figure}

Figs.~\ref{f:BDbatch}--\ref{fig:BDbatch2} imply that the growth
pattern can be changed by altering the initial substrate concentrations.
We show below that this conclusion is consistent with the data.

As a first example, consider a cell type that prefers to consume $S_{1}$
under substrate-excess conditions ($\sigma_{0,1},\sigma_{0,2}\approx1$).
Its growth pattern at any $\sigma_{0,1}$, $\sigma_{0,2}$ is then
described by Fig.~\ref{fig:BDbatch2}a, which implies that if $\sigma_{0,1}$
is decreased sufficiently from 1, both substrates will be consumed
simultaneously. Such behavior has been observed in experiments. If
the initial concentrations of glucose and galactose are 20~mg/L and
5~mg/L, respectively, \emph{E.~coli} ML308 consumes glucose before
galactose (Fig.~\ref{fig:BatchTransitions}a). If the initial concentration
of glucose is reduced to 5~mg/L or lower, glucose and galactose are
consumed simultaneously (Fig.~\ref{fig:BatchTransitions}b).

The data in Figs.~\ref{fig:BatchTransitions}a,b are open to question
because the initial cell densities are so large ($\sim10^{9}$ cells~L$^{-1}$)
that the substrates may have been exhausted before the enzymes reached
the quasisteady state corresponding to balanced growth. However, similar
results have also been obtained in studies performed with very small
initial cell densities ($10^{3}$--$10^{5}$ cells~L$^{-1}$). Indeed,
Schmidt and Alexander observed that {}``simultaneous use of two substrates
is concentration dependent \ldots{} \emph{Pseudomonas} sp.~ANL 50
mineralized glucose and aniline simultaneously when present at 3~$\mu$g
L$^{-1}$ but metabolized them diauxically at 300 $\mu$g~L$^{-1}$''
\citep[Figs.~3--4]{Schmidt1985b}. Likewise, \emph{Pseudomonas acidovorans}
consumed acetate before phenol when the concentrations of these substrates
were >70 and 2 $\mu$g~L$^{-1}$, respectively. If the initial concentration
of acetate was reduced to 13~$\mu$g~L$^{-1}$, both substrates
were consumed simultaneously.

In terms of the model, these transitions from sequential to simultaneous
substrate consumption can be understood as follows. Sequential substrate
consumption occurs at high concentrations of the preferred substrate
because it supports such a large specific growth rate that the enzymes
associated with the secondary substrate are diluted to extinction.
Reducing the initial concentration of the preferred substrate decreases
its ability to support growth, and thus dilute the secondary substrate
enzymes to extinction. Consequently, both substrates are consumed.

The growth of \emph{Pseudomonas }sp.~on p-nitrophenol (PNP) and glucose
furnishes additional examples of growth pattern transitions driven
by changes in the initial substrate concentrations. Schmidt et al.~found
that no PNP was consumed if its initial concentration was 10 $\mu$g~L$^{-1}$,
and significant mineralization occurred at the higher initial concentrations
of 50 and 100 $\mu$g~L$^{-1}$ (Fig.~\ref{fig:BatchTransitions}c).
It follows that the threshold concentration for PNP lies between 10
and 50 $\mu$g~L$^{-1}$. Importantly, the addition of 20 mg~L$^{-1}$
glucose to a culture containing 10 $\mu$g~L$^{-1}$ PNP failed to
stimulate PNP consumption. The authors concluded that {}``the threshold
was not a result of the fact that the concentration of PNP was too
low to meet maintenance energy requirements of the organism, but rather
supports the hypothesis that the concentration was too low to induce
degradative enzymes.'' Finally, it was observed that when the initial
concentrations of PNP and glucose were increased to saturating levels
of 3 mg~L$^{-1}$ and 10 mg~L$^{-1}$, respectively, both substrates
were consumed simultaneously (Fig.~\ref{fig:BatchTransitions}d).
All these results are consistent with the bifurcation diagram shown
in Fig.~\ref{f:BDbatch}.

Figs.~\ref{f:BDbatch}--\ref{fig:BDbatch2} extend the results obtained
in~\citealp{Narang2007a}. There, we constructed bifurcation diagrams
describing the variation of the growth patterns in response to changes
in the physiological parameters at (fixed) saturating substrate concentrations.
These \emph{genetic} bifurcation diagrams explained the phenotypes
of many different mutants and recombinant cells. Figs.~\ref{f:BDbatch}--\ref{fig:BDbatch2}
can be viewed as \emph{epigenetic} bifurcation diagrams, since they
describe the variation of the growth pattern when a given cell type,
characterized by fixed physiological parameters, is exposed to various
initial substrate concentrations.

\subsubsection{Continuous cultures}

In the experimental literature, the growth pattern of mixed-substrate
continuous cultures is classified based on the manner in which the
two substrates are consumed when the dilution rate is increased at
a given pair of feed concentrations. Either both substrates are consumed
at all dilution rates up to washout, or both substrates are consumed
up to an intermediate dilution rate beyond which only one of the substrates
is consumed. We show below that the model predicts these growth patterns.
Moreover, it can quantitatively capture the variation of the steady
state enzyme levels, substrate concentrations, and cell density with
respect to the dilution rate and the feed concentrations.

\paragraph{Mathematical classification of the growth patterns}

The mathematical correlate of the experimentally observed growth patterns
corresponds to the bifurcation diagram for eqs.~\eqref{eq:ContS}--\eqref{eq:ContRg},
which describes the existence and stability of the steady states in
the 3-dimensional $\sigma_{f,1},\sigma_{f,2},D$-space.%
\begin{table}
\caption{\label{t:StabilityConditions}Necessary and sufficient conditions
for existence and stability of the steady states for mixed-substrate
growth in continuous cultures (see Appendix C for details).}

\begin{tabular}{|c|c|c|c|c|}
\hline
Steady  & Defining  & Biological  & Existence  & Stability \tabularnewline
state & property & meaning & condition(s) & condition(s)\tabularnewline
\hline
\hline
$E_{000}$ & $e_{1}=0,e_{2}=0,$ & No substrate  & Always  & $D_{t,1}<0,$\tabularnewline
 & $c=0$ & consumed & exists & $D_{t,2}<0$\tabularnewline
\hline
$E_{101}$ & $e_{1}>0,e_{2}=0,$ & Only $S_{1}$ & $D_{t,1}>0,$ & $D>D_{t,2}$\tabularnewline
 & $c>0$ & consumed & $0<D<D_{c,1}$ & \tabularnewline
\hline
$E_{100}$ & $e_{1}>0,e_{2}=0,$ & Washout & $D_{t,1}>0$ & $D_{t,2}<D_{c,1},$\tabularnewline
 & $c=0$ & of $E_{101}$ &  & $D>D_{c,1}$\tabularnewline
\hline
$E_{011}$ & $e_{1}=0,e_{2}>0$ & Only $S_{2}$ & $D_{t,2}>0,$ & $D>D_{t,1}$\tabularnewline
 & $c>0$ & consumed & $0<D<D_{c,2}$ & \tabularnewline
\hline
$E_{010}$ & $e_{1}=0,e_{2}>0,$ & Washout & $D_{t,2}>0$ & $D_{t,1}<D_{c,2},$\tabularnewline
 & $c=0$ & of $E_{011}$ &  & $D>D_{c,2}$\tabularnewline
\hline
$E_{111}$ & $e_{1}>0,e_{2}>0,$ & Both $S_{1},S_{2}$ & $D_{t,1},D_{t,2}>0,$ & Whenever it\tabularnewline
 & $c>0$ & consumed & $D<D_{t,1},D_{t,2},D_{c}$ & exists\tabularnewline
\hline
$E_{110}$ & $e_{1}>0,e_{2}>0,$ & Washout & $D_{t,2}>D_{c,1}$, & $D>D_{c}$\tabularnewline
 & $c=0$ & of $E_{111}$ & $D_{t,1}>D_{c,2}$ & \tabularnewline
\hline
\end{tabular}
\end{table}

Eqs.~\eqref{eq:ContS}--\eqref{eq:ContRg} have 7 steady states,
but 5 of them are {}``single-substrate steady states'' (see rows
1--5 of Table~\ref{t:StabilityConditions}). Indeed, $E_{000}$,
$E_{101}$, and $E_{100}$ (resp., $E_{000}$, $E_{011}$, and $E_{010}$)
are observed in chemostats limited by $S_{1}$ (resp., $S_{2}$).
Their existence in mixed-substrate growth implies that even if both
substrates are supplied to the chemostat, there are steady states
at which only one of the substrates is consumed. We shall see below
that these steady states can become stable under certain conditions.
Here, it suffices to observe that only two of 7 steady states, namely,
$E_{111}$ ($e_{1}>0$, $e_{2}>0$, $c>0$) and $E_{110}$ ($e_{1}>0$,
$e_{2}>0$, $c>0$), are uniquely associated with mixed-substrate
growth. They correspond to simultaneous consumption of both substrates
and its washout.

The existence and stability of the steady states are completely determined
by five special dilution rates, namely, $D_{t,i}$, $D_{c,i}$, and
$D_{c}$ (Table~1, columns~4--5). These dilution rates are the analogs
of the special specific growth rates considered in the analysis of
batch cultures. Indeed, the \emph{transition} dilution rate for $S_{i}$,
\begin{equation}
D_{t,i}\equiv\frac{V_{e,i}\sigma_{f,i}}{\bar{K}_{e,i}}-k_{e,i},\label{eq:Dt}\end{equation}
is the analog of the extinction growth rate, $r_{g,i}^{*}$. It is
the maximum dilution rate up to which synthesis of $E_{i}$ can be
sustained. The \emph{critical} dilution rate, $D_{c,i}$, given by
the relation\[
D_{c,i}=Y_{i}V_{s,i}\frac{-\left(\bar{K}_{e,i}+\frac{k_{e,i}}{Y_{i}V_{s,i}}\right)+\sqrt{\left(\bar{K}_{e,i}+\frac{k_{e,i}}{Y_{i}V_{s,i}}\right)^{2}+\frac{4K_{e,i}D_{t,i}}{Y_{1}V_{s,i}}}}{2},\]
is the analog of $r_{g,i}$. It is the maximum dilution rate up to
which growth can be sustained in a chemostat supplied with only $S_{i}$,
and satisfies the relation, $D_{ci}\le D_{t,i}$ with equality being
obtained precisely when $\sigma_{i}=\sigma_{i}^{*}$, in which case,
$D_{c,i}=D_{t,i}=0$ (see dashed and blue lines in Fig.~\ref{fig:BDss}).
Finally, the critical dilution rate, $D_{c}$, is defined as\[
D_{c}\equiv Y_{1}V_{s,1}e_{1}\sigma_{f,1}+Y_{2}V_{s,2}e_{2}\sigma_{f,2},\]
where $e_{1},e_{2}$ are the unique positive steady state solutions
of \eqref{eq:BatchReduced1}--\eqref{eq:BatchReduced2} with $\sigma_{0,i}$
replaced by $\sigma_{f,i}$. It is the analog of the function, $r_{g,12}$.
Thus, $r_{g,i}^{*}$, $r_{g,i}$, $r_{g,12}$, and their analogs,
$D_{t,i}$, $D_{c,i}$, $D_{c}$, are defined by formally identical
expressions, the only difference being that $\sigma_{0,i}$ is replaced
by $\sigma_{f,i}$.

\begin{figure}
\noindent \begin{centering}
\includegraphics[width=5in]{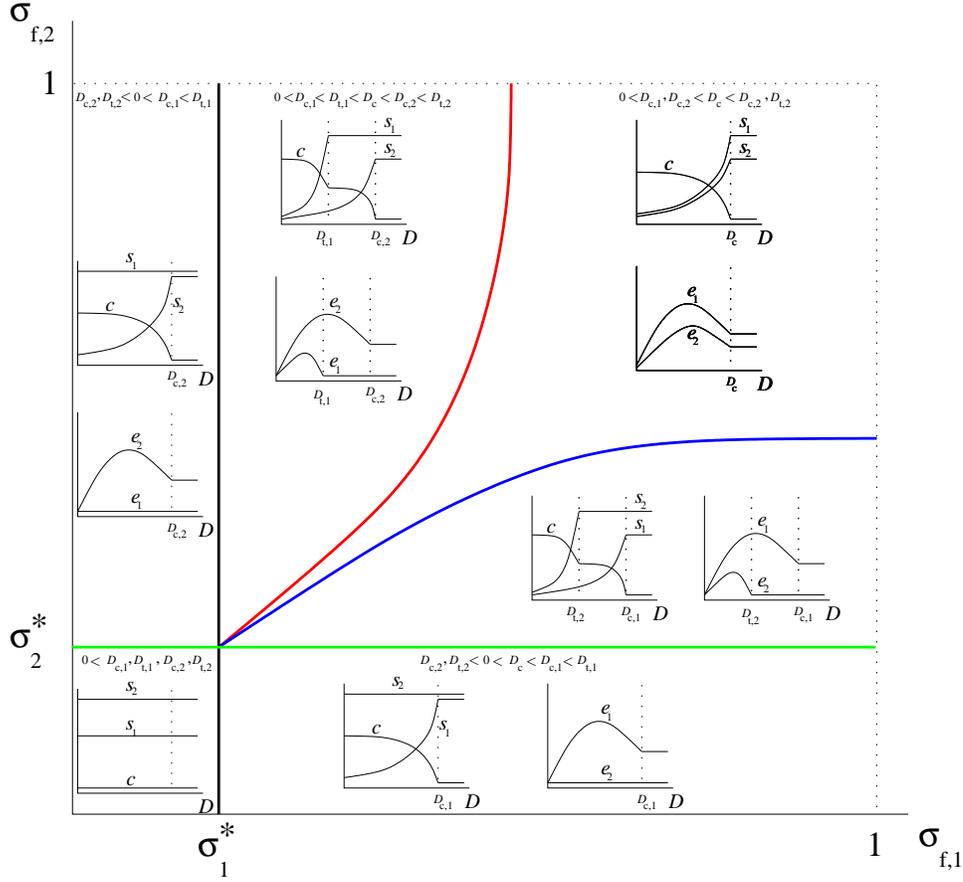}
\par\end{centering}

\caption{\label{f:BDcont}Bifurcation diagram classifying the growth patterns
of mixed-substrate continuous cultures at various feed concentrations,
$\sigma_{f,1}$ and $\sigma_{f,2}$. The green and black lines represent
the threshold concentrations for induction of $E_{1}$ and $E_{2}$,
respectively. The blue and red curves represent the locus of initial
substrate concentrations defined by the equations, $D_{c,1}=D_{t,2}$
and $D_{c,2}=D_{t,1}$, respectively. The insets in each of the six
regions show the variation of the steady states with $D$ at the corresponding
feed concentrations.}

\end{figure}

The formal similarity of the expressions for $D_{t,i}$, $D_{c,i}$,
$D_{c}$ and $r_{g,i}^{*}$, $r_{g,i}$, $r_{g,12}$ implies that
for a given set of physiological parameters, the relations, $\sigma_{f,i}=\sigma_{i}^{*}$,
$D_{t,1}=D_{c,2}$ and $D_{t,2}=D_{c,1}$, define curves on the $\sigma_{f,1},\sigma_{f,2}$-plane
that are identical to the curves on the $\sigma_{0,1},\sigma_{0,2}$-plane
defined by the equations, $\sigma_{0,i}=\sigma_{i}^{*}$, $r_{g,1}^{*}=r_{g,2}$
and $r_{g,2}^{*}=r_{g,1}$. It follows that the relations, $\sigma_{f,i}=\sigma_{i}^{*}$,
$D_{t,1}=D_{c,2}$, $D_{t,2}=D_{c,1}$, determine a partitioning of
the $\sigma_{f,1},\sigma_{f,2}$-plane into 6 regions, which is identical
to the partitioning of the bifurcation diagram for batch cultures,
the only difference being that the coordinates are $\sigma_{f,1},\sigma_{f,2}$,
rather than $\sigma_{0,1},\sigma_{0,2}$ (Fig.~\ref{f:BDcont}).

Each of the six regions in Fig.~\ref{f:BDcont} corresponds to unique
growth pattern that can be inferred from the existence and stability
properties summarized in columns 4--5 of Table~1. For instance, if
$\left(\sigma_{f,1},\sigma_{f,2}\right)$ lies in the region between
the green and blue curves in Fig.~\ref{f:BDcont}, both substrates
are consumed for all $0<D<D_{t,2}$, only $S_{1}$ is consumed for
all $D_{t,2}\le D<D_{c,1}$, and neither substrate is consumed because
the cells are washed out for all $D\ge D_{c,1}$. To see this, observe
that in the region of interest (between the green and blue curves),
the special dilution rates stand in the relation\[
0<D_{c,2}<D_{t,2}<D_{c}<D_{c,1}<D_{t,1}.\]
The existence and stability conditions listed in Table~\ref{t:StabilityConditions}
therefore imply that as $D$ increases, 4 of the 7 steady states play
no role because they are unstable whenever they exist, or do not exist
at all. Indeed:

\begin{enumerate}
\item $E_{000}$ and $E_{010}$ exist at all $D>0$, but they are always
unstable because the stability conditions for these two steady states
($D_{t,1},D_{t,2}>0$ and $D_{t,1}<D_{c,2}$, respectively) are violated
in the region of interest .
\item $E_{011}$ exists for all $0<D<D_{c,2}$, but it is unstable whenever
it exists because the stability condition, $D>D_{t,1}$, is not satisfied
in the region of interest.
\item $E_{110}$ does not exist at all since one of the existence conditions,
$D_{t,2}>D_{c,1}$, is violated in the region of interest.
\end{enumerate}
It follows that only the remaining three steady states are relevant.
One can check by appealing to Table~1 that $E_{111}$ is stable for
$0<D<D_{t,2}$, $E_{101}$ is stable for $D_{t,2}<D<D_{c,1}$, and
$E_{100}$ is stable for $D>D_{c,1}$. Hence, both substrates are
consumed for all $0<D<D_{t,2}$. At $D=D_{t,2}$, $e_{2}$ becomes
zero. Only $S_{1}$ is consumed for all $D_{t,2}\le D<D_{c,1}$ until
$c$ becomes zero at $D=D_{c,1}$. Neither substrate is consumed for
$D>D_{c,1}$.

Similar arguments, applied to all the regions in Fig.~\ref{f:BDcont},
show that there are 6 distinct growth patterns. Three of these growth
patterns occur when at least one of the feed concentrations is below
its threshold level.

\begin{enumerate}
\item If both feed concentrations are below the threshold level, $E_{000}$
is always stable, so that neither substrate is consumed at any dilution
rate.
\item If $\sigma_{f,2}<\sigma_{2}^{*}$, but $\sigma_{f,1}>\sigma_{1}^{*}$,
$E_{101}$ is stable (i.e., only $S_{1}$ is consumed) for all $0<D<D_{c,1}$,
and $E_{100}$ is stable (i.e., the cells wash out) whenever $D\ge D_{c,1}$.
This is qualitatively similar to the behavior observed during single-substrate
growth on $S_{1}$ (Fig.~\ref{fig:BDss}). Thus, if the feed concentration
of $S_{2}$ is at sub-threshold levels, it has no effect on the behavior
of the chemostat.
\item If $\sigma_{f,1}<\sigma_{1}^{*}$, and $\sigma_{f,2}>\sigma_{2}^{2}$,
$E_{011}$ is stable (i.e., only $S_{2}$ is consumed) for all $0<D<D_{c,2}$,
and $E_{010}$ is stable (i.e., the cells wash out) for $D\ge D_{c,2}$.
This is qualitatively similar to the behavior of the chemostat during
single-substrate growth on $S_{2}$.
\end{enumerate}
If both feed concentrations are above their respective threshold levels,
there are three additional growth patterns.

\begin{enumerate}
\item If $\sigma_{f,1}$ and $\sigma_{f,2}$ lie between the red and blue
curves, $E_{111}$ (i.e., both substrates are consumed) for $0<D<D_{c}$,
and $E_{110}$ is stable (i.e., cells wash out) for $D\ge D_{c}$.
Thus, both substrates are consumed at all dilution rates up to washout.
\item If $\sigma_{f,1}$ and $\sigma_{f,2}$ lie between the green and blue
curves, $E_{111}$ is stable (i.e., both substrates are consumed)
for all $0<D<D_{t,2}$, $E_{101}$ is stable (i.e., only $S_{1}$
is consumed) for all $D_{t,1}\le D<D_{c,1}$, and $E_{100}$ is stable
(the cells wash out) for all $D>D_{c,1}$.
\item If $\sigma_{f,1}$ and $\sigma_{f,2}$ lie between the black and red
curves, $E_{111}$ is stable (i.e., both substrates are consumed)
for all $0<D<D_{t,1}$, $E_{011}$ is stable (i.e., only $S_{2}$
is consumed) for all $D_{t,2}\le D<D_{c,2}$, and $E_{100}$ is stable
(the cells wash out) for all $D>D_{c,2}$.
\end{enumerate}
These growth patterns are the mathematical correlates of the growth
patterns shown in Fig.~\ref{fig:MSgrowthPatterns}.

The model predicts that the growth pattern in continuous cultures
can be changed by altering the feed concentrations. Unfortunately,
there is no experimental data at subsaturating feed concentrations
($\sigma_{f,i}\lesssim K_{s,i}$) due to technical difficulties associated
with wall growth. Henceforth, we shall confine our attention to growth
at saturating feed concentrations ($\sigma_{f,i}\approx1$).

We have shown above that for a given cell type, the bifurcation diagrams
for batch and continuous cultures are formally identical --- one can
be generated from the other by a mere relabeling of the coordinate
axes. This formal identity provides a precise mathematical explanation
for the empirically observed correlation between the batch and continuous
growth patterns of a given cell type. To see this, suppose the cells
in question consume $S_{1}$ preferentially in substrate-excess batch
cultures ($\sigma_{0,i}\approx1$). The bifurcation diagram for this
system then has the form shown in Fig.~\ref{fig:BDbatch2}a. The
continuous growth patterns of this system are given by the very same
figure, the only difference being that the coordinates, $\sigma_{0,1}$,
$\sigma_{0,2}$, are replaced by $\sigma_{f,1}$, $\sigma_{f,2}$,
respectively. It follows that if the cells are grown in continuous
cultures fed with saturating substrate concentrations ($\sigma_{f,i}\approx1$),
they will consume both substrates at low dilution rates, and only
$S_{1}$ at all high dilution rates up to washout (Fig.~\ref{fig:MSgrowthPatterns}b).
A similar argument shows that if the cells consume both substrates
in substrate-excess batch culture (bifurcation diagram given by a
relabeled Fig.~\ref{f:BDbatch}), their growth in continuous cultures
will be such that both substrates are consumed at all dilution rates
up to washout (Fig.~\ref{fig:MSgrowthPatterns}a).

\paragraph{The model predicts the observed variation of the steady states with
$D$ and $s_{f,i}$}

Although the model predicts the \emph{existence} of the observed growth
patterns, it remains to determine if the variation of the steady states
with $D$ and $s_{f,i}$ is in \emph{quantitative} agreement with
the data. Since the variation of the {}``single-substrate steady
states'' with $D$ and $s_{f,i}$ is consistent with the data (Section~\ref{sec:SingleSubstrateGrowth}),
it suffices to focus on $E_{111}$, the only growth-supporting steady
state uniquely associated with mixed-substrate growth. This steady
state can be observed in systems exhibiting both the simultaneous
and the preferential growth patterns. However, since the data for
simultaneous systems is limited, we shall focus on preferential systems.
In our discussion of these systems, we shall assume, furthermore,
that $S_{1}$ is the {}``preferred'' substrate. This entails no
loss of generality since the equations for $S_{1}$ and $S_{2}$ (and
the associated physiological variables) are formally identical: Interchanging
the indices does not change the form of the equations.

Two types of experiments can be found in the literature.

\begin{enumerate}
\item Either the dilution rate was changed at fixed feed concentrations.
\item Alternatively, the mass fraction of the substrates in the feed, $\psi_{i}\equiv s_{f,i}/(s_{f,1}+s_{f,2})$,
was changed at fixed total feed concentration, $s_{f,t}\equiv s_{f,1}+s_{f,2}$,
and (sufficiently small) dilution rate.
\end{enumerate}
The model predicts that in the first case, $E_{111}$ is stable at
all dilution rates below $D_{t,2}$, the transition dilution rate
for the {}``less preferred'' substrate. In the second case, $E_{111}$
is stable at all feed fractions, except those near the extreme values,
0 and 1, where the model predicts threshold effects. Such threshold
effects have been observed when the feed contained a very small fraction
of one of the substrates~\citep{Kovar2002,Ng1973,Rudolph2002}. Nevertheless,
we shall restrict our attention to intermediate values of $\psi_{i}$
at which $E_{111}$ is the stable steady state.

The steady state, $E_{111}$, satisfies the equations\begin{align}
0 & =D(s_{f,i}-s_{i})-r_{s,i}c,\label{eq:E111_s}\\
0 & =V_{e,i}\frac{\sigma_{i}}{\bar{K}_{e,i}+e_{i}\sigma_{i}}-\left(D+k_{e,i}\right),\label{eq:E111_e}\\
0 & =Y_{1}r_{s,1}+Y_{2}r_{s,2}-D,\label{eq:E111_c}\\
x_{i} & \approx\frac{r_{s,i}}{k_{x,i}}.\label{eq:E111_x}\end{align}
These equations do not yield an analytical solution valid for all
$D$ and $s_{f,i}$. However, we show below that they can be solved
for the two limiting cases of low and high $D$, from which the entire
solution can be pieced together.

Before considering these limiting cases, it is useful to note that
eqs.~\eqref{eq:E111_s} and \eqref{eq:E111_c} imply that \begin{align}
c & =Y_{1}\left(s_{f,1}-s_{1}\right)+Y_{2}\left(s_{f,2}-s_{2}\right),\label{eq:E111_cSum}\\
\frac{Y_{i}r_{s,i}}{D} & =\beta_{i}\equiv\frac{Y_{i}(s_{f,i}-s_{i})}{Y_{1}(s_{f,1}-s_{1})+Y_{2}(s_{f,2}-s_{2})}.\label{eq:E111_beta}\end{align}
The first relation says that the steady state cell density is the
sum of the cell densities derived from the two substrates. The second
relation states that $\beta_{i}$, the fraction of biomass derived
from $S_{i}$, equals the fraction of the total specific growth rate,
$D$, supported by $S_{i}$. Evidently, $\beta_{i}\le1$ with equality
being attained only during single-substrate growth.

Eq.~\eqref{eq:E111_beta} immediately yields \begin{equation}
r_{s,i}=\frac{\beta_{i}D}{Y_{i}},\label{eq:E111_rS}\end{equation}
which explains the following well-known empirical observation: At
any given $D$, the mixed-substrate specific uptake rate of $S_{i}$
is always lower than the single-substrate specific uptake rate of
$S_{i}$~\citep[reviewed in][]{egli95}. This correlation is a natural
consequence of the mass balances for the substrates and cells, together
with the constant yield property. The mixed-substrate specific uptake
rate is smaller simply because both substrates are consumed to support
the specific growth rate, $D$, whereas only one substrate supports
the very same specific growth rate during single-substrate growth.

\paragraph{Low dilution rates}

The first limiting case corresponds to dilution rates so small that
both substrates are at subsaturating levels, i.e., $s_{i}\lesssim K_{s,i}\ll s_{f,i}$.
Since both substrates are almost completely consumed, \eqref{eq:E111_cSum}--\eqref{eq:E111_beta}
become\begin{align}
c & \approx Y_{1}s_{f,1}+Y_{2}s_{f,2},\label{eq:E111_c_LowD}\\
r_{s,i} & =\frac{\beta_{i}D}{Y_{i}},\;\beta_{i}\approx\frac{Y_{i}s_{f,i}}{Y_{1}s_{f,1}+Y_{2}s_{f,2}}=\frac{Y_{i}\psi_{i}}{Y_{1}\psi_{1}+Y_{2}\psi_{2}}.\label{eq:E111_rS_LowD}\end{align}
Evidently, $\beta_{i}$ is completely determined by $\psi_{i}$, the
fraction of $S_{i}$ in the feed. Moreover, $\beta_{i}(\psi_{i})$
is an increasing function of $\psi_{i}$ with $\beta_{i}(0)=0$ and
$\beta_{i}(1)=1$. It is identical to $\psi_{i}$ when the yields
on both substrates are the same.

\begin{figure}
\noindent \begin{centering}
\subfigure[]{\includegraphics[width=2.6in]{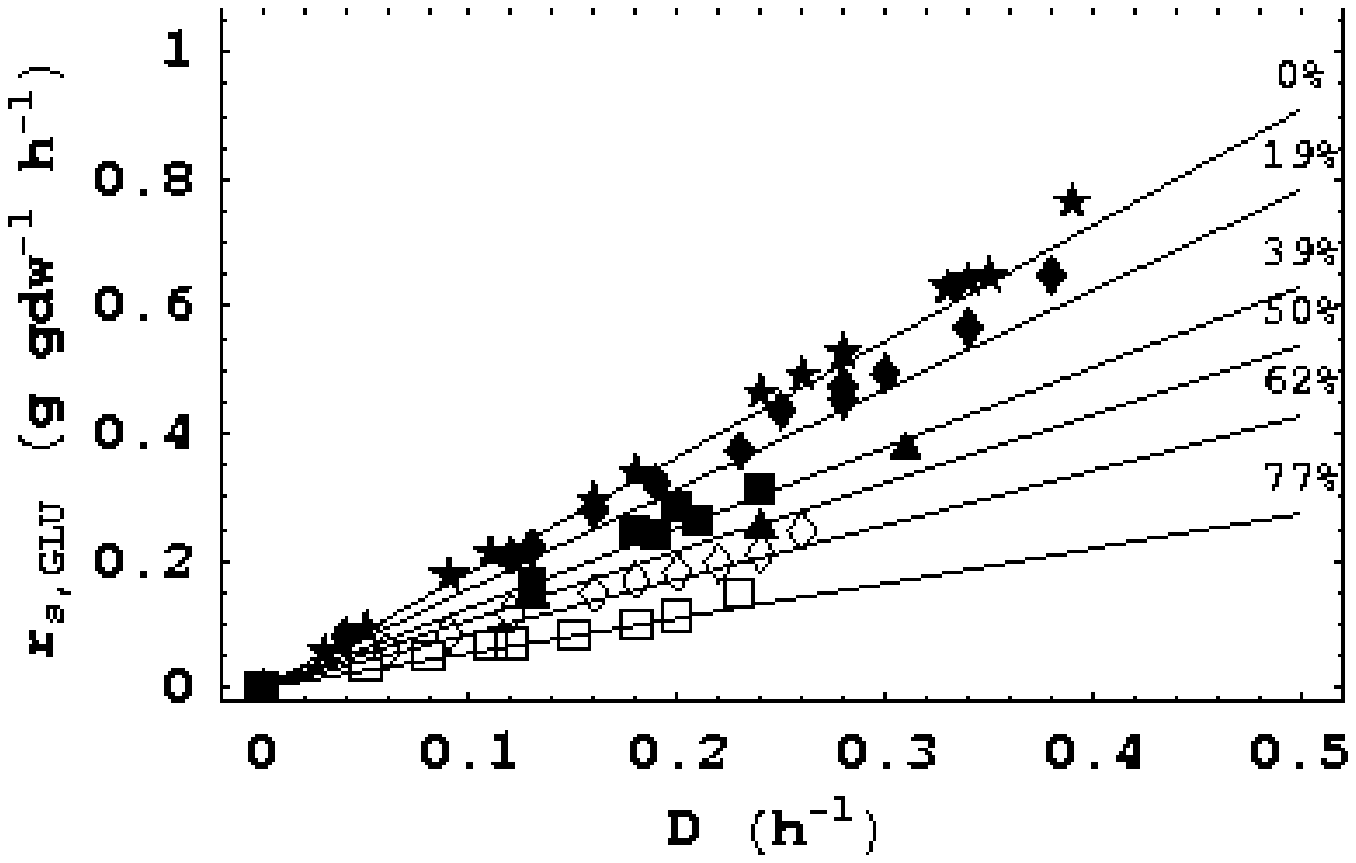}}\hspace*{0.1in}\subfigure[]{\includegraphics[width=2.6in]{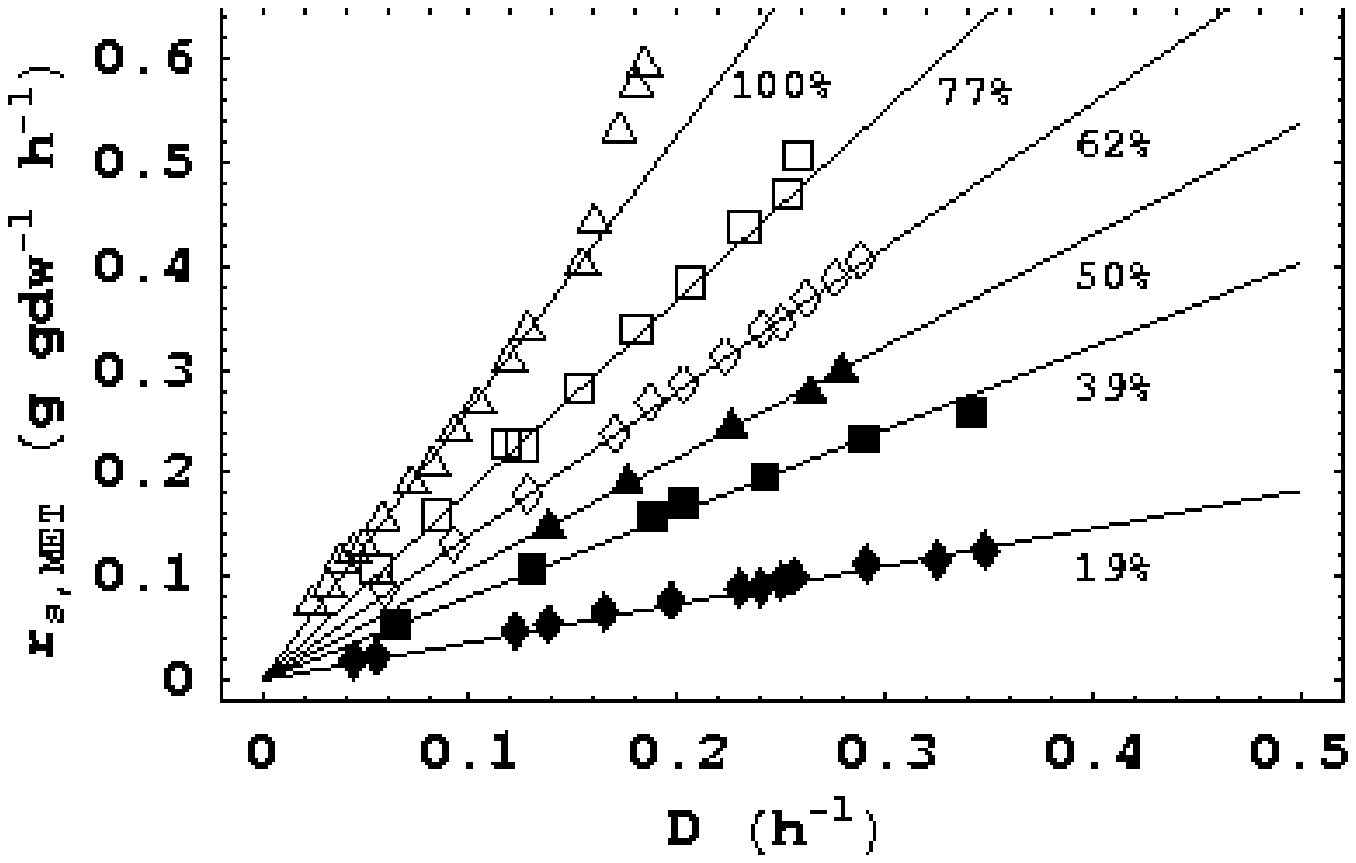}}
\par\end{centering}

\noindent \begin{centering}
\subfigure[]{\includegraphics[width=2.6in]{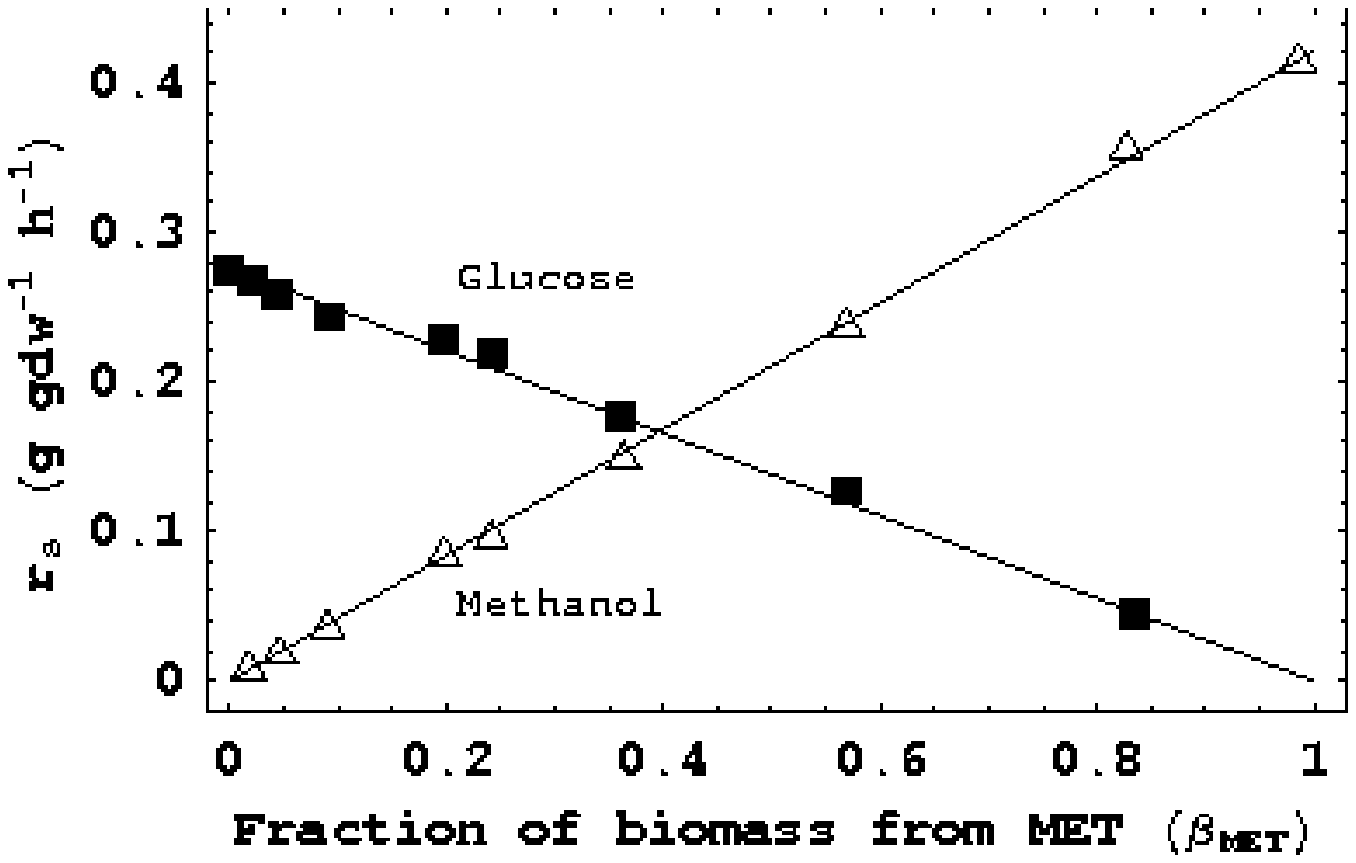}}\hspace*{0.1in}\subfigure[]{\includegraphics[width=2.6in]{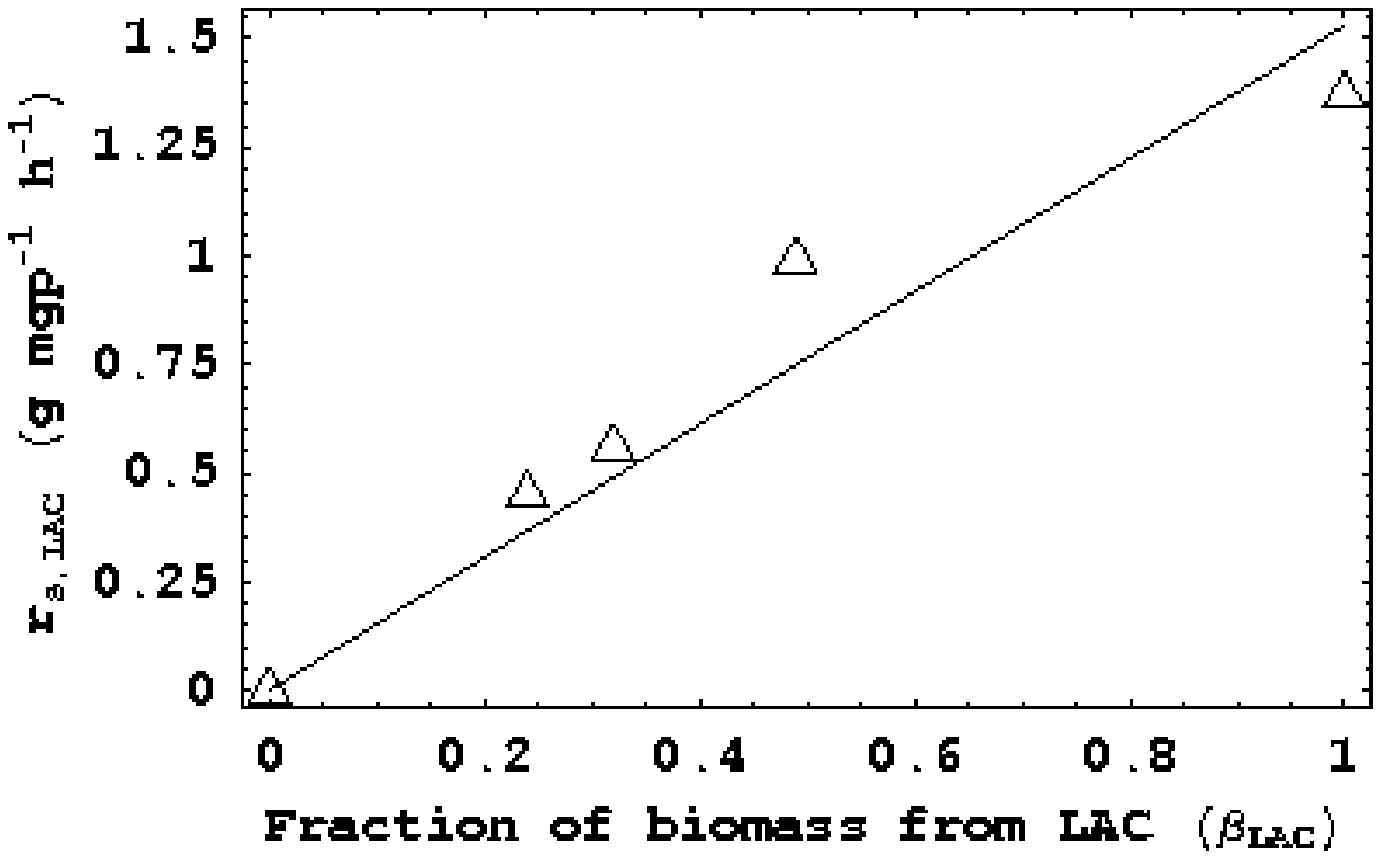}}
\par\end{centering}

\caption{\textbf{\label{fig:E111_rS}Upper panel:} The specific substrate uptake
rate increases linearly when the dilution rate is increased at fixed
feed concentrations. Specific uptake rates of (a)~glucose and (b)~methanol
during growth of \emph{H.~polymorpha} on various mixtures of glucose
and methanol~\citep[Figs.~4--5]{egli86b}. The percentages show the
percent methanol in the feed. The total feed concentration of glucose
and methanol was 5 g~L$^{-1}$. \textbf{Lower panel:} If the fraction
of $S_{i}$ in the feed is increased at a fixed dilution rate, the
specific substrate uptake rate increases linearly with $\beta_{i}$.
(c)~Growth of \emph{H. polymorpha} on a mixture of glucose and methanol
at $D=0.145$~h$^{-1}$~\citep[Fig.~5]{egli82a}. (d)~Growth of
\emph{E. coli} K12 on various mixtures of lactose and glucose at $D=0.4$~h$^{-1}$
\citep[calculated from Fig.~4 of][]{Smith1980}. The lines in the
figures show the specific substrate uptake rates predicted by eq.~(\ref{eq:E111_rS_LowD})
with the experimentally measured single-substrate yields, $Y_{\textnormal{GLU}}=0.55$,
$Y_{\textnormal{MET}}=0.38$ in (a)--(c), and $Y_{\textnormal{GLU}}=Y_{\textnormal{LAC}}$
in (d).}

\end{figure}

Eq.~\eqref{eq:E111_rS_LowD} implies that if $D$ is increased at
fixed feed concentrations, $r_{s,i}$ increases linearly. Likewise,
if $\psi_{i}$, and hence, $\beta_{i}$, is increased at a fixed dilution
rate, $r_{s,i}$ increases linearly with $\beta_{i}$. Both conclusions
are consistent with the data. Fig.~\ref{fig:E111_rS} shows that
eq.~\eqref{eq:E111_rS_LowD} provides excellent fits to the data
for growth of \emph{H. polymorpha} on glucose + methanol and \emph{E.
coli} on glucose + lactose. The validity of \eqref{eq:E111_rS_LowD}
at low dilution rates has been demonstrated for many different systems~\citep[reviewed in][]{kovarova98}.

\begin{figure}
\noindent \begin{centering}
\subfigure[]{\includegraphics[width=2.6in]{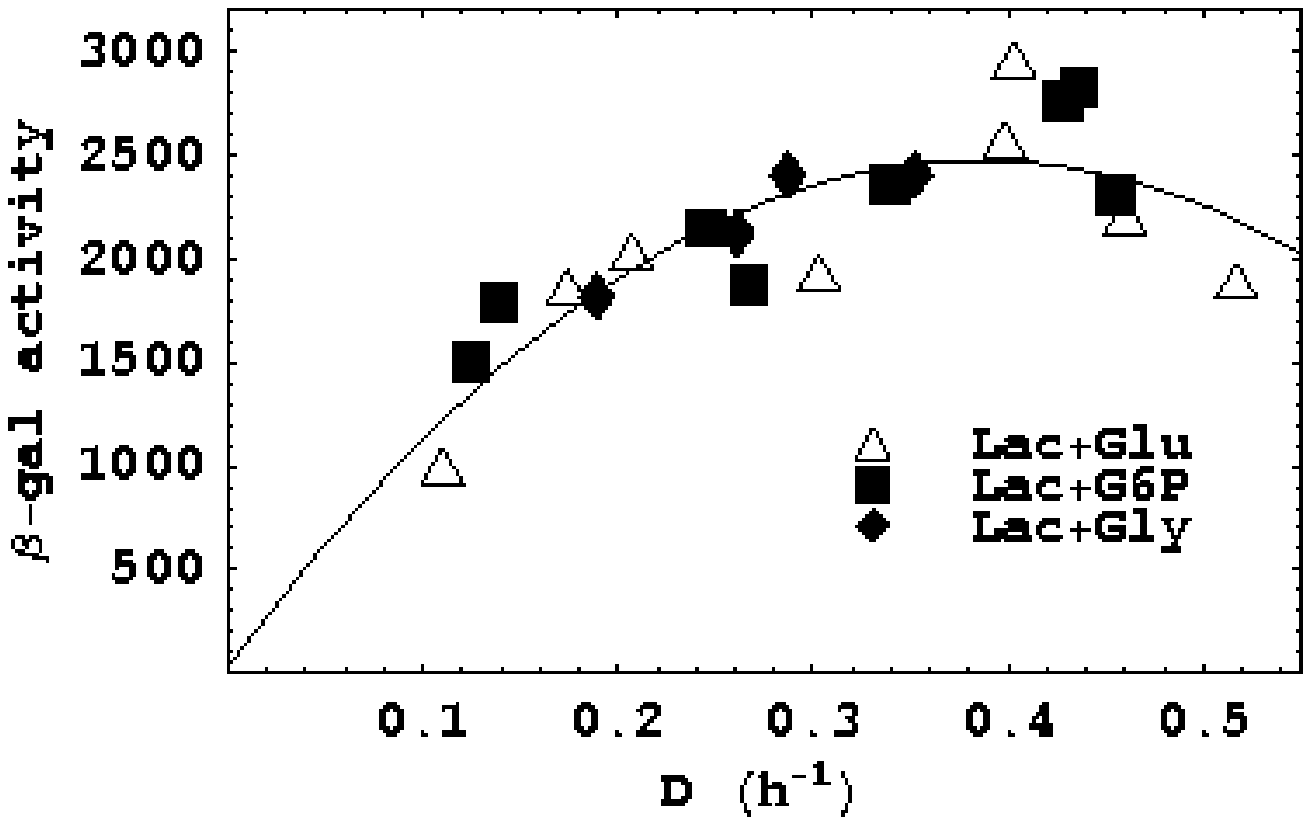}}\hspace*{0.1in}\subfigure[]{\includegraphics[width=2.6in]{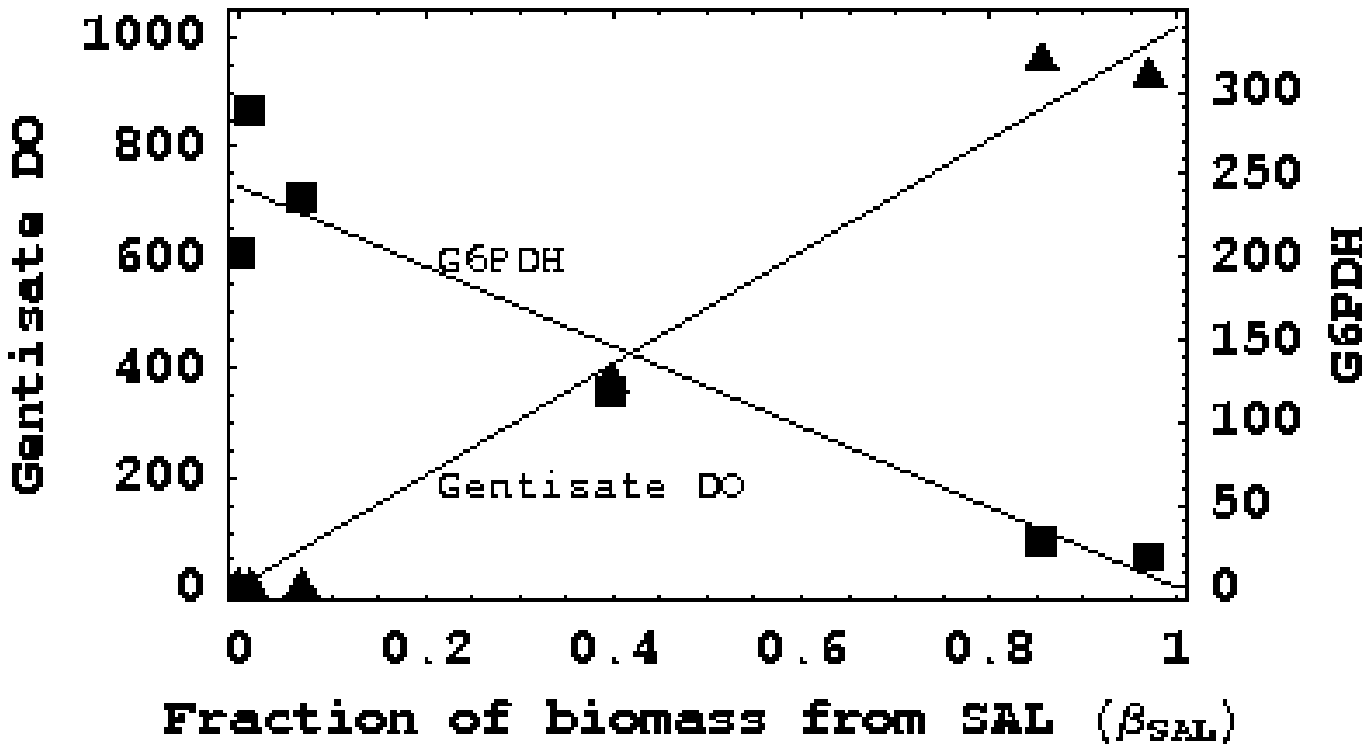}}
\par\end{centering}

\noindent \begin{centering}
\subfigure[]{\includegraphics[width=2.6in]{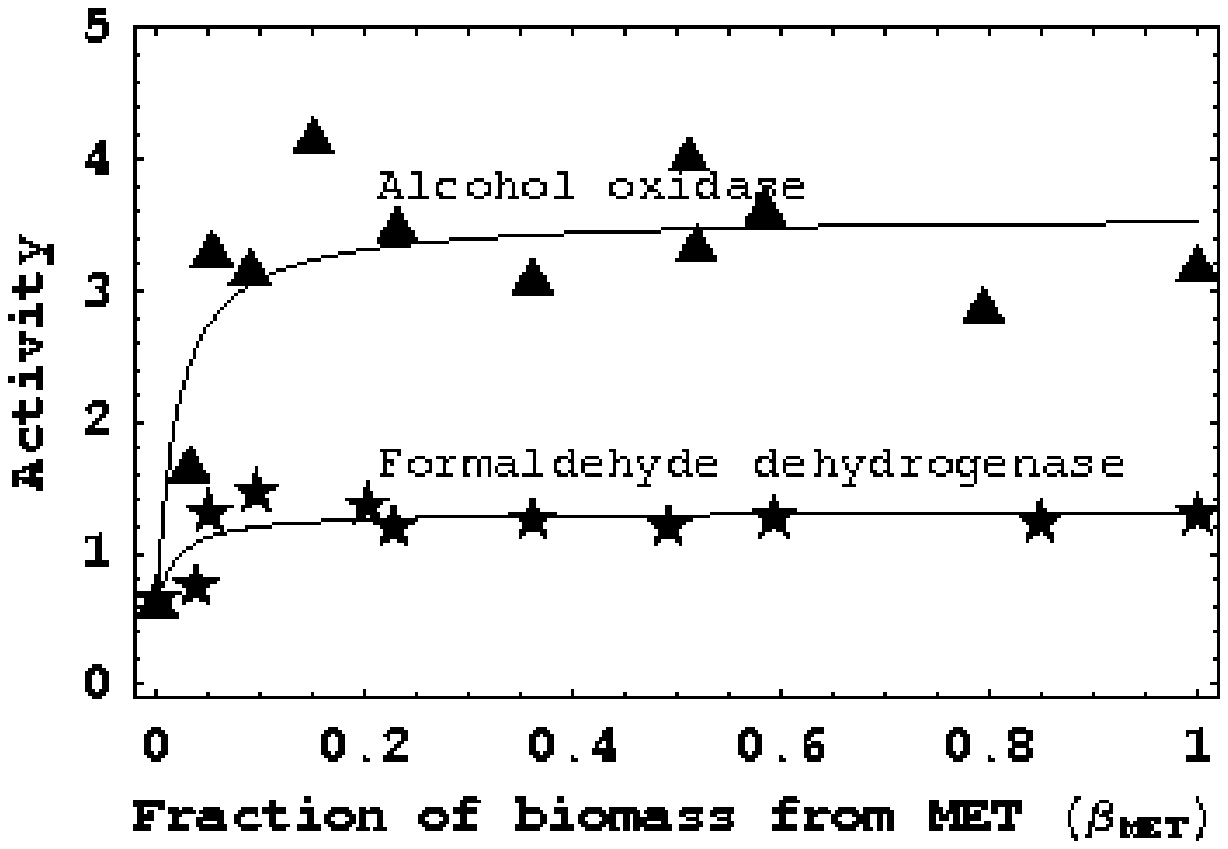}}\hspace*{0.1in}\subfigure[]{\includegraphics[width=2.6in]{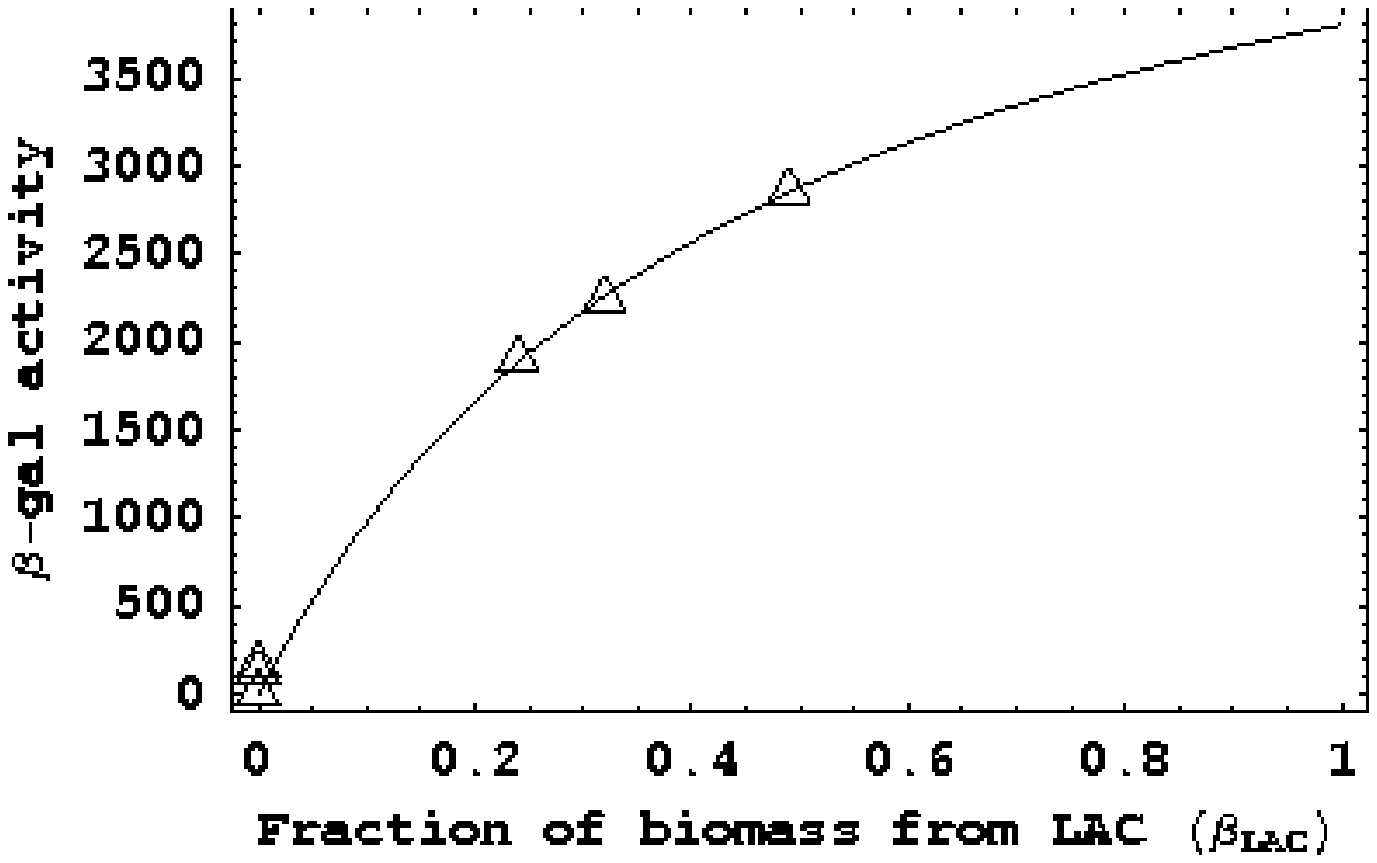}}
\par\end{centering}

\caption{\textbf{\label{fig:E111_e}}Variation of the peripheral enzyme levels:
(a)~Activity of $\beta$-galactosidase during growth of \emph{E.
coli} K12 on mixtures of lactose (1mM) plus glucose (2mM), glucose-6-phosphate
(2mM), or glycerol (4mM)~\citep[Fig.~3]{Smith1980}. (b)~Activity
of gentisate-1,2-dioxygenase and glucose-6-phosphate dehydrogenase
(G6PDH) during growth of \emph{P. aeruginosa} on salicylate + glucose
at $D=0.02$~h$^{-1}$~\citep[Fig.~3]{Rudolph2002}. (c)~Activities
of alcohol oxidase and formaldehyde dehydrogenase during growth of
\emph{H. polymorpha} on various mixtures of glucose and methanol at
$D=0.15$~h$^{-1}$ and $s_{f,t}=5$~g L$^{-1}$~\citep[Figs.~7A,C]{egli82a}.
(d)~Activity of $\beta$-galactosidase during growth of \emph{E.
coli} K12 on various mixtures of lactose and glucose at $D=0.4$~h$^{-1}$
\citep[calculated from Fig~5 of][]{Smith1980}.}

\end{figure}

The physiological variables, $x_{i}$ and $e_{i}$, are completely
determined by the specific uptake rate of $S_{i}$. Indeed, \eqref{eq:E111_x}
and \eqref{eq:E111_e} imply that\begin{align}
x_{i} & \approx\frac{r_{s,i}}{k_{x,i}}=\frac{\beta_{i}D}{Y_{i}k_{x,i}},\label{eq:E111_x_LowD}\\
e_{i} & =\frac{V_{e,i}}{D+k_{e,i}}\frac{\beta_{i}D}{Y_{i}V_{s,i}\bar{K}_{e,i}+\beta_{i}D}.\label{eq:E111_e_LowD}\end{align}
It follows from \eqref{eq:E111_e_LowD} that:

\begin{enumerate}
\item If the dilution rate is increased at fixed feed concentrations, the
steady state enzyme level passes through a maximum. However, at every
dilution rate, the mixed-substrate enzyme level is always lower than
the single-substrate enzyme level.
\item If the fraction of $S_{i}$ in the feed is increased at a fixed dilution
rate, the activity of $E_{i}$ increases with $\beta_{i}$. The increase
is linear if $D$ is small, and hyperbolic if $D$ is large.
\end{enumerate}
Both results are consistent with the data (Fig.~\ref{fig:E111_e}).

Based on studies with batch cultures of \emph{E. coli}, it is widely
believed that glucose and glucose-6-phosphate (G6P) are strong inhibitors
of $\beta$-galactosidase synthesis, whereas glycerol is a weak inhibitor.
It is therefore remarkable that the $\beta$-galactosidase activity
at any given dilution rate is the same in three different continuous
cultures fed with lactose + glucose, lactose + G6P, and lactose +
glycerol (Fig.~\ref{fig:E111_e}a). The model provides a simple explanation
for this data. To see this, let $S_{1}$ and $S_{2}$ denote lactose
and glucose/G6P/glycerol, respectively. Now, \eqref{eq:E111_e_LowD}
implies that any given dilution rate, the influence of $S_{2}$ on
the $\beta$-galactosidase activity, $e_{1}$, is completely determined
by the parameter, \[
\beta_{1}\approx\frac{Y_{1}s_{f,1}}{Y_{1}s_{f,1}+Y_{2}s_{f,2}},\]
which is identical to $\psi_{1}$, the mass fraction of lactose in
the feed, because the yields on lactose, glucose, G6P, and glycerol
are similar. It turns out that the feed concentrations used in the
experiments were such that $\psi_{1}\approx0.45$ in all the experiments.
Consequently, the $\beta$-galactosidase activity at any $D$ is independent
of the chemical identity of $S_{2}$.

\begin{figure}
\noindent \begin{centering}
\subfigure[]{\includegraphics[width=2.6in]{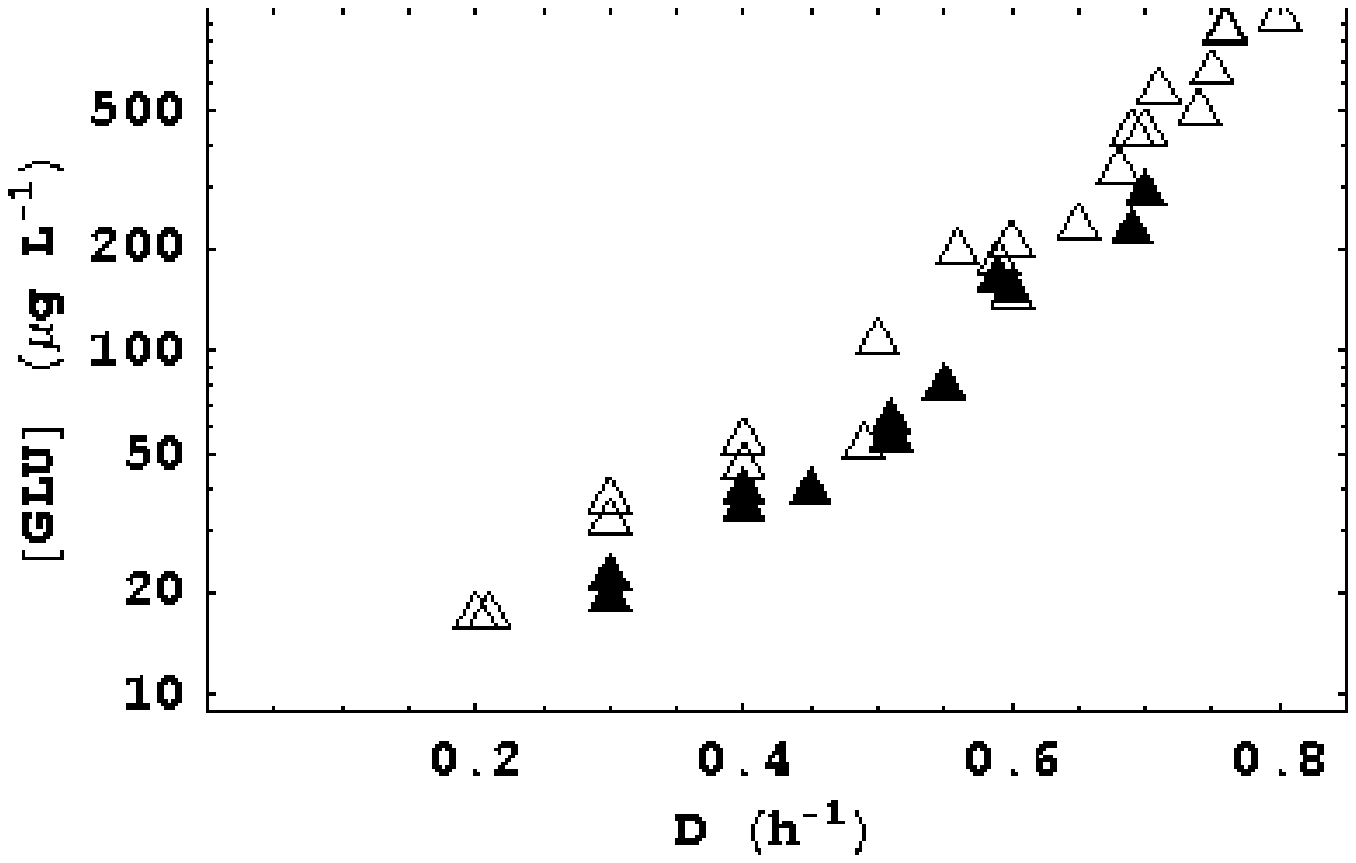}}\hspace*{0.1in}\subfigure[]{\includegraphics[width=2.6in]{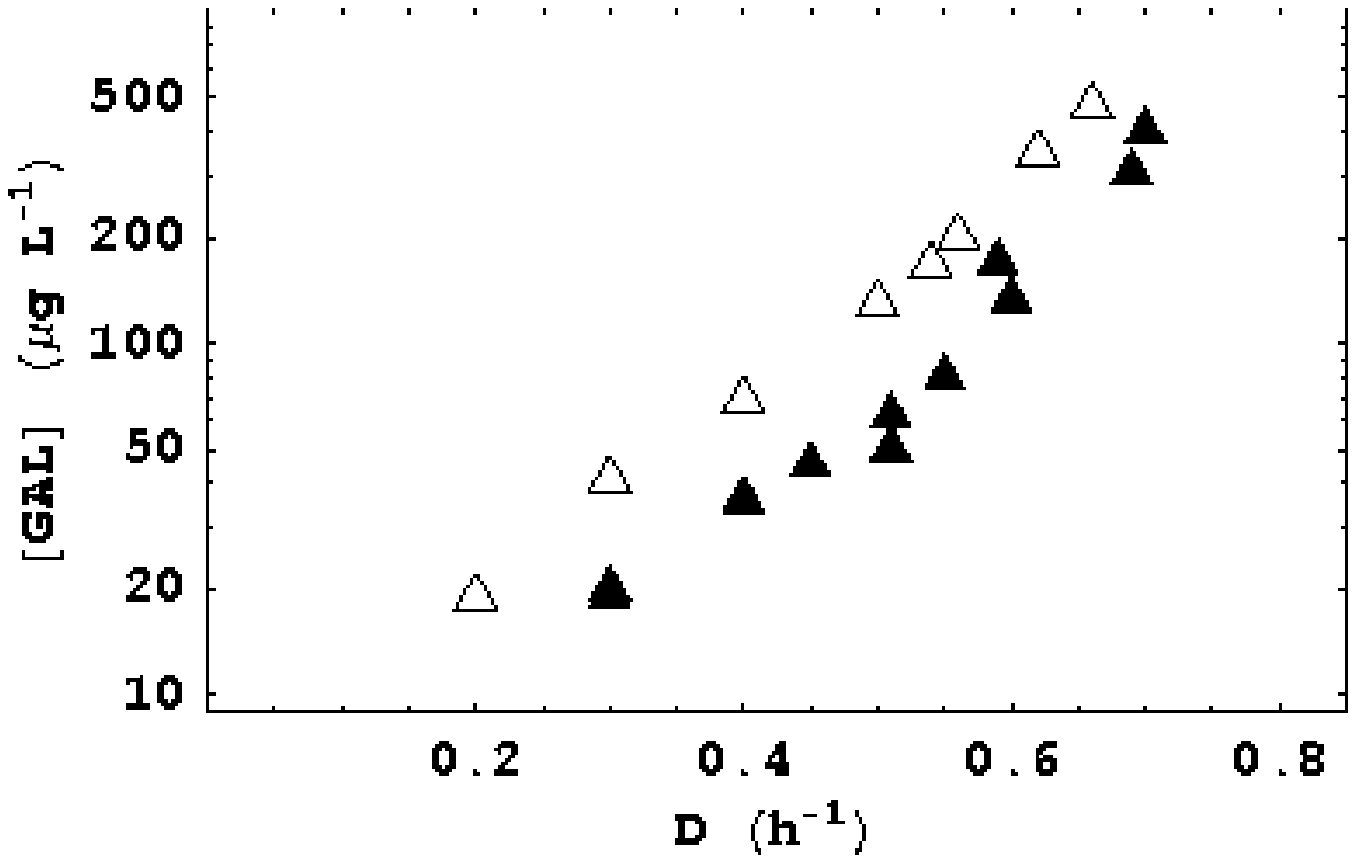}}
\par\end{centering}

\noindent \begin{centering}
\subfigure[]{\includegraphics[width=2.6in]{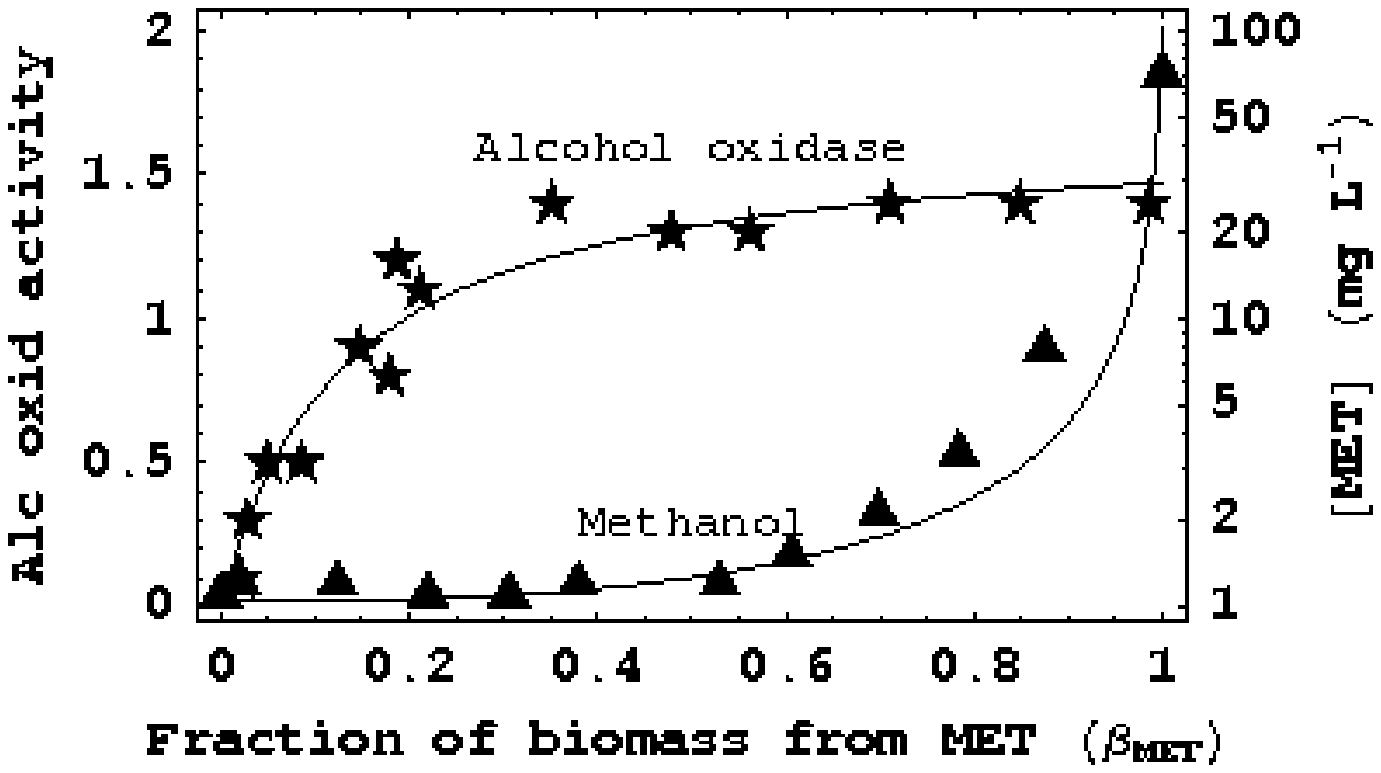}}\hspace*{0.1in}\subfigure[]{\includegraphics[width=2.6in]{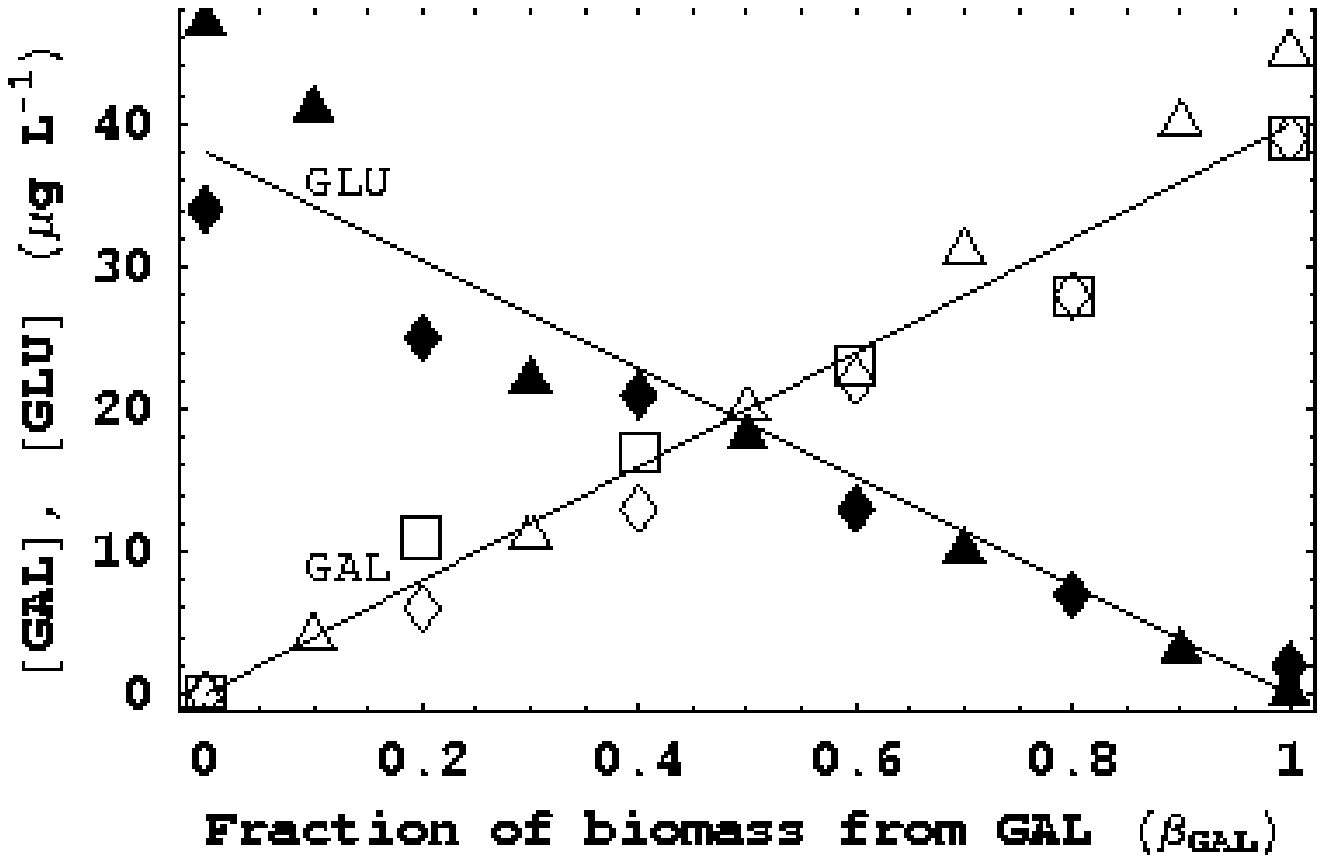}}
\par\end{centering}

\caption{\label{fig:E111_s}\textbf{Upper panel:} Variation of the glucose
and galactose concentrations with the dilution rate during single-
and mixed-substrate growth of \emph{E. coli} ML308 on glucose or/and
galactose~\citep[Appendix~B]{lendenmann94,lendenmann96}. The residual
concentrations of (a)~glucose and (b)~galactose during mixed-substrate
growth ($\blacktriangle$) are lower than the corresponding levels
during single-substrate growth ($\vartriangle$). \textbf{Lower panel:}
The relationship between the substrate concentration and the peripheral
enzyme level. (a)~Growth of \emph{C. boidinii} on various mixtures
of methanol and glucose at $D=0.14$~h$^{-1}$ and $s_{f,t}=5.0$
g~L$^{-1}$ \citep[Fig.~7]{egli93}. When the activity of alcohol
oxidase increases ($0\le\beta_{\textnormal{MET}}\lesssim0.4$), the
methanol concentration is constant. When the activity of alcohol oxidase
saturates ($\beta_{\textnormal{MET}}\gtrsim0.4$), the methanol concentration
increases. (b)~Residual concentrations of galactose (open symbols)
and glucose (closed symbols) during growth on \emph{E. coli} ML308
on various mixtures of galactose and glucose at $D=0.3$~h$^{-1}$
and $s_{f,t}=1$~($\vartriangle,\blacktriangle$), 10~($\lozenge,\blacklozenge$),
and 100~($\square$) mg~L$^{-1}$~\citep{lendenmann94,lendenmann96}.}

\end{figure}

The steady state substrate concentrations are given by the expression\begin{equation}
\sigma_{i}=\frac{r_{s,i}}{V_{s,i}e_{i}}=\frac{1}{Y_{i}V_{s,i}V_{e,i}}\left(D+k_{e,i}\right)\left(\beta_{i}D+Y_{i}V_{s,i}\bar{K}_{e,i}\right),\label{eq:E111_s_LowD}\end{equation}
which follows immediately from \eqref{eq:E111_rS_LowD} and \eqref{eq:E111_e_LowD}.
Evidently, if the dilution rate is increased at fixed feed concentrations,
the substrate concentrations increase. However, comparison with \eqref{eq:E101_s1}
shows that at every dilution rate, the substrate concentrations during
mixed-substrate growth are lower than the corresponding levels during
single-substrate growth. This agrees with the data (Figs.~\ref{fig:E111_s}a,b).
Eq.~\eqref{eq:E111_s_LowD} also implies that if the fraction of
$S_{i}$ in the feed, and hence, $\beta_{i}$, is increased at a fixed
dilution rate, the substrate concentration is constant when $\beta_{i}$
is small, and increases when $\beta_{i}$ is large. This occurs because
at small $\beta_{i}$, the linear increase of $r_{s,i}$ is driven
entirely by the linear increase of $e_{i}$. Since the enzyme level
saturates at sufficiently large $\beta_{i}$, further improvement
of $r_{s,i}$ is obtained by a corresponding increase of the substrate
concentration. Fig~\ref{fig:E111_s}c shows that this conclusion
is consistent with the data. To be sure, there are instances in which
the substrate concentrations appear to increase for all $0\le\beta_{i}\le1$
(Fig.~\ref{fig:E111_s}d). The model implies that this occurs because
$Y_{i}V_{s,i}\bar{K}_{e,i}$ is so small that the peripheral enzymes
are saturated at relatively small values of $\beta_{i}D$.

In general, one expects the mixed-substrate steady states to depend
on three independent parameters, namely, $D$, $\sigma_{f,1}$, and
$\sigma_{f,2}$. However, the model implies that at sufficiently small
dilution rates (such that $s_{i}\ll s_{f,i}$), the steady state specific
uptake rates, and hence, the steady state enzyme levels and substrate
concentrations, are completely determined by only two independent
parameters, $D$ and $\beta_{i}(\psi_{i})$. Experiments provide direct
evidence supporting this conclusion. Indeed, the data in Fig.~\ref{fig:E111_s}d
were obtained by performing experiments with three different values
of the total feed concentration ($s_{f,t}=1,10,100$~mg/L). It was
observed that at any given fraction of galactose in the feed, the
residual sugar concentrations were the same, regardless of the total
feed concentration. Thus, the residual substrate concentrations are
completely determined by the dilution rate and the fraction of the
substrates in the feed (rather than the absolute values of the feed
concentrations). The same is true of the enzyme levels (Fig.~\ref{fig:E111_e}b).
Athough three different substrates (glucose, G6P, glycerol) were used
in the experiments, the $\beta$-galactosidase activity at any given
$D$ was the same regardless of the identity of the substrate, since
the fraction of lactose in the feed was identical ($\sim$0.45) in
all three experiments.

\paragraph{High dilution rates}

Eq.~\eqref{eq:E111_s_LowD} implies that if the dilution is increased
at fixed feed concentrations, the substrate concentrations increase
monotonically. At a sufficiently high dilution rate, $s_{2}$ approaches
saturating levels ($s_{2}\gg K_{s,2}$), while $s_{1}$ remains at
subsaturating levels ($s_{1}\lesssim K_{s,1}$). The dilution rate
at which $s_{2}$ switches from subsaturating to saturating levels,
denoted $D_{s,2}$, can be roughly estimated from the relation\begin{equation}
1\approx\frac{1}{Y_{2}V_{s,2}V_{e,2}}\left(D_{s,2}+k_{e,2}\right)\left(D_{s,2}\beta_{2}+Y_{2}V_{s,2}\bar{K}_{e,2}\right),\label{eq:SwitchingD}\end{equation}
obtained by letting $\sigma_{2}\approx1$ in eq.~\eqref{eq:E111_s_LowD}.
At dilution rates exceeding $D_{s,2}$, growth is limited by $S_{1}$
only. Based on our earlier analysis of single-substrate growth, we
expect that the steady state concentrations of $S_{1}$ and the physiological
variables are completely determined by $D$, and the cell density
increases linearly with the feed concentration of $S_{1}$ (Fig.~\ref{fig:SSdata}).
We show below that this is indeed the case.

\begin{figure}
\noindent \begin{centering}
\subfigure[]{\includegraphics[width=2.6in]{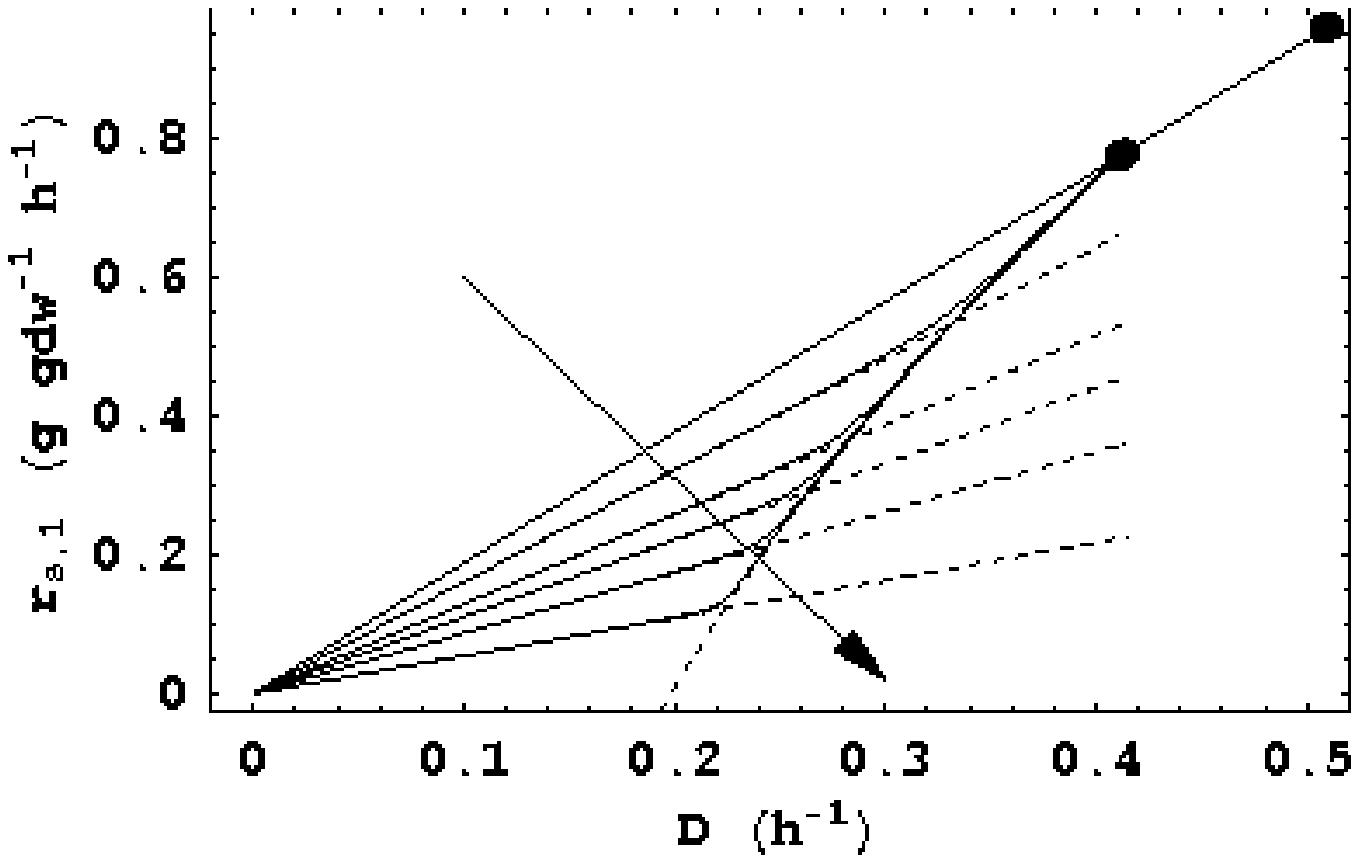}}
\hspace*{0.1in} \subfigure[]{\includegraphics[width=2.6in]{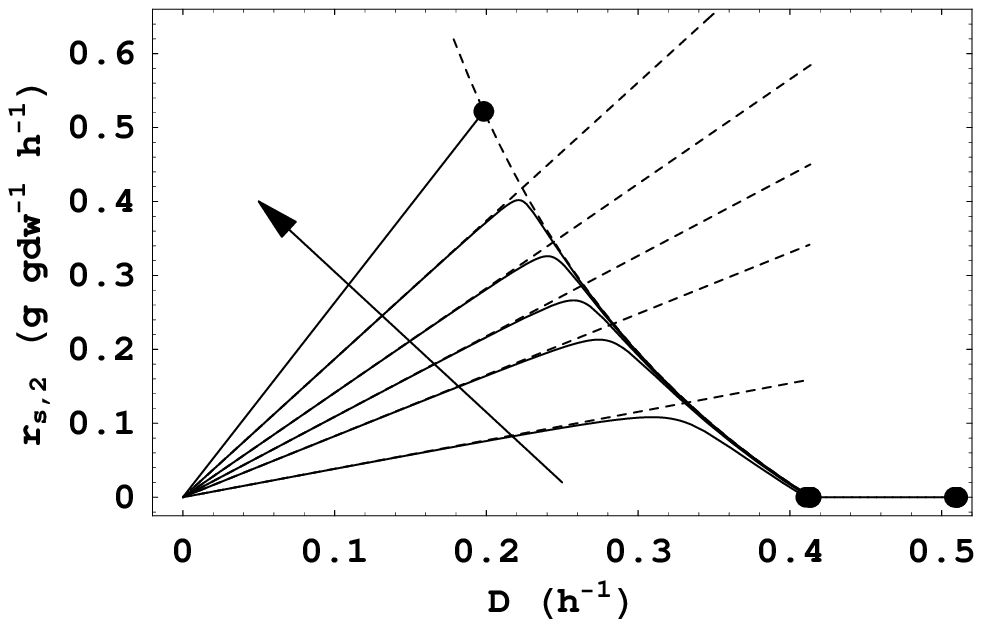}}
\par\end{centering}

\noindent \begin{centering}
\subfigure[]{\includegraphics[width=2.6in]{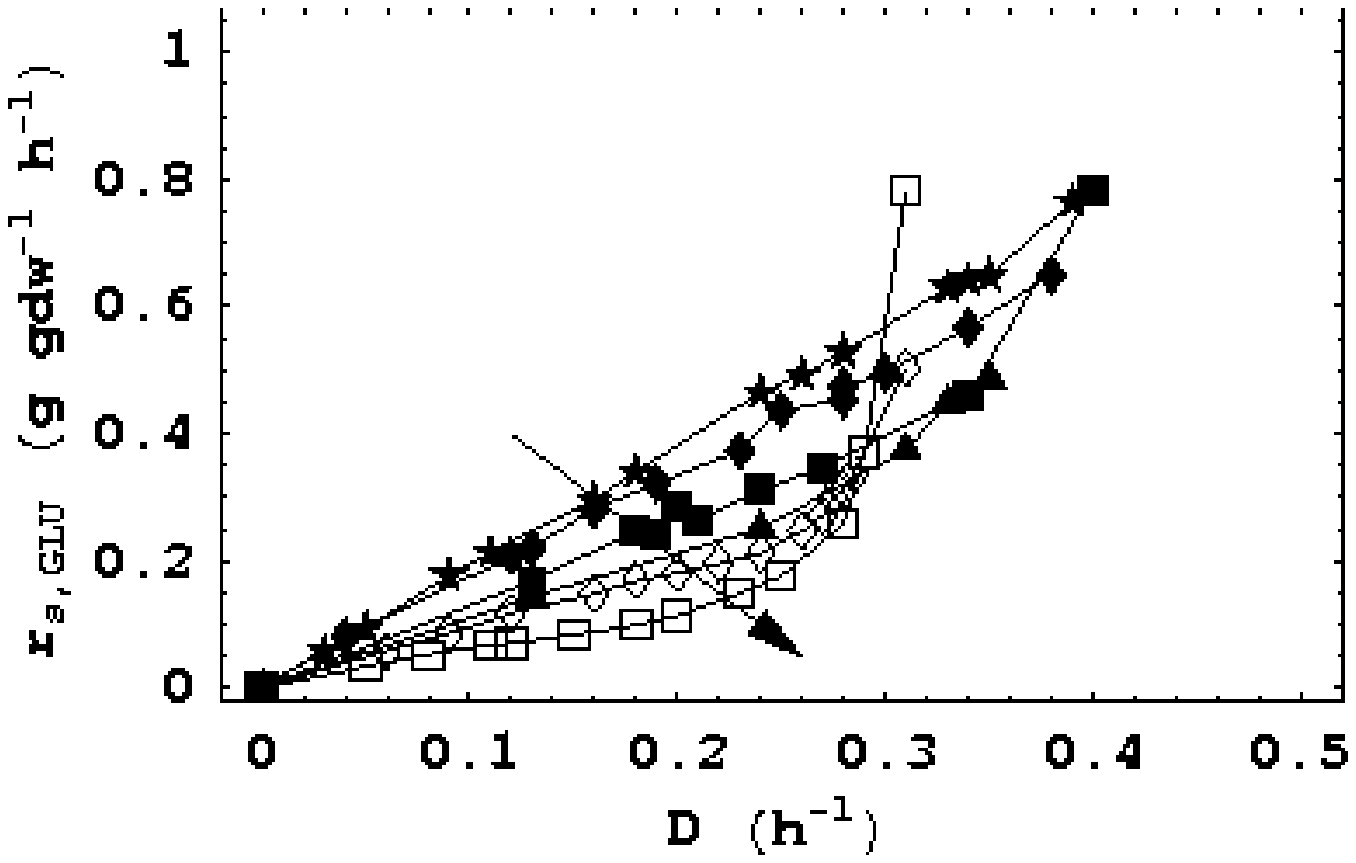}}
\hspace*{0.1in}
\subfigure[]{\includegraphics[width=2.6in]{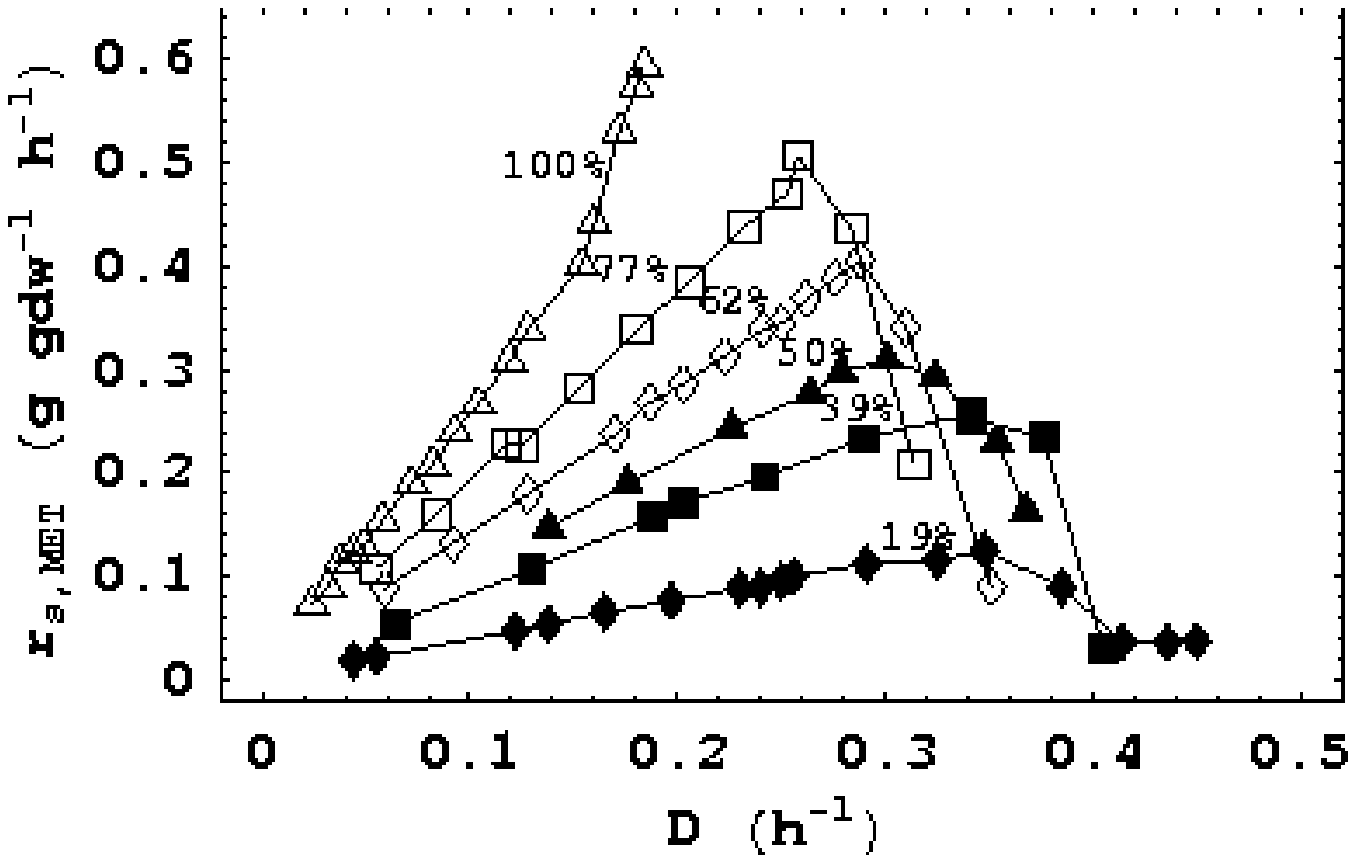}}
\par\end{centering}

\caption{\label{fig:MSUptakeRates}Near the transition dilution rate, the specific
substrate uptake rates are independent of the feed composition. \textbf{Upper
panel:} Model simulations. The arrows point in the direction of increasing
$S_{2}$ in the feed. The full lines show the specific uptake rates
predicted by numerical simulations of the model. The dashed lines
show the specific uptake rates predicted by eqs.~\eqref{eq:E111_rS_LowD}
and \eqref{eq:E111_rS2_HighD}--\eqref{eq:E111_rS1_HighD}. \textbf{Lower
panel:} Specific uptake rates of glucose and methanol during mixed-substrate
growth of \emph{H. polymorpha}~\citep[Figs.~4--5]{egli86b}.}

\end{figure}

When $s_{2}\gg K_{s,2}$, eqs.~\eqref{eq:E111_x} and \eqref{eq:E111_e}
yield \begin{align}
x_{2} & =\frac{r_{s,2}}{k_{x,2}}\approx\frac{V_{s,2}e_{2}}{k_{x,2}},\nonumber \\
e_{2} & =\frac{V_{e,2}}{D+k_{e,2}}-\frac{\bar{K}_{e,2}}{\sigma_{2}}\approx\frac{\bar{K}_{e,2}}{D+k_{e,2}}\left(D_{t,2}-D\right),\label{eq:E111_e2_HighD}\end{align}
where the second relation was obtained by appealing to the approximations,
$\sigma_{2},\sigma_{f,2}\approx1$. Evidently, \begin{equation}
r_{s,2}\approx V_{s,2}e_{2}\label{eq:E111_rS2_HighD}\end{equation}
is a function of $D$. Since $Y_{1}r_{s,1}+Y_{2}r_{s,2}=D$, we conclude
that\begin{equation}
r_{s,1}=\frac{D-Y_{2}r_{s,2}}{Y_{1}},\label{eq:E111_rS1_HighD}\end{equation}
and\begin{align}
x_{1} & =\frac{r_{s,1}}{k_{x,1}},e_{1}=\frac{V_{e,1}}{D+k_{e,1}}\frac{x_{1}}{K_{e,1}+x_{1}},\label{eq:E111_e1_HighD}\end{align}
are also functions of $D$.

\begin{table}
\caption{\label{tab:ParamValues}Parameter values used to simulate the growth
of \emph{H. polymorpha} on mixtures of glucose ($S_{1}$) and methanol
($S_{2}$). The yields are based on experimental measurements~(Egli
et al, 1986). The remaining parameters, which have not been measured
experimentally, are based on order-of-magnitude estimates obtained
from well-characterized systems, such as the \emph{lac} operon~(Shoemaker
et al, 2003, Appendix A)}

\noindent \begin{centering}
\begin{tabular}{|c|c|c|c|c|c|c|c|}
\hline
$i$ & $Y_{i}$ & $V_{s,i}$ (h$^{-1}$) & $K_{s,i}$ (g L$^{-1}$) & $k_{x,i}$ (h$^{-1}$) & $V_{e,i}$ (h$^{-1}$) & $K_{e,i}$ & $k_{e,i}$ (h$^{-1}$)\tabularnewline
\hline
\hline
1 & 0.55 & $10^{3}$ & $10^{-2}$ & $10^{3}$ & $9\times10^{-4}$ & $7\times10^{-4}$ & 0.05\tabularnewline
\hline
2 & 0.38 & $10^{3}$ & $10^{-2}$ & $10^{3}$ & $3\times10^{-4}$ & $6\times10^{-4}$ & 0.05\tabularnewline
\hline
\end{tabular}
\par\end{centering}
\end{table}

Since $r_{s,1}$ and $r_{s,2}$ are completely determined by $D$,
the curves representing the specific substrate uptake rates at various
feed compositions must collapse into a single curve. This conclusion
is consistent with the data for growth of \emph{H. polymorpha} on
glucose + methanol (Figs.~\ref{fig:MSUptakeRates}c,d). Indeed, simulations
of the model, performed with the parameter values in Table~2, are
in good agreement with the data, with the exception of two minor discrepancies
(Figs.~\ref{fig:MSUptakeRates}a,b). The specific uptake rates of
glucose corresponding to 77\%, and 62\% methanol in the feed are discrepant
for $D\gtrsim0.25$~h$^{-1}$ (curves labeled $\square$ and $\lozenge$
in Fig.~\ref{fig:MSUptakeRates}c). Likewise, the single-substrate
specific uptake rate of methanol deviates from the model prediction
when $D\gtrsim0.15$~h$^{-1}$ (curve labeled $\triangle$ in Fig.~\ref{fig:MSUptakeRates}d).
Under these exceptional conditions, the residual methanol concentrations
are so high that toxic effects arise, thus reducing the yields on
glucose and methanol. The observed specific uptake rates are therefore
higher than the predicted rates.

\begin{figure}
\noindent \begin{centering}
\subfigure[]{\includegraphics[width=2.6in]{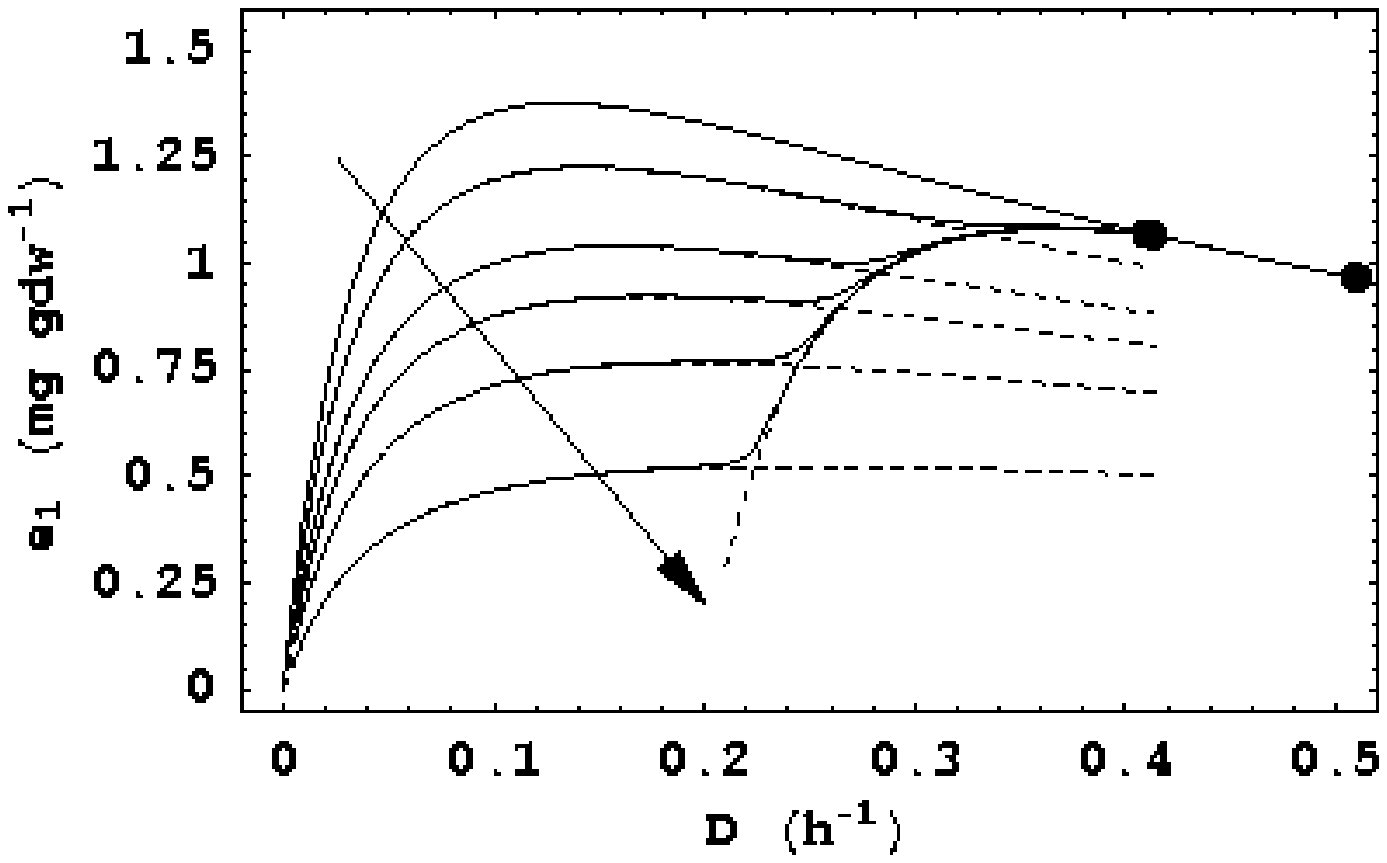}}\hspace*{0.1in}\subfigure[]{\includegraphics[width=2.6in]{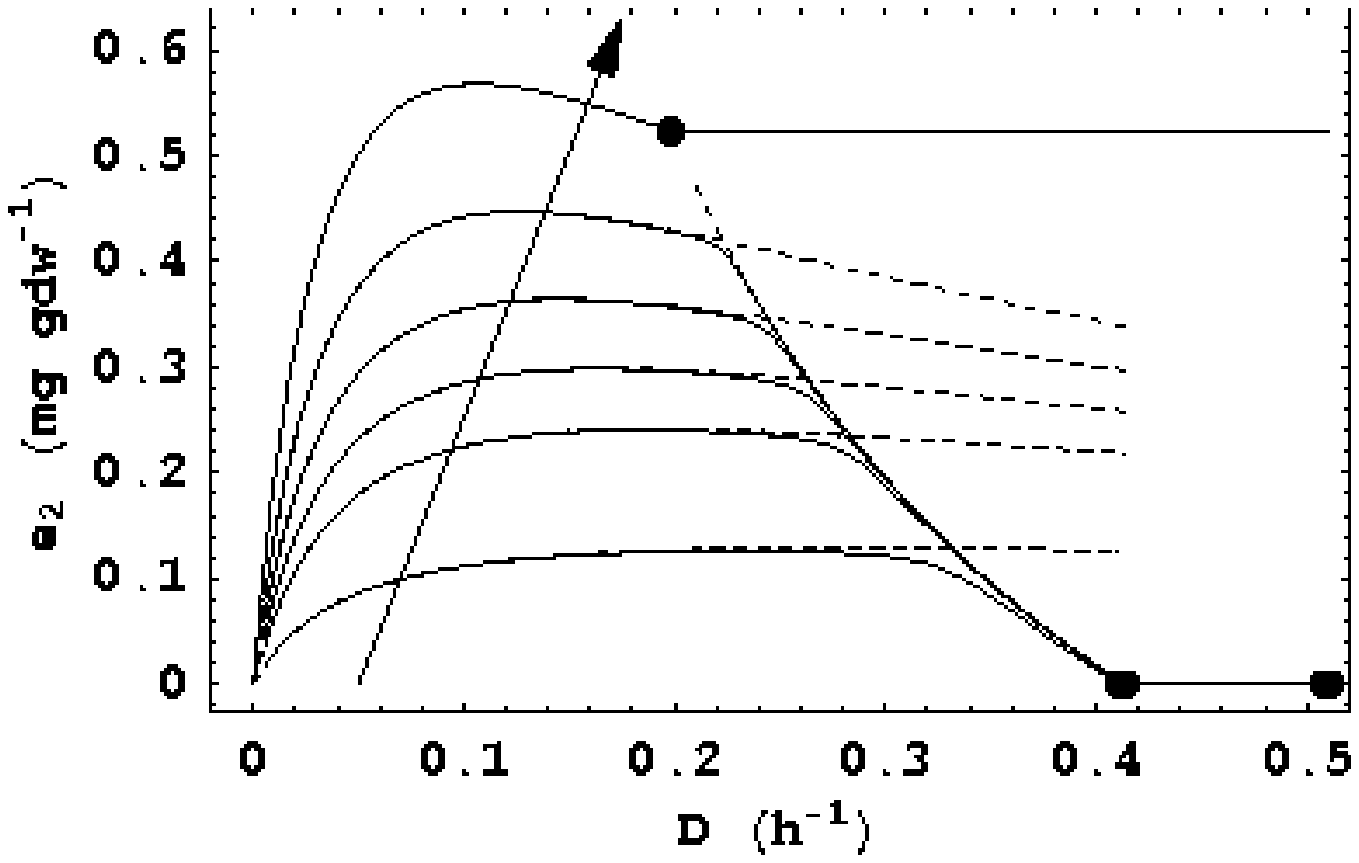}}
\par\end{centering}

\noindent \begin{centering}
\subfigure[]{\includegraphics[width=2.6in]{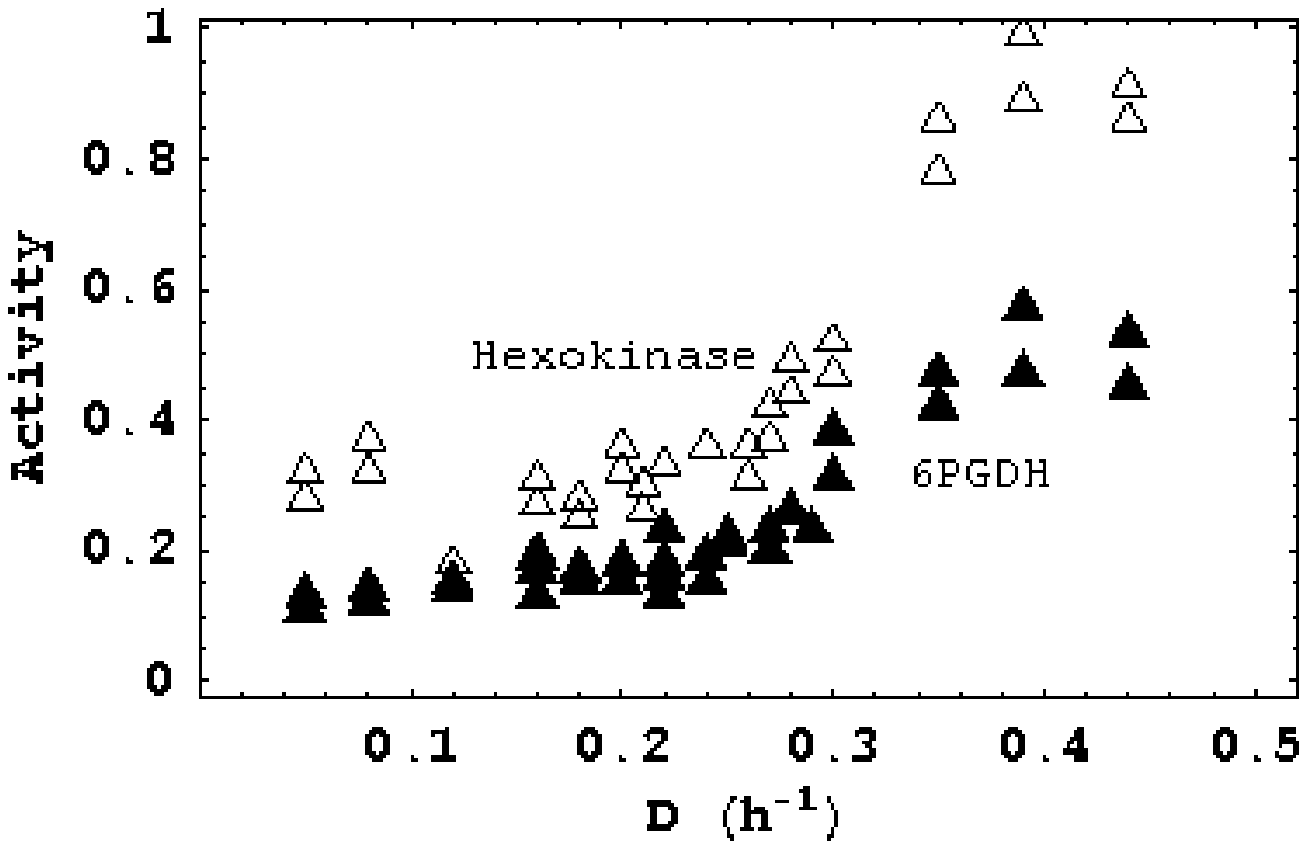}}\hspace*{0.1in}\subfigure[]{\includegraphics[width=2.6in]{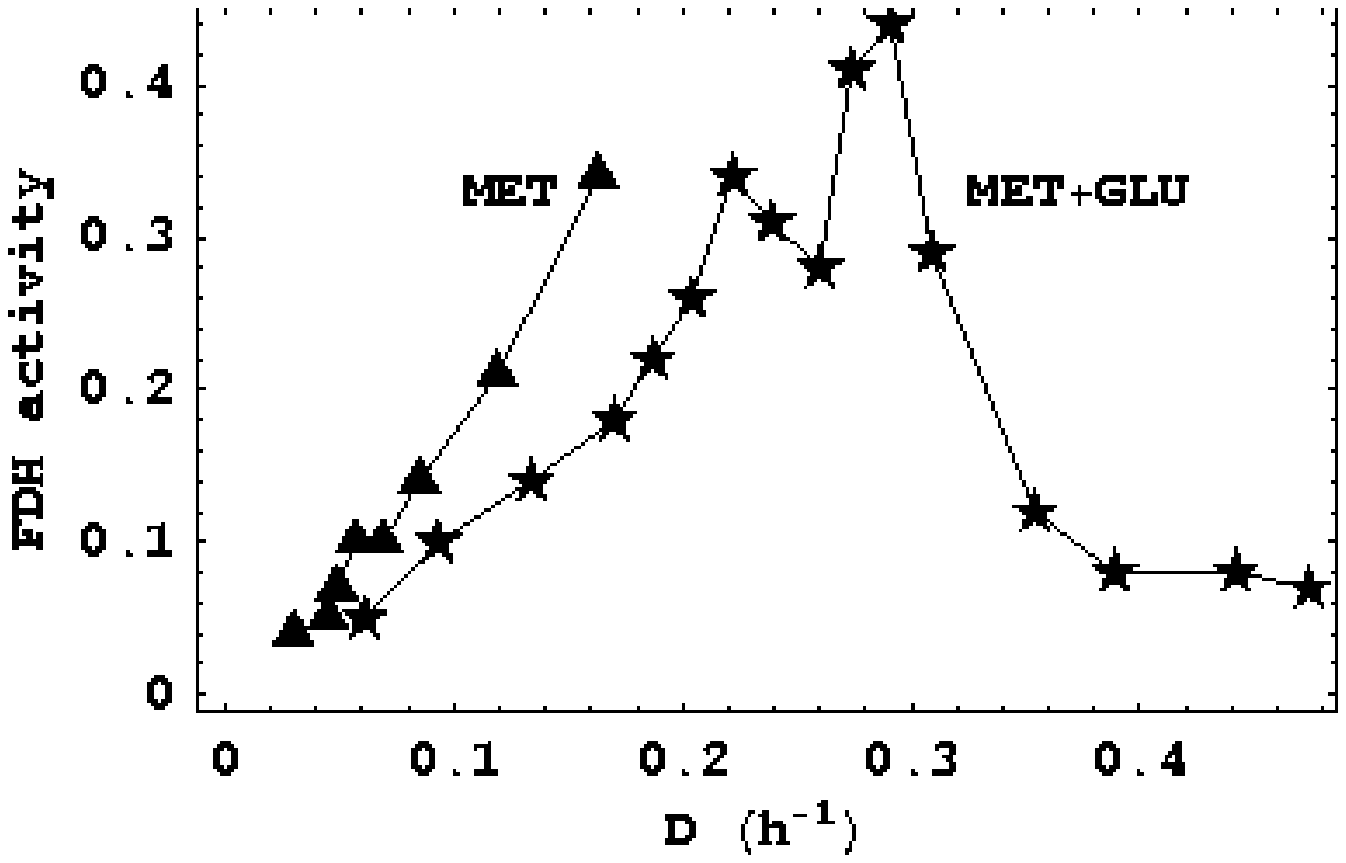}}
\par\end{centering}

\caption{\label{fig:EnzymeLevels}Variation of the peripheral enzyme levels
for glucose and methanol. \textbf{Upper panel:} Model simulations.
The arrows point in the direction of increasing $S_{2}$ in the feed.
The full lines show the specific uptake rates predicted by numerical
simulations of the model. The dashed lines show the enzyme levels
predicted by eqs.~\eqref{eq:E111_e_LowD} and \eqref{eq:E111_e2_HighD},
\eqref{eq:E111_e1_HighD}. \textbf{Lower panel:} Activities of (c)~the
glucose enzymes, hexokinase and 6-phoshogluconate dehydrogenase (6PGDH),
and (d) the methanol enzyme, formaldehyde dehydrogenase, during mixed-substrate
growth of \emph{H. polymorpha }on 3 g~L$^{-1}$ glucose and 2 g~L$^{-1}$
methanol~\citep[Fig.~4]{egli82b}.}

\end{figure}

Eqs.~\eqref{eq:E111_e2_HighD}--\eqref{eq:E111_rS2_HighD} imply
that at sufficiently high dilution rates, $e_{2}$ and $r_{s,2}$
decrease with $D$ until they become zero at $D=D_{t,2}$ (Fig.~\ref{fig:EnzymeLevels}b).
Consequently, $r_{s,1}$ and $e_{1}$ increase with $D$ to ensure
that the total specific growth rate remains equal to $D$ (Fig.~\ref{fig:EnzymeLevels}a).
Both conclusions are consistent with the data for growth of \emph{H.
polymorpha} on glucose + methanol. The activities of the glucose enzymes,
hexokinase and 6-phosphogluconate dehydrogenase, increase with $D$
(Fig.~\ref{fig:EnzymeLevels}c), while the activity of the methanol
enzyme, formaldehyde dehydrogenase, decreases with $D$ (Fig.~\ref{fig:EnzymeLevels}d).

\begin{figure}
\noindent \begin{centering}
\subfigure[]{\includegraphics[width=2.6in]{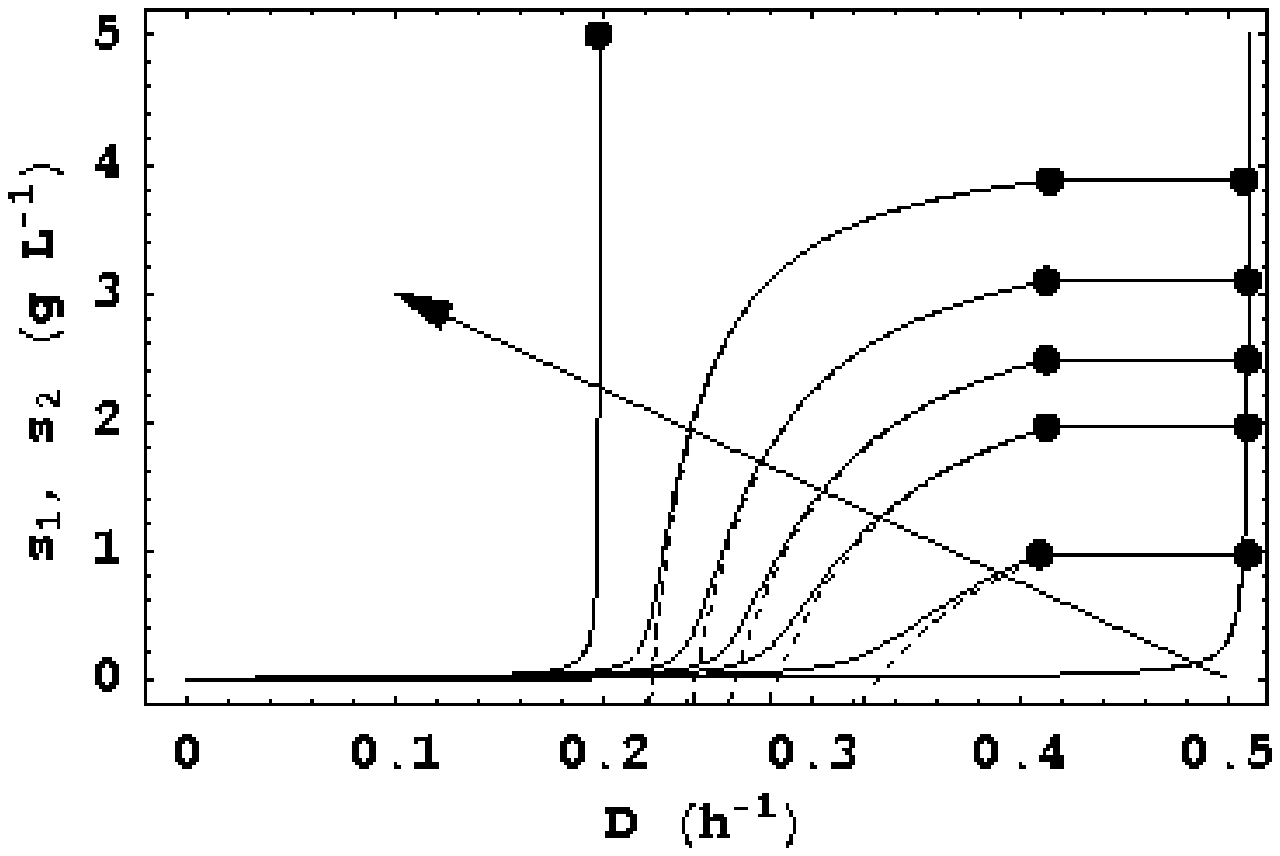}}\hspace*{0.1in}\subfigure[]{\includegraphics[width=2.6in]{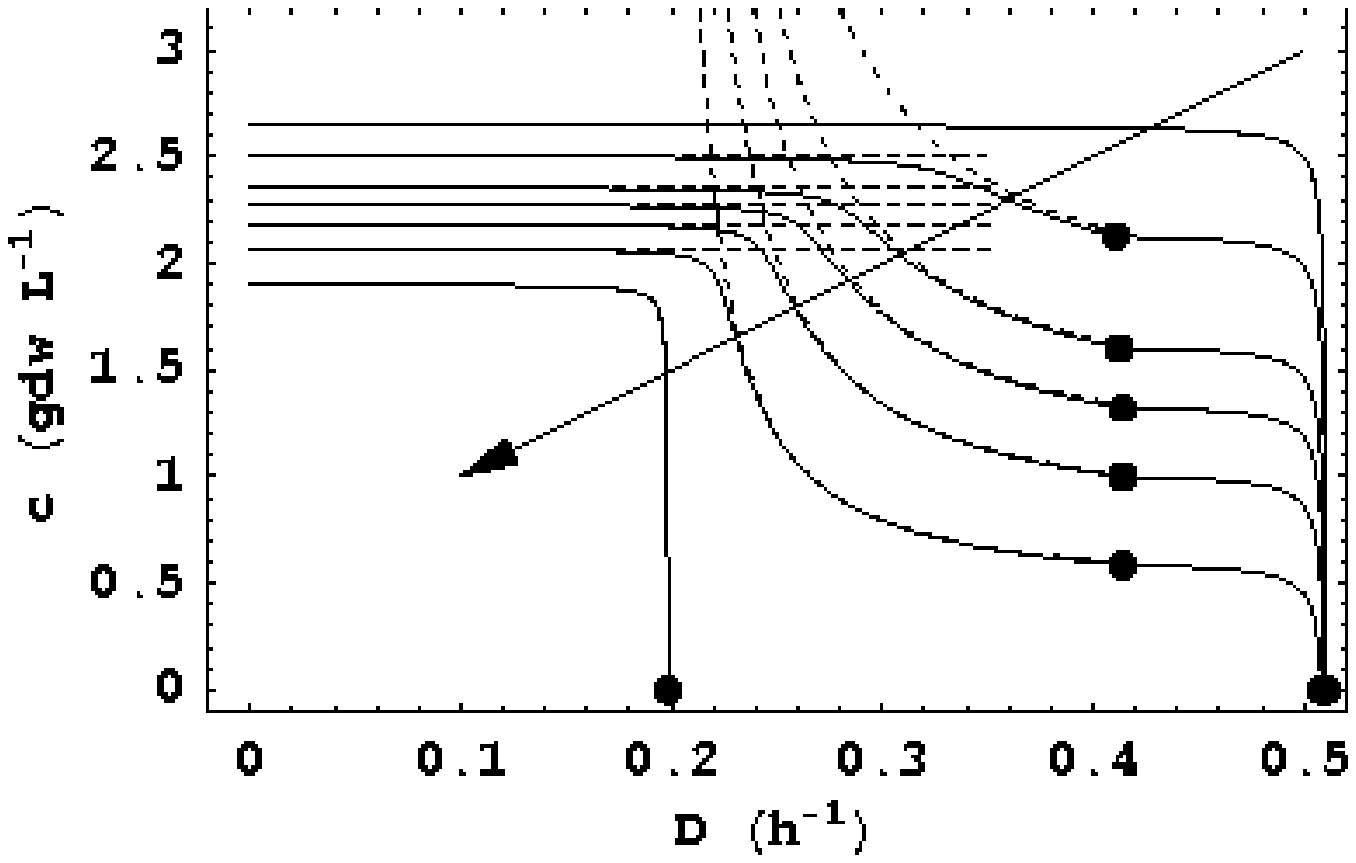}}
\par\end{centering}

\caption{\label{fig:SubstrateCellConcs}The model simulations are in good agreement
with the data for growth of \emph{H. polymorpha} on glucose + methanol
shown in Fig.~\ref{f:GlucoseMethanol}. The arrows point in the direction
of increasing $S_{2}$ in the feed. The full lines show the substrate
concentrations and cell density predicted by numerical simulations
of the model. The dashed lines show the substrate concentrations and
cell density predicted by eqs.~\eqref{eq:E111_c_LowD}, \eqref{eq:E111_s_LowD}
and \eqref{eq:E111_s1_HighD}--\eqref{eq:E111_s2_HighD}.}

\end{figure}

Since $r_{s,1}$ and $e_{1}$ are functions of $D$, so is\begin{equation}
\sigma_{1}=\frac{r_{s,1}}{V_{s,1}e_{1}}.\label{eq:E111_s1_HighD}\end{equation}
In fact, the only variables that depend on the feed concentrations
are the cell density and the concentration of $S_{2}$. To see this,
it suffices to observe that eq.~\eqref{eq:E111_s} yields\begin{align}
c & =\frac{D(s_{f,1}-s_{1})}{r_{s,1}}\approx\left(\frac{D}{r_{s,1}}\right)s_{f,1},\label{eq:E111_c_HighD}\\
s_{f,2}-s_{2} & =\frac{r_{s,2}c}{D}\approx\left(\frac{r_{s,2}}{r_{s,1}}\right)s_{f,1},\label{eq:E111_s2_HighD}\end{align}
where the terms in parentheses are functions of $D$. As expected,
$c$ increases linearly with $s_{f,1}$ because growth is limited
by $S_{1}$ only. Fig.~\ref{fig:SubstrateCellConcs} shows that numerical
simulations of the model with the parameter values in Table~2 are
in good agreement with the data for growth of \emph{H. polymorpha}
on glucose + methanol (Fig.~\ref{f:GlucoseMethanol}).

\paragraph{The patched limiting solutions approximate the exact solution}

Taken together, the two approximate solutions corresponding to low
and high dilution rates coincide with the (exact) numerical solution
at all dilution rates, except in a small neighborhood of the dilution
rate at which the two approximate solutions intersect (compare the
dashed and full lines in Figs.~\ref{fig:MSUptakeRates}--\ref{fig:SubstrateCellConcs}).
This remarkable agreement between the approximate and exact solution
obtains because as $D$ increases, $s_{2}$ changes from subsaturating
to saturating levels so rapidly that the change is, for all practical
purposes, discontinuous. The switching dilution rate, $D_{s,2}$,
provides a good approximation to the dilution rate at which this near-discontinuous
change occurs. The model therefore provides explicit formulas that
approximate the specific uptake rates, substrate concentrations, cell
density, and enzyme levels at all dilution rates and saturating feed
concentrations. The approximate solutions corresponding to low and
high dilution rates are valid for $D<D_{s,2}$ and $D>D_{s,2}$, respectively.

\paragraph{Egli's transition dilution rate differs from our transition dilution
rate}

There is an important difference between the transition dilution rates
defined by Egli and us. In our model, $D_{t,2}$ is the dilution rate
at which $e_{2}$, the activity of methanol enzymes becomes zero.
This transition dilution rate, defined by eq.~\eqref{eq:Dt}, \emph{increases}
with the feed concentration of methanol. However, at the saturating
feed concentrations used in the experiments, it is essentially independent
of the feed concentration. On the other hand, Egli defined the transition
dilution rate empirically as the dilution rate at which the residual
methanol concentration achieved a sufficiently high level (e.g., half
the feed concentration of methanol). The mathematical correlate of
this empirically defined transition dilution rate is the switching
dilution rate, $D_{s,2}$ (at which $s_{2}$ switches from subsaturating
to saturating levels). Consistent with Egli's observations, $D_{s,2}$
\emph{decreases} with the fraction of methanol in the feed. Indeed,
\eqref{eq:SwitchingD} yields\[
\beta_{2}=Y_{2}V_{s,2}\bar{K}_{e,2}\frac{D_{t,2}-D_{s,2}}{D_{s,2}(D_{s,2}+k_{e,2})},\]
where $D_{t,2}$ is essentially constant at saturating feed concentrations.
It follows that $D_{s,2}$ is a decreasing function of $\beta_{2}$,
and, hence, $\psi_{2}$. Moreover, in the limiting cases of feeds
consisting of almost pure methanol ($\beta_{2}\approx1$) and pure
glucose ($\beta_{2}\approx0$), the switching dilution rate, $D_{s,2}$,
tends to $D_{c,2}$ and $D_{t,2}$, respectively.

\section{Discussion}

We have shown above that a simple model accounting for only induction
and growth captures the experimental data under a wide variety of
conditions. This does not prove that the model is correct unless the
underlying mechanism is consistent with experiments. Now, the model
is based on following 3 assumptions. (a)~The yield of biomass on
a substrate is constant. (b)~The induction rate follows hyperbolic
kinetics, and regulatory mechanisms have a negligible effect on the
induction rate. (c)~Each substrate is transported by a unique system
of lumped peripheral enzymes. In what follows, we discuss the validity
of these assumptions.

\subsection{Constant yields}

\begin{figure}
\noindent \begin{centering}
\subfigure[]{\includegraphics[width=2.6in]{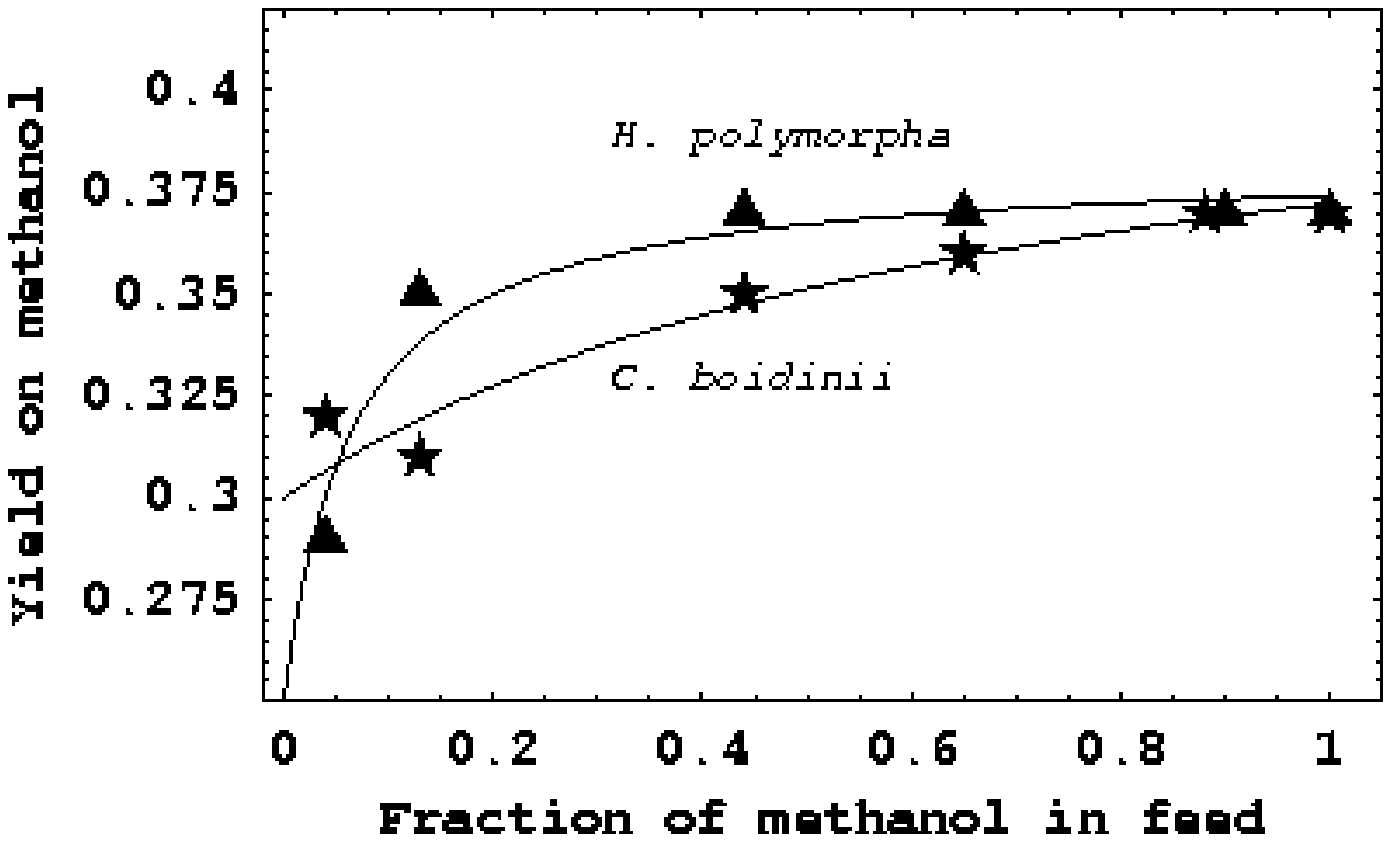}}\hspace*{0.1in}\subfigure[]{\includegraphics[width=2.6in]{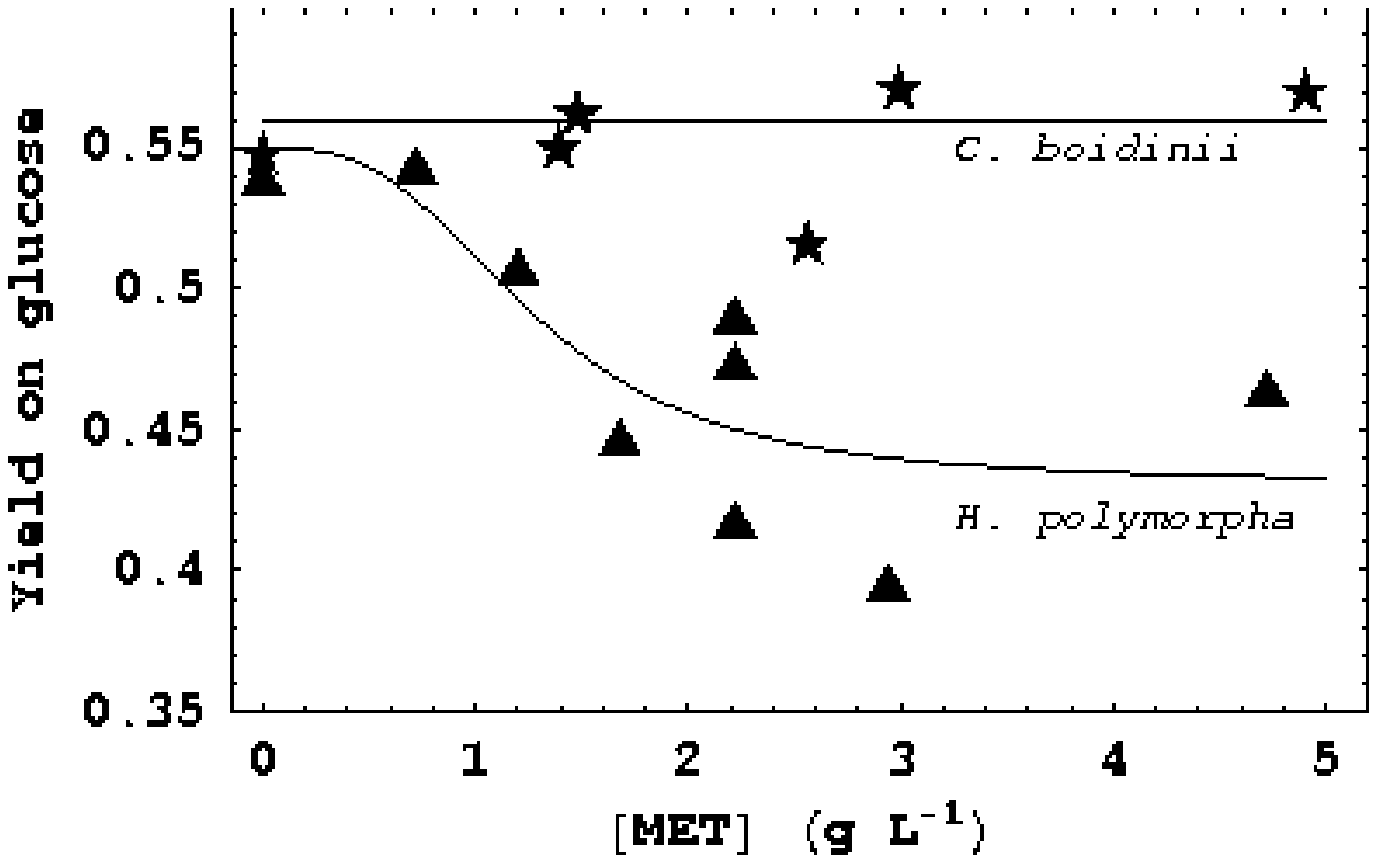}}
\par\end{centering}

\caption{\label{fig:Yields}Yields on the substrates during growth of \emph{H.
polymorpha} and \emph{C. boidinii} on glucose + methanol. (a)~The
yield on methanol is nearly constant at almost all feed concentrations~\citep[Fig.~4]{egli82a}.
The dilution rate is fixed at $\sim$0.15~h$^{-1}$. (b)~In \emph{H.
polymorpha}, the yield on glucose decreases $\sim$20\% in the presence
of high methanol concentrations~\citep[Fig.~3]{egli82b}.}

\end{figure}

Both direct and indirect evidence show that in many cases, the yield
is, for the most part, constant.

Indirect evidence is obtained from experiments in which the specific
growth and substrate uptake rates were measured during mixed-substrate
growth at various dilution rates and feed concentrations. Given these
rates, eq.~\eqref{eq:ContRg} can be satisfied by infinitely many
$Y_{1}$ and $Y_{2}$. It is therefore striking that in all these
experiments, \eqref{eq:ContRg} is satisfied by the single-substrate
yields on the two substrates. These include growth of \emph{E. coli}
on various sugars~\citep{lendenmann96,Smith1980}, organic acids~\citep{narang97a},
glucose + 3-phenylpropionic acid~\citep{kovarova97}; growth of \emph{Chelabacter
heintzii} on glucose + nitrilotriacetic acid~\citep{bally94a}; and
growth of methylotrophic yeasts on methanol + glucose (Fig.~\ref{fig:E111_rS}c)
and methanol + sorbitol~\citep{eggeling81}.

Direct evidence is obtained from experiments in which at least one
of the yields was measured by using radioactively labeled substrates~(Fig.~\ref{fig:Yields}).
These measurements show that the yield on a substrate remains essentially
equal to the single-substrate yield, except when toxic effects arise,
or the fraction of the substrate in the feed is below $\sim$10\%
~\citep{Rudolph2001}.

\subsection{Induction kinetics}

The induction kinetics and regulatory mechanisms can influence the
behavior of the model. Specifically, if the induction kinetics are
sigmoidal, multiple stable steady states arise, resulting in the manifestation
of the \emph{maintenance} or \emph{pre-induction} effect~\citep{narang98b,Narang2007c}.
Likewise, introduction of regulatory effects in the model will alter
its quantitative predictions (although the analysis of the data for
the \emph{lac} operon shows that regulatory effects play a minor role).
However, the essential property of the model, namely, the competitive
(Lotka-Volterra) interaction between the enzymes, is unaffected by
these changes --- the model still predicts the existence of extinction
and coexistence steady states obtained with the minimal model~\citep[see Section~4.1 of][]{Narang2007a}.

\subsection{Transport kinetics}

Many substrates are imported into the cell by two (or more) transport
systems~\citep[Table~1]{Ferenci1996}. In general, these transport
systems dominate substrate import at different substrate concentrations.
For instance, galactose is transported by the low-affinity galactose
permease under substrate-excess conditions, and the high-affinity
methylgalactosidase system under substrate-limiting conditions~\citep{Weickert1993}.
If these transport systems dominate in distinct regimes of the dilution
rate, they are unlikely to have a significant effect on the model.
However, if the two regimes overlap, it is difficult to predict the
outcome without analyzing a suitably extended model. This will be
the focus of future work.

Lumping of peripheral enzymes has a profound effect if the inducer
is synthesized and consumed by coordinately expressed enzymes. Following
Savageau, we illustrate this point by appealing to the \emph{lac}
operon~\citep[Fig.~10]{Savageau2001}. It turns out that the concentration
of the inducer, allolactose, is completely determined by the concentration
of intracellular lactose. Moreover, the dynamics of intracellular
lactose, denoted $x$, are given by the equation\[
\frac{dx}{dt}=V_{s}e_{p}\frac{s}{K_{s}+s}-k_{x}e_{b}x-r_{g}x,\]
where $s$ denotes extracellular lactose, $e_{p}$ denotes permease,
and $e_{b}$ denotes $\beta$-galactosidase. Since permease and $\beta$-galactosidase
are encoded by the same operon, intuition suggests that they must
be coordinately expressed, i.e., $e_{p}=e_{b}$. It follows that the
quasisteady state concentration of intracellular lactose, $x\approx(V_{s}/k_{x})s/(K_{s}+s)$,
is independent of the enzyme level. Thus, positive feedback, a key
component of the model disappears if the inducer (or its precursor)
is synthesized by coordinately expressed enzymes. Under this condition,
the positive feedback generated by one of the enzymes (e.g., permease)
is neutralized by the other enzyme (e.g., $\beta$-galactosidase).

The relative rates of permease and $\beta$-galactosidase expression
are not known, but there is good evidence that galactosidase and transacetylase
are not expressed coordinately~\citep[reviewed in][]{Adhya2003}.
Indeed, Ullmann and coworkers have shown that the higher the protein
synthesis rate, the higher the relative rate of transacetylase synthesis~\citep{Danchin1981}.
Furthermore, this uncoupling occurs at the level of transcription,
and involves the transcription termination factor, Rho~\citep{Guidi-Rontani1984}.
It seems likely that permease and $\beta$-galactosidase are also
uncoupled, but this remains a hypothesis until direct evidence is
obtained.

\section{Conclusions}

We formulated a minimal model of mixed-substrate growth that accounts
for only induction and growth (but no regulation), the two processes
that occur in all microbes including bacteria and yeasts.

\begin{enumerate}
\item Quantitative analysis of the data for both bacteria and yeasts shows
that the decline of the enzyme activities in constitutive and wild-type
cells, typically attributed to specific regulatory mechanisms, is
almost entirely due to dilution.
\item We analyzed the model by constructing the bifurcation diagrams for
both batch and continuous cultures.

\begin{enumerate}
\item The bifurcation diagram for batch cultures describes the substrate
consumption patterns at any given initial substrate concentrations.
It shows that: (i)~When the initial concentrations of one or more
substrates are sufficiently small, threshold effects arise because
the enzymes cannot be induced. (ii)~Even if the initial concentrations
of both substrates are at supra-threshold levels, induction of the
enzymes for one of the substrates may not occur because of dilution.
Thus, diauxic growth is feasible even if there are no regulatory mechanisms.
(c)~The substrate consumption pattern can be switched from sequential
to simultaneous (or \emph{vice versa}) by altering the initial substrate
concentrations. All these conclusions are consistent with the data
in the literature.
\item The bifurcation diagram for continuous cultures, which describes the
substrate consumption patterns at any given dilution rate and feed
concentrations, is formally identical to the bifurcation diagram for
batch cultures. This gives a precise mathematical basis for the well-known
empirical correlation between the growth patterns in batch and continuous
cultures. Specifically, substrates that are simultaneously consumed
in substrate-excess batch cultures are consumed simultaneously in
continuous cultures at all dilution rates up to washout. However,
substrates that are sequentially consumed in substrate-excess batch
cultures are consumed simultaneously in continuous cultures only if
the dilution rate is sufficiently small --- at large dilution rates,
only the preferred substrate is consumed. This switch occurs precisely
at the so-called transition dilution rate because synthesis of the
secondary substrate enzymes is abolished by dilution. The transition
dilution rate is significantly higher the maximum growth rate on the
secondary substrate. This {}``enhanced growth effect'' occurs because
the enzyme levels are positive at the critical dilution rate of the
secondary substrate --- they become vanishingly small only if the
cells are subjected to even higher dilution rates.
\end{enumerate}
\item At the saturating feed concentrations typically used in experiments,
the physiological steady states are completely determined by only
two independent parameters, the dilution rate and the mass fraction
of the substrate in the feed. We derived analytical expressions for
the steady state values of the substrates, cell density, and enzymes
that provide good fits to the extensive data for the growth of methylotrophic
yeasts on glucose and methanol.
\end{enumerate}
\begin{ack}
This research was supported in part with funds from the National Science
Foundation under contract NSF DMS-0517954. One of us (A.N.) is grateful
to Dr.~Stefan Oehler (IMBB-FoRTH) for discussions regarding the regulation
of the \emph{lac} operon.
\end{ack}
\bibliographystyle{C:/texmf/bibtex/bst/elsevier/elsart-harv}

\appendix

\section{\label{sec:AnalysisBatch}Bifurcation diagram for batch cultures}

\subsection{Conditions for existence and stability of the steady states}

The steady states that can be attained during the first few hours
of batch growth are given by the equations\begin{align}
0=\frac{de_{1}}{dt} & =V_{e,1}\frac{e_{1}\sigma_{0,1}}{\bar{K}_{e,1}+e_{1}\sigma_{0,1}}-\left(Y_{1}V_{s,1}e_{1}\sigma_{0,1}+Y_{2}V_{s,2}e_{2}\sigma_{0,2}+k_{e,1}\right)e_{1},\label{eq:App2E1}\\
0=\frac{de_{2}}{dt} & =V_{e,2}\frac{e_{2}\sigma_{0,2}}{\bar{K}_{e,2}+e_{2}\sigma_{0,2}}-\left(Y_{1}V_{s,1}e_{1}\sigma_{0,1}+Y_{2}V_{s,2}e_{2}\sigma_{0,2}+k_{e,2}\right)e_{2}.\label{eq:App2E2}\end{align}
The Jacobian has the form\begin{equation}
J(e_{1},e_{2})=\left[\begin{array}{cc}
J_{11} & -Y_{2}V_{s,2}e_{1}\sigma_{0,2}\\
-Y_{1}V_{s,1}e_{2}\sigma_{0,1} & J_{22}\end{array}\right],\label{eq:App2J}\end{equation}
where\begin{align}
J_{11} & \equiv V_{e,1}\frac{\bar{K}_{e,1}\sigma_{0,1}}{\left(\bar{K}_{e,1}+e_{1}\sigma_{0,1}\right)^{2}}-2Y_{1}V_{s,1}e_{1}\sigma_{0,1}-Y_{2}V_{s,2}e_{2}\sigma_{0,2}-k_{e,1},\label{eq:App2J11}\\
J_{22} & \equiv V_{e,2}\frac{\bar{K}_{e,2}\sigma_{0,2}}{\left(\bar{K}_{e,2}+e_{2}\sigma_{0,2}\right)^{2}}-Y_{1}V_{s,1}e_{1}\sigma_{0,1}-2Y_{2}V_{s,2}e_{2}\sigma_{0,2}-k_{e,2}.\label{eq:App2J22}\end{align}
In what follows, we derive the existence and stability conditions
for each of the 4 steady states.

\begin{enumerate}
\item $E_{00}$ ($e_{1}=0,e_{2}=0$): It is evident from (\ref{eq:App2E1})--(\ref{eq:App2E2})
that this steady always exists. Since\[
J(E_{00})=\left[\begin{array}{cc}
\frac{V_{e,1}\sigma_{0,1}}{\bar{K}_{e,1}}-k_{e,1} & 0\\
0 & \frac{V_{e,2}\sigma_{0,2}}{\bar{K}_{e,2}}-k_{e,2}\end{array}\right],\]
$E_{00}$ is stable if and only if\begin{align*}
r_{g,i}^{*}\equiv\frac{V_{e,i}\sigma_{0,i}}{\bar{K}_{e,i}}-k_{e,i}<0 & \Leftrightarrow\sigma_{0,i}<\sigma_{i}^{*}\equiv\frac{k_{e,i}}{V_{e,i}/\bar{K}_{e,i}}\end{align*}
for $i=1,2$.
\item $E_{10}$ ($e_{1}>0,e_{2}=0$): This steady state exists if and only
if the equation\begin{equation}
V_{e,1}\frac{\sigma_{0,1}}{\bar{K}_{e,1}+e_{1}\sigma_{0,1}}=Y_{1}V_{s,1}e_{1}\sigma_{0,1}+k_{e,1}\label{eq:App2E10Eqn}\end{equation}
has a positive solution, $e_{1}>0$. Such a solution exists precisely
when\[
r_{g,1}^{*}>0\Leftrightarrow\sigma_{0,1}>\sigma_{1}^{*},\]
in which case\begin{align}
\left.e_{1}\right|_{E_{10}} & =\frac{-\left(\bar{K}_{e,1}+\frac{k_{e,1}}{Y_{1}V_{s,1}}\right)+\sqrt{\left(\bar{K}_{e,1}+\frac{k_{e,1}}{Y_{1}V_{s,1}}\right)^{2}+\frac{4\bar{K}_{e,1}r_{g,1}^{*}}{Y_{1}V_{s,1}}}}{2\sigma_{0,1}},\nonumber \\
r_{g,1} & =Y_{1}V_{s,1}\frac{-\left(\bar{K}_{e,1}+\frac{k_{e,1}}{Y_{1}V_{s,1}}\right)+\sqrt{\left(\bar{K}_{e,1}+\frac{k_{e,1}}{Y_{1}V_{s,1}}\right)^{2}+\frac{4\bar{K}_{e,1}r_{g,1}^{*}}{Y_{1}V_{s,1}}}}{2},\label{eq:App2E10_rg1Max}\end{align}
where $r_{g,1}\equiv\left.Y_{1}V_{s,1}e_{1}\sigma_{0,1}\right|_{E_{10}}$
is the exponential growth rate at $E_{10}$.\\
To determine the stability conditions, observe that \[
J(E_{10})=\left[\begin{array}{cc}
J_{11} & -Y_{2}V_{s,2}e_{1}\sigma_{0,2}\\
0 & J_{22}\end{array}\right],\]
where\begin{align*}
J_{11} & =V_{e,1}\frac{\bar{K}_{e,1}\sigma_{0,1}}{\left(\bar{K}_{e,1}+e_{1}\sigma_{0,1}\right)^{2}}-2Y_{1}V_{s,1}e_{1}\sigma_{0,1}-k_{e,1},\\
J_{22} & =V_{e,2}\frac{\sigma_{0,2}}{\bar{K}_{e,2}}-Y_{1}V_{s,1}e_{1}\sigma_{0,1}-k_{e,2}.\end{align*}
Now, $J_{11}$ is always negative because (\ref{eq:App2E10Eqn}) implies
that at $E_{10}$,\[
V_{e,1}\frac{\bar{K}_{e,1}\sigma_{0,1}}{\left(\bar{K}_{e,1}+e_{1}\sigma_{0,1}\right)^{2}}=\frac{\bar{K}_{e,1}}{\bar{K}_{e,1}+e_{1}\sigma_{0,1}}\left(Y_{1}V_{s,1}e_{1}\sigma_{0,1}+k_{e,1}\right),\]
so that\[
J_{11}=-\left(Y_{1}V_{s,1}e_{1}\sigma_{0,1}+k_{e,1}\right)\frac{e_{1}\sigma_{0,1}}{\bar{K}_{e,1}+e_{1}\sigma_{0,1}}-Y_{1}V_{s,1}e_{1}\sigma_{0,1}<0.\]
Hence, $E_{10}$ is stable if and only if\begin{align*}
J_{22}<0 & \Leftrightarrow V_{e,2}\frac{\sigma_{0,2}}{\bar{K}_{e,2}}<\left.Y_{1}V_{s,1}e_{1}\sigma_{0,1}\right|_{E_{10}}+k_{e,2},\\
 & \Leftrightarrow r_{g,2}^{*}<r_{g,1}.\end{align*}

\item $E_{01}$ ($e_{1}=0,e_{2}>0$): Arguments similar to those used above
for $E_{10}$ show that $E_{01}$ exists if and only if \[
r_{g,2}^{*}>0\Leftrightarrow\sigma_{0,2}>\sigma_{2}^{*},\]
in which case\begin{align}
\left.e_{2}\right|_{E_{01}} & =\frac{-\left(\bar{K}_{e,2}+\frac{k_{e,2}}{Y_{2}V_{s,2}}\right)+\sqrt{\left(\bar{K}_{e,2}+\frac{k_{e,2}}{Y_{2}V_{s,2}}\right)^{2}+\frac{4\bar{K}_{e,2}r_{g,2}^{*}}{Y_{2}V_{s,2}}}}{2\sigma_{0,2}},\nonumber \\
r_{g,2}^{\textnormal{max}} & =Y_{2}V_{s,2}\frac{-\left(\bar{K}_{e,2}+\frac{k_{e,2}}{Y_{2}V_{s,2}}\right)+\sqrt{\left(\bar{K}_{e,2}+\frac{k_{e,2}}{Y_{2}V_{s,2}}\right)^{2}+\frac{4\bar{K}_{e,2}r_{g,2}^{*}}{Y_{2}V_{s,2}}}}{2},\label{eq:App2E01_rg2Max}\end{align}
where $r_{g,2}\equiv\left.Y_{2}V_{s,2}e_{2}\sigma_{0,2}\right|_{E_{01}}$
is the exponential growth rate at $E_{01}$.
\item $E_{11}$ ($e_{1}>0,e_{2}>0$): This steady state exists if and only
if the equations\begin{align}
0 & =V_{e,1}\frac{\sigma_{0,1}}{\bar{K}_{e,1}+e_{1}\sigma_{0,1}}-\left(Y_{1}V_{s,1}e_{1}\sigma_{0,1}+Y_{2}V_{s,2}e_{2}\sigma_{0,2}+k_{e,1}\right),\label{eq:App2E11_1}\\
0 & =V_{e,2}\frac{\sigma_{0,2}}{\bar{K}_{e,2}+e_{2}\sigma_{0,2}}-\left(Y_{1}V_{s,1}e_{1}\sigma_{0,1}+Y_{2}V_{s,2}e_{2}\sigma_{0,2}+k_{e,2}\right),\label{eq:App2E11_2}\end{align}
have positive solutions, $e_{1},e_{2}>0$. The conditions for such
solutions can be determined by eliminating one of the variables, $e_{1},e_{2}$,
from the above equations. To this end, observe that eqs.~(\ref{eq:App2E11_1})--(\ref{eq:App2E11_2})
imply the relation \begin{equation}
V_{e,1}\frac{\sigma_{0,1}}{\bar{K}_{e,1}+e_{1}\sigma_{0,1}}-k_{e,1}=V_{e,2}\frac{\sigma_{0,2}}{\bar{K}_{e,2}+e_{2}\sigma_{0,2}}-k_{e,2}.\label{eq:E11elim}\end{equation}
Hence, we can eliminate $e_{2}$ from (\ref{eq:App2E11_1})--(\ref{eq:App2E11_2})
by solving (\ref{eq:E11elim}) for $e_{2}$, and substituting it in
(\ref{eq:App2E11_1}) to obtain the equation\begin{align*}
V_{e,1}\frac{\sigma_{0,1}}{\bar{K}_{e,1}+e_{1}\sigma_{0,1}}-k_{e,1} & =Y_{1}V_{s,1}e_{1}\sigma_{0,1}+Y_{2}V_{s,2}\frac{V_{e,2}\sigma_{0,2}}{V_{e,1}\frac{\sigma_{0,1}}{\bar{K}_{e,1}+\sigma_{0,1}}-k_{e,1}+k_{e,2}}\\
 & \quad\quad\quad-Y_{2}V_{s,2}\bar{K}_{e,2},\end{align*}
which has a (unique) positive solution, $e_{1}>0$, if and only if\[
\frac{V_{e,1}\sigma_{0,1}}{\bar{K}_{e,1}}-k_{e,1}>Y_{2}V_{s,2}\frac{V_{e,2}\sigma_{0,1}}{\left(\frac{V_{e,1}\sigma_{0,1}}{\bar{K}_{e,1}}-k_{e,1}\right)+k_{e,2}}-Y_{2}V_{s,2}\bar{K}_{e,2}.\]
One can check that this is equivalent to the condition\[
r_{g,1}^{*}\equiv\frac{V_{e,1}\sigma_{0,1}}{\bar{K}_{e,1}}-k_{e,1}>\left.Y_{2}V_{s,2}e_{2}\sigma_{0,2}\right|_{E_{01}}\equiv r_{g,2},\]
i.e., $E_{10}$ is unstable.\\
Similarly, eliminating $e_{1}$ from (\ref{eq:App2E11_1})-(\ref{eq:App2E11_2})
yields\begin{align*}
V_{e,2}\frac{\sigma_{0,2}}{\bar{K}_{e,1}+e_{2}\sigma_{0,2}}-k_{e,2} & =Y_{2}V_{s,2}e_{2}\sigma_{0,2}+Y_{1}V_{s,1}\frac{V_{e,1}\sigma_{0,1}}{V_{e,2}\frac{\sigma_{0,2}}{\bar{K}_{e,2}+\sigma_{0,2}}-k_{e,2}+k_{e,1}}\\
 & \quad\quad\quad-Y_{1}V_{s,1}\bar{K}_{e,1},\end{align*}
which has a positive solution, $e_{2}>0$, if and only if $E_{10}$
is unstable. We conclude that $E_{11}$ exists and is unique if and
only if both $E_{10}$ and $E_{01}$ (exist and) are unstable.\\
To determine the stability condition, observe that (\ref{eq:App2E11_1})
implies\[
V_{e,1}\frac{\bar{K}_{e,1}\sigma_{0,1}}{\left(\bar{K}_{e,1}+e_{1}\sigma_{0,1}\right)^{2}}=\left(Y_{1}V_{s,1}e_{1}\sigma_{0,1}+Y_{2}V_{s,2}e_{2}\sigma_{0,2}+k_{e,1}\right)\frac{\bar{K}_{e,1}}{\bar{K}_{e,1}+e_{1}\sigma_{0,1}}.\]
Hence,\begin{align*}
J_{11} & =\left(Y_{1}V_{s,1}e_{1}\sigma_{0,1}+Y_{2}V_{s,2}e_{2}\sigma_{0,2}+k_{e,1}\right)\frac{\bar{K}_{e,1}}{\bar{K}_{e,1}+e_{1}\sigma_{0,1}}\\
 & \quad\quad\quad-2Y_{1}V_{s,1}e_{1}\sigma_{0,1}-Y_{2}V_{s,2}e_{2}\sigma_{0,2}-k_{e,2},\\
 & =-\left(Y_{1}V_{s,1}e_{1}\sigma_{0,1}+Y_{2}V_{s,2}e_{2}\sigma_{0,2}+k_{e,1}\right)\frac{e_{1}\sigma_{0,1}}{\bar{K}_{e,1}+e_{1}\sigma_{0,1}}-Y_{1}V_{s,1}e_{1}\sigma_{0,1},\end{align*}
and similarly,\[
J_{22}=-\left(Y_{1}V_{s,1}e_{1}\sigma_{0,1}+Y_{2}V_{s,2}e_{2}\sigma_{0,2}+k_{e,1}\right)\frac{e_{2}\sigma_{0,2}}{\bar{K}_{e,2}+e_{2}\sigma_{0,2}}-Y_{2}V_{s,2}e_{2}\sigma_{0,2}.\]
It follows immediately that $\textnormal{tr}\, J<0$ and $\det J>0$.
Hence, $E_{11}$ is stable whenever it exists.
\end{enumerate}

\subsection{Relative magnitudes of $r_{g,i}$ and $r_{g,i}^{*}$}

The exponential growth rate on $S_{i}$, $r_{g,i}(\sigma_{0,i})$,
is always less than $r_{g,i}^{*}(\sigma_{0,i})$, the specific growth
rate at which $E_{i}$ becomes extinct. To see this, observe that
(\ref{eq:App2E10_rg1Max})--(\ref{eq:App2E01_rg2Max}) can be rewritten
as\begin{equation}
r_{g,i}^{*}=\left(1+\frac{k_{e,i}}{Y_{i}V_{s,i}\bar{K}_{e,i}}\right)r_{g,i}+\left(\frac{1}{Y_{i}V_{s,i}\bar{K}_{e,i}}\right)r_{g,i}^{2},\label{eq:App2rgiStar}\end{equation}
which implies that\[
r_{g,i}\le r_{g,i}^{*},\]
with equality being attained when $\sigma_{i}=\sigma_{i}^{*}$, in
which case, $r_{g,i}=r_{g,i}^{*}=0$.

\subsection{Disposition of the bifurcation curves}

The lower (blue) bifurcation curve in Fig.~\ref{f:BDbatch} is defined
by the condition\begin{align}
r_{g,1} & =r_{g,2}^{*}.\label{eq:App2Curve1Defn}\end{align}
Since\begin{equation}
r_{g,1}^{*}=\left(1+\frac{k_{e,1}}{Y_{1}V_{s,1}\bar{K}_{e,1}}\right)r_{g,1}+\left(\frac{1}{Y_{1}V_{s,1}\bar{K}_{e,1}}\right)r_{g,1}^{2},\label{eq:App2rg1Star}\end{equation}
the bifurcation curve defined by (\ref{eq:App2Curve1Defn}) satisfies
the equation\[
r_{g,1}^{*}=\left(1+\frac{k_{e,1}}{Y_{1}V_{s,1}\bar{K}_{e,1}}\right)r_{g,2}^{*}+\frac{1}{Y_{1}V_{s,1}\bar{K}_{e,1}}\left(r_{g,2}^{*}\right)^{2},\]
obtained from (\ref{eq:App2rg1Star}) by replacing $r_{g,1}$ with
$r_{g,2}^{*}$. Substituting the definition of $r_{g,i}^{*}$ into
the above equation yields \begin{align*}
\frac{V_{e,1}}{\bar{K}_{e,1}}\left(\sigma_{0,1}-\sigma_{1}^{*}\right) & =\left(1+\frac{k_{e,1}}{Y_{1}V_{s,1}\bar{K}_{e,1}}\right)\left[\frac{V_{e,2}}{\bar{K}_{e,2}}\left(\sigma_{0,2}-\sigma_{2}^{*}\right)\right]\\
 & \qquad+\frac{1}{Y_{1}V_{s,1}\bar{K}_{e,1}}\left[\frac{V_{e,2}}{\bar{K}_{e,2}}\left(\sigma_{0,2}-\sigma_{2}^{*}\right)\right]^{2},\end{align*}
which defines an increasing curve on the $\sigma_{0,1},\sigma_{0,2}$-plane
passing through the point $\left(\sigma_{1}^{*},\sigma_{2}^{*}\right)$.
Moreover, since the points on this curve satisfy the relation\[
\frac{V_{e,1}}{\bar{K}_{e,1}}\left(\sigma_{0,1}-\sigma_{1}^{*}\right)>\frac{V_{e,2}}{\bar{K}_{e,2}}\left(\sigma_{0,2}-\sigma_{2}^{*}\right),\]
they never go \emph{above} the line\begin{equation}
\frac{V_{e,1}}{\bar{K}_{e,1}}\left(\sigma_{0,1}-\sigma_{1}^{*}\right)=\frac{V_{e,2}}{\bar{K}_{e,2}}\left(\sigma_{0,2}-\sigma_{2}^{*}\right).\label{eq:App2LineDefn}\end{equation}
A similar argument shows that the upper (red) bifurcation curve in
Fig.~\ref{f:BDbatch}, which is defined by the equation \begin{equation}
r_{g,2}=r_{g,1}^{*},\label{eq:App2Curve2Defn}\end{equation}
satisfies the relation\begin{align*}
\frac{V_{e,2}}{\bar{K}_{e,2}}\left(\sigma_{0,2}-\sigma_{2}^{*}\right) & =\left(1+\frac{k_{e,2}}{Y_{2}V_{s,2}\bar{K}_{e,2}}\right)\left[\frac{V_{e,1}}{\bar{K}_{e,1}}\left(\sigma_{0,2}-\sigma_{1}^{*}\right)\right]\\
 & \qquad+\frac{1}{Y_{2}V_{s,2}\bar{K}_{e,12}}\left[\frac{V_{e,1}}{\bar{K}_{e,1}}\left(\sigma_{0,1}-\sigma_{1}^{*}\right)\right]^{2}.\end{align*}
This relation also defines an increasing curve passing through $\left(\sigma_{1}^{*},\sigma_{2}^{*}\right)$,
but the curve never goes \emph{below} the line defined by (\ref{eq:App2LineDefn})
because\[
\frac{V_{e,2}}{\bar{K}_{e,2}}\left(\sigma_{0,2}-\sigma_{2}^{*}\right)>\frac{V_{e,1}}{\bar{K}_{e,1}}\left(\sigma_{0,1}-\sigma_{1}^{*}\right).\]
Hence, the curve defined by (\ref{eq:App2Curve2Defn}) always lies
above the curve defined by (\ref{eq:App2Curve1Defn}).

Finally, we note that as the two bifurcation curves approach the point,
$(\sigma_{1}^{*},\sigma_{2}^{*})$, they are approximated by the lines\begin{align*}
\frac{V_{e,1}}{\bar{K}_{e,1}}\left(\sigma_{0,1}-\sigma_{1}^{*}\right) & \approx\left(1+\frac{k_{e,1}}{Y_{1}V_{s,1}\bar{K}_{e,1}}\right)\left[\frac{V_{e,2}}{\bar{K}_{e,2}}\left(\sigma_{0,2}-\sigma_{2}^{*}\right)\right],\\
\frac{V_{e,2}}{\bar{K}_{e,2}}\left(\sigma_{0,2}-\sigma_{2}^{*}\right) & \approx\left(1+\frac{k_{e,2}}{Y_{2}V_{s,2}\bar{K}_{e,2}}\right)\left[\frac{V_{e,1}}{\bar{K}_{e,1}}\left(\sigma_{0,1}-\sigma_{1}^{*}\right)\right],\end{align*}
which almost coincide with the line defined by (\ref{eq:App2LineDefn})
because $k_{e,i}\ll Y_{i}V_{s,i}\bar{K}_{e,i}$. Thus, for typical
parameter values, the curves have the quasicusp-shaped geometry shown
in Fig.~\ref{f:BDbatch}.

\section{\label{a:StabilityContinuous}Bifurcation diagram for continuous
cultures}

The steady states of continuous cultures satisfy the equations\begin{align}
0=\frac{ds_{1}}{dt} & =D\left(s_{f,1}-s_{1}\right)-r_{s,1}c,\label{eq:App3S1}\\
0=\frac{ds_{2}}{dt} & =D\left(s_{f,2}-s_{1}\right)-r_{s,2}c,\label{eq:App3S2}\\
0=\frac{dc}{dt} & =\left(r_{g}-D\right)c,\label{eq:App3C}\\
0=\frac{de_{1}}{dt} & =R_{1}\equiv V_{e,1}\frac{e_{1}\sigma_{1}}{\bar{K}_{e,1}+e_{1}\sigma_{1}}-\left(r_{g}+k_{e,1}\right)e_{1},\label{eq:App3E1}\\
0=\frac{de_{2}}{dt} & =R_{2}\equiv V_{e,2}\frac{e_{2}\sigma_{2}}{\bar{K}_{e,2}+e_{2}\sigma_{2}}-\left(r_{g}+k_{e,2}\right)e_{2},\label{eq:App3E2}\end{align}
where $r_{g}=Y_{1}r_{s,1}+Y_{2}r_{s,2}$. The Jacobian is\begin{equation}
J=\left[\begin{array}{ccccc}
-D-c\frac{\partial r_{s,1}}{\partial s_{1}} & 0 & -r_{s,1} & -c\frac{\partial r_{s,1}}{\partial e_{1}} & 0\\
0 & -D-c\frac{\partial r_{s,2}}{\partial s_{2}} & -r_{s,2} & 0 & -c\frac{\partial r_{s,2}}{\partial e2_{1}}\\
c\frac{\partial r_{g}}{\partial s_{1}} & c\frac{\partial r_{g}}{\partial s_{2}} & r_{g}-D & c\frac{\partial r_{g}}{\partial e_{1}} & c\frac{\partial r_{g}}{\partial e_{2}}\\
\frac{\partial R_{1}}{\partial s_{1}} & \frac{\partial R_{1}}{\partial s_{2}} & 0 & \frac{\partial R_{1}}{\partial e_{1}} & \frac{\partial R_{1}}{\partial e_{2}}\\
\frac{\partial R_{2}}{\partial s_{1}} & \frac{\partial R_{2}}{\partial s_{2}} & 0 & \frac{\partial R_{2}}{\partial e_{1}} & \frac{\partial R_{2}}{\partial e_{2}}\end{array}\right]\label{eq:App3Jgen}\end{equation}
We begin by considering the washout steady states ($c=0$), since
the existence and stability conditions for these steady states follow
immediately from the foregoing analysis of batch cultures.

\subsection{Existence and stability of washout steady states}

At a washout steady state, $c=0,s_{1}=s_{f,1},s_{2}=s_{f,2}$. Such
a steady state exists if and only if there exist $e_{1},e_{2}>0$
satisfying the equations\begin{align*}
0=\frac{de_{1}}{dt} & =V_{e,1}\frac{e_{1}\sigma_{f,1}}{\bar{K}_{e,1}+e_{1}\sigma_{f,1}}-\left(Y_{1}V_{s,1}e_{1}\sigma_{f,1}+Y_{2}V_{s,2}e_{2}\sigma_{f,2}+k_{e,1}\right)e_{1},\\
0=\frac{de_{2}}{dt} & =V_{e,2}\frac{e_{2}\sigma_{f,2}}{\bar{K}_{e,2}+e_{2}\sigma_{f,2}}-\left(Y_{1}V_{s,1}e_{1}\sigma_{f,1}+Y_{2}V_{s,2}e_{2}\sigma_{f,2}+k_{e,2}\right)e_{2},\end{align*}
which are formally similar to eqs.~(\ref{eq:App2E1})--(\ref{eq:App2E2}),
the only difference being that $\sigma_{0,i}$ is replaced by $\sigma_{f,i}$.

It follows from (\ref{eq:App3Jgen}) that the Jacobian at a washout
steady steady is

\[
J=\left[\begin{array}{ccccc}
-D & 0 & -r_{s,1} & 0 & 0\\
0 & -D & -r_{s,2} & 0 & 0\\
0 & 0 & r_{g}-D & 0 & 0\\
\frac{\partial R_{1}}{\partial s_{1}} & \frac{\partial R_{1}}{\partial s_{2}} & 0 & \frac{\partial R_{1}}{\partial e_{1}} & \frac{\partial R_{1}}{\partial e_{2}}\\
\frac{\partial R_{2}}{\partial s_{1}} & \frac{\partial R_{2}}{\partial s_{2}} & 0 & \frac{\partial R_{2}}{\partial e_{1}} & \frac{\partial R_{2}}{\partial e_{2}}\end{array}\right].\]
Thus, a washout steady state is stable if and only if at this steady
state, $r_{g}<D$, and the submatrix\[
R\equiv\left[\begin{array}{cc}
\frac{\partial R_{1}}{\partial e_{1}} & \frac{\partial R_{1}}{\partial e_{2}}\\
\frac{\partial R_{2}}{\partial e_{1}} & \frac{\partial R_{2}}{\partial e_{2}}\end{array}\right]\]
has negative eigenvalues. This submatrix is formally similar to the
matrix defined by eqs.~(\ref{eq:App2J})--(\ref{eq:App2J22}), the
only difference being that $\sigma_{0,i}$ is replaced by $\sigma_{f,i}$.

Given the above results, we expect the existence and stability conditions
for the washout steady states, $E_{000}$, $E_{100}$, $E_{010}$,
$E_{110}$, to be formally similar to the existence and stability
conditions for the corresponding batch culture steady states, namely,
$E_{00}$, $E_{10}$, $E_{01}$, and $E_{11}$. We show below that
this is indeed the case.

\begin{enumerate}
\item $E_{000}$ ($e_{1}=0$, $e_{2}=0$, $c=0$): It is evident that this
steady state always exists. Since $\left.r_{g}\right|_{E_{000}}=0<D$,
$E_{000}$ is stable if and only if $R(E_{000})$ has negative eigenvalues.
Since $R(E_{000})$ is formally similar to $J(E_{00})$, we conclude
that $E_{000}$ is stable if and only if\begin{equation}
D_{t,i}\equiv\frac{V_{e,i}\sigma_{f,i}}{\bar{K}_{e,i}}-k_{e,i}<0\Leftrightarrow\sigma_{f,i}<\sigma_{i}^{*}\equiv\frac{k_{e,i}}{V_{e,i}/K_{e,i}}\label{eq:StabilityE00_2}\end{equation}
for $i=1,2$.
\item $E_{100}$ ($e_{1}>0$, $e_{2}=0$, $c=0$): This steady state exists
if and only if\begin{equation}
V_{e,1}\frac{\sigma_{f,1}}{\bar{K}_{e,1}+e_{1}\sigma_{f,1}}=Y_{1}V_{s,1}e_{1}\sigma_{f,1}+k_{e,1}\label{eq:App3E10ss}\end{equation}
which is formally similar to eq.~(\ref{eq:App2E10Eqn}). It follows
that $E_{100}$ exists if and only if \[
D_{t,1}>0\Leftrightarrow\sigma_{f,1}>\sigma_{1}^{*},\]
in which case\begin{align}
\left.e_{1}\right|_{E_{100}} & =\frac{-\left(\bar{K}_{e,1}+\frac{k_{e,1}}{Y_{1}V_{s,1}}\right)+\sqrt{\left(\bar{K}_{e,1}+\frac{k_{e,1}}{Y_{1}V_{s,1}}\right)^{2}+\frac{4K_{e,1}D_{t,1}}{Y_{1}V_{s,1}}}}{2\sigma_{f,1}},\nonumber \\
D_{c,1} & =Y_{1}V_{s,1}\frac{-\left(\bar{K}_{e,1}+\frac{k_{e,1}}{Y_{1}V_{s,1}}\right)+\sqrt{\left(\bar{K}_{e,1}+\frac{k_{e,1}}{Y_{1}V_{s,1}}\right)^{2}+\frac{4K_{e,1}D_{t,1}}{Y_{1}V_{s,1}}}}{2},\label{eq:App3Dc1}\end{align}
where $D_{c,1}\equiv\left.r_{g}\right|_{E_{100}}=\left.Y_{1}V_{s,1}e_{1}\sigma_{1}\right|_{E_{100}}$
is the critical dilution rate during single-substrate growth on $S_{1}$.\\
$E_{100}$ is stable if and only if $D>D_{c,1}$ and $D_{t,2}<D_{c,1}$,
where the latter relation ensures that the submatrix, $R(E_{100})$,
which is formally similar to $J(E_{10})$, has negative eigenvalues.
In the particular case of single-substrate growth on $S_{1}$, the
second stability condition is always satisfied ($D_{t,2}=-k_{e,2}<0<D_{c,1}$),
so that $E_{100}$ is stable whenever $D>D_{c,1}$.
\item $E_{010}$ ($e_{1}=0$, $e_{2}>0$, $c=0$): Arguments similar to
those used above for $E_{100}$ show that $E_{010}$ exists if and
only if \[
D_{t,2}>0\Leftrightarrow\sigma_{f,2}>\sigma_{2}^{*},\]
in which case \begin{align}
\left.e_{2}\right|_{E_{010}} & =\frac{-\left(\bar{K}_{e,2}+\frac{k_{e,2}}{Y_{2}V_{s,2}}\right)+\sqrt{\left(\bar{K}_{e,2}+\frac{k_{e,2}}{Y_{2}V_{s,2}}\right)^{2}+\frac{4K_{e,2}D_{t,2}}{Y_{2}V_{s,2}}}}{2\sigma_{f,2}},\nonumber \\
D_{c,2} & =Y_{2}V_{s,2}\frac{-\left(\bar{K}_{e,2}+\frac{k_{e,2}}{Y_{2}V_{s,2}}\right)+\sqrt{\left(\bar{K}_{e,2}+\frac{k_{e,2}}{Y_{2}V_{s,2}}\right)^{2}+\frac{4K_{e,2}D_{t,2}}{Y_{2}V_{s,2}}}}{2},\label{eq:App3Dc2}\end{align}
where $D_{c,2}\equiv\left.r_{g}\right|_{E_{010}}=\left.Y_{2}V_{s,2}e_{2}\sigma_{2}\right|_{E_{010}}$
is the critical dilution rate during single-substrate growth on $S_{2}$.
It is stable if and only if $D>D_{c,2}$ and $D_{t,1}<D_{c,2}$.
\item $E_{110}$ ($e_{1}=0$, $e_{2}>0$, $c>0$): This steady state exists
if and only if the equations\begin{align}
0 & =V_{e,1}\frac{\sigma_{f,1}}{\bar{K}_{e,1}+e_{1}\sigma_{f,1}}-\left(Y_{1}V_{s,1}e_{1}\sigma_{f,1}+Y_{2}V_{s,2}e_{2}\sigma_{f,2}+k_{e,1}\right),\label{eq:App3E11_1}\\
0 & =V_{e,2}\frac{\sigma_{f,2}}{\bar{K}_{e,2}+e_{2}\sigma_{f,2}}-\left(Y_{1}V_{s,1}e_{1}\sigma_{f,1}+Y_{2}V_{s,2}e_{2}\sigma_{f,2}+k_{e,2}\right),\label{eq:App3E11_2}\end{align}
have positive solutions. These equations are formally similar to eqs.~(\ref{eq:App2E11_1})--(\ref{eq:App2E11_2}).
Hence, $E_{110}$ exists and is unique if and only if \[
0<D_{t,2}<D_{c,1},\;0<D_{t,1}<D_{c,2}.\]
Since $R(E_{110})$ is formally similar to $J(E_{11})$, the eigenvalues
of $R(E_{110})$ are always negative. It follows that $E_{110}$ is
stable if and only if \[
D>D_{c}\equiv\left.r_{g}\right|_{E_{110}}=\left.Y_{1}V_{s,1}e_{1}\sigma_{1}+Y_{2}V_{s,2}e_{2}\sigma_{2}\right|_{E_{110}}.\]
We cannot solve explicitly for $D_{c}$, but its properties can be
inferred from an implicit representation, which can be derived as
follows. At $E_{110}$, the steady state enzyme balances imply that\begin{align*}
V_{e,i}\frac{\sigma_{f,i}}{\bar{K}_{e,i}+e_{1}\sigma_{f,i}}=D_{c}+k_{e,1} & \Leftrightarrow e_{i}\sigma_{f,i}=\frac{V_{e,i}}{D_{c}+k_{e,i}}-\bar{K}_{e,i}\\
 & \Leftrightarrow e_{i}\sigma_{f,i}=\frac{\bar{K}_{e,i}}{D_{c}+k_{e,i}}\left(D_{t,i}-D_{c}\right).\end{align*}
Hence, $D_{c}$ satisfies the equation \begin{equation}
D_{c}=\frac{Y_{1}V_{s,1}\bar{K}_{e,1}}{D_{c}+k_{e,1}}\left(D_{t,1}-D_{c}\right)+\frac{Y_{2}V_{s,2}\bar{K}_{e,2}}{D_{c}+k_{e,2}}\left(D_{t,2}-D_{c}\right).\label{eq:App3DcDefn}\end{equation}
We shall appeal to this equation below.
\end{enumerate}
Before proceeding to the persistence steady states, we pause to consider
the geometry and disposition of the surfaces.

\subsubsection{Geometry and disposition of the surfaces of $D_{t,i}(\sigma_{f,i})$
and $D_{c,i}(\sigma_{f,i})$:}

The geometry and disposition of the surfaces of $D_{c,i}$ and $D_{t,i}$
follow from arguments similar to those used above in the analysis
of batch cultures. Indeed, (\ref{eq:App3Dc1})--(\ref{eq:App3Dc2})
imply that\[
D_{t,i}=\left(1+\frac{k_{e,i}}{Y_{1}V_{s,i}\bar{K}_{e,i}}\right)D_{c,i}+\left(\frac{1}{Y_{i}V_{s,i}\bar{K}_{e,i}}\right)D_{c,i}^{2},\]
which is formally similar to (\ref{eq:App2rgiStar}). It follows that
$D_{t,i}\ge D_{c,i}$, and the bifurcation curves defined by the equations,
$D_{t,2}=D_{c,1}$ and $D_{t,1}=D_{c,2}$, are identical to the bifurcation
curves in Fig.~\ref{f:BDbatch}, the only difference being that they
lie on the $\sigma_{f,1},\sigma_{f,2}$-plane (as opposed to the $\sigma_{0,1},\sigma_{0,2}$-plane).

\subsubsection{Geometry and disposition of the surface of $D_{c}(\sigma_{f,1},\sigma_{f,2})$}

The surface of $D_{t,1}$ intersects the surfaces of both $D_{c}$
and $D_{c,2}$ along the very same curves. Indeed, it follows from
(\ref{eq:App3DcDefn}) that $D_{t,1}$ intersects the surfaces of
$D_{c}$ and $D_{t,1}$ along the curve\[
D_{t,1}=\frac{Y_{2}V_{s,2}\bar{K}_{e,2}}{D_{t,1}+k_{e,2}}\left(D_{t,2}-D_{t,1}\right),\]
which is identical to the equation defining the curve, $D_{t,1}=D_{c,2}$.
A similar argument shows that the surface of $D_{t,2}$ intersects
the surfaces of $D_{c}$ and $D_{c,1}$ along the same curve.

Differentiation of (\ref{eq:App3DcDefn}) yields\[
\frac{\partial D_{c}}{\partial\sigma_{f,i}}=\frac{\frac{Y_{i}V_{s,i}V_{e,i}}{D+k_{e,i}}}{1+\frac{Y_{1}V_{s,1}\bar{K}_{e,1}}{(D+k_{e,1})^{2}}+\frac{Y_{2}V_{s,2}\bar{K}_{e,2}}{(D+k_{e,2})^{2}}}>0.\]
This immediately implies that the magnitude of $D_{c}$ (relative
to the magnitudes of $D_{t,i}$ and $D_{c,i}$) is as shown in Fig.~\ref{f:BDcont}.

\subsection{Persistence steady states}

There are three persistence steady states.

\subsubsection{$E_{101}$ ($e_{1}>0$, $e_{2}=0$, $c>0$) }

\begin{figure}
\noindent \begin{centering}
\includegraphics[width=2.6in]{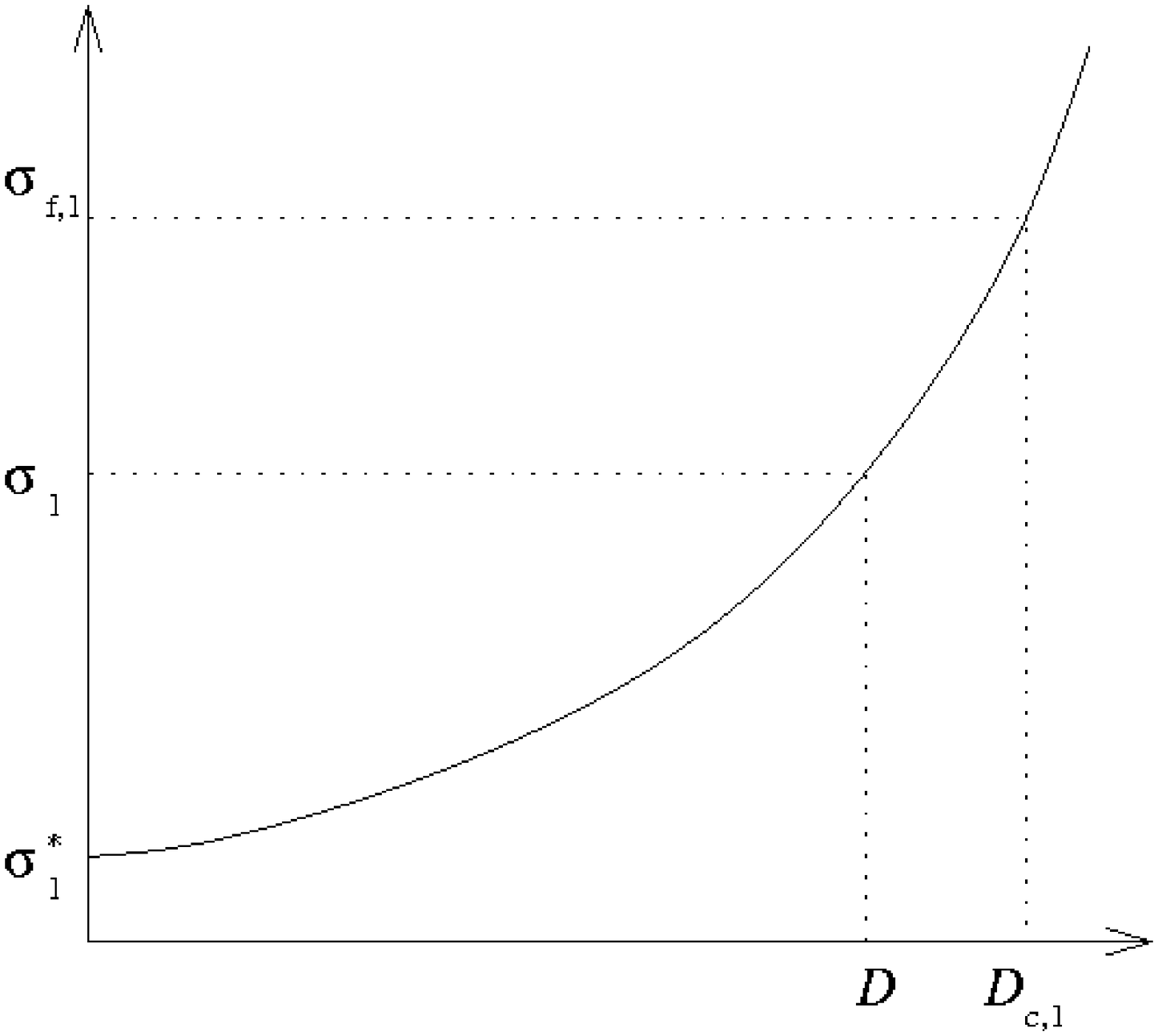}\hspace*{0.1in}\includegraphics[width=2.6in]{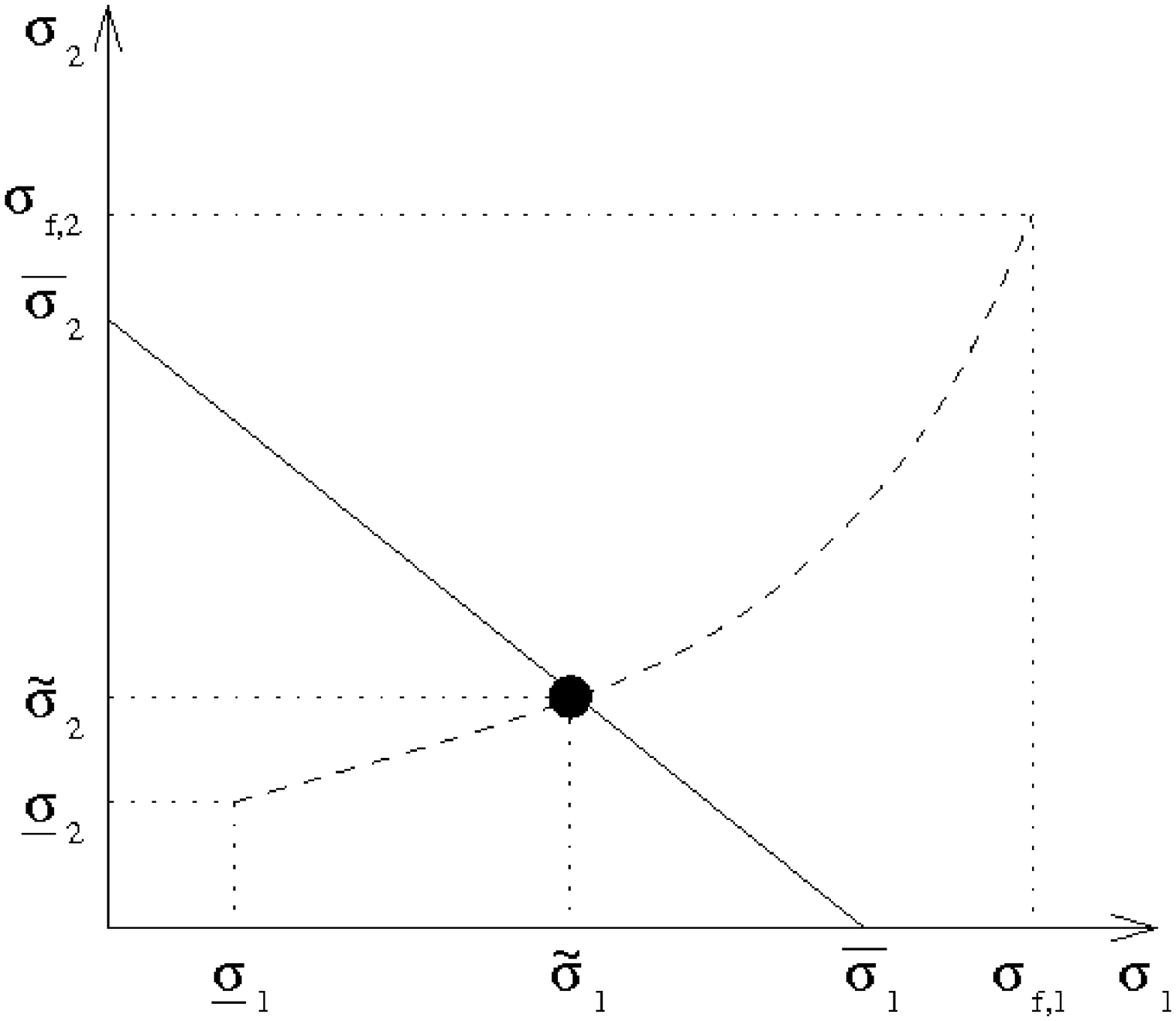}
\par\end{centering}

\caption{\label{fig:StabilityProofs}(a)~The residual substrate concentration
at $E_{101}$ satisfies the relation, $0<\sigma_{1}<\sigma_{f,1}$,
precisely when $\sigma_{f,1}>\sigma_{1}^{*}$ and $0<D<D_{c,1}$.
(b)~Graphical depiction of the conditions for existence of $E_{111}$.}

\end{figure}

This steady state exists if and only if the equations\begin{align}
0 & =D(s_{f,1}-s_{1})-V_{s,1}e_{1}\sigma_{1}c,\label{eq:E101s1}\\
0 & =V_{e,1}\frac{\sigma_{1}}{\bar{K}_{e,1}+e_{1}\sigma_{1}}-\left(Y_{1}V_{s,1}e_{1}\sigma_{1}+k_{e,1}\right),\label{eq:E101e1}\\
0 & =Y_{1}V_{s,1}e_{1}\sigma_{1}-D.\label{eq:E101c}\end{align}
have positive solutions. These equations can be explicitly solved
for $s_{1}$, $e_{1}$, and $c$. Indeed, (\ref{eq:E101c}) implies
that\[
\sigma_{1}=\frac{D}{Y_{1}V_{s,1}e_{1}},\]
which can be substituted in (\ref{eq:E101s1})--(\ref{eq:E101e1})
to obtain\begin{align}
c & =Y_{1}(s_{f,1}-s_{1}),\label{eq:App2E101c}\\
e_{1} & =\frac{V_{e,1}}{D+k_{e,1}}\frac{D}{Y_{1}V_{s,1}\bar{K}_{e,1}+D},\label{eq:App2E101e1}\\
\sigma_{1}= & \frac{D}{Y_{1}V_{s,1}e_{1}}=\frac{1}{Y_{1}V_{s,1}V_{e,1}}\left(D+k_{e,1}\right)\left(D+Y_{1}V_{s,1}\bar{K}_{e,1}\right).\label{eq:App2E101s1}\end{align}
Now, $e_{1}>0$ for all $D>0$, and $c>0$ whenever $s_{1}<s_{f,1}$.
Hence, $E_{101}$ exists if and only if (\ref{eq:App2E101s1}) has
a solution satisfying $0<\sigma_{1}<\sigma_{f,1}$. It follows from
the graph of $\sigma_{1}$ (Fig.~\ref{fig:StabilityProofs}a) that
such a solution exists if and only if \[
\sigma_{f,1}>\sigma_{1}^{*},\;0<D<D_{c,1}.\]
Here, $D_{c,1}$ satisfies the relation\[
\sigma_{f,1}==\frac{1}{Y_{1}V_{s,1}V_{e,1}}\left(D_{c,1}+k_{e,1}\right)\left(D_{c,1}+Y_{1}V_{s,1}\bar{K}_{e,1}\right),\]
which has the positive solution given by (\ref{eq:App3Dc1}).

To determine the stability of $E_{101}$, observe that the Jacobian
is similar to the matrix\[
\left[\begin{array}{ccccc}
-c\frac{\partial r_{s,1}}{\partial s_{1}} & -c\frac{\partial r_{s,1}}{\partial e_{1}} & -r_{s,1} & 0 & Y_{2}r_{s,1}\\
\frac{\partial R_{1}}{\partial s_{1}} & \frac{\partial R_{1}}{\partial e_{1}} & 0 & \frac{\partial R_{1}}{\partial e_{2}} & \frac{\partial R_{1}}{\partial s_{2}}\\
0 & 0 & -D & 0 & 0\\
0 & 0 & 0 & \frac{\partial R_{2}}{\partial e_{2}} & 0\\
0 & 0 & 0 & -c\frac{\partial r_{s,2}}{\partial e_{2}} & -D\end{array}\right],\]
which represents the Jacobian of the new system of equations obtained
if we make the linear coordinate change\[
\left\{ s_{1},s_{2},e_{1},e_{2},c\right\} \longrightarrow\left\{ s_{1},e_{1},Y_{1}s_{1}+Y_{1}s_{2}+c,e_{2},s_{2}\right\} .\]
It follows that $E_{101}$ is stable if and only if $\partial R_{2}/\partial e_{2}<0$
and the submatrix\[
\left[\begin{array}{cc}
-c\frac{\partial r_{s,1}}{\partial s_{1}} & -c\frac{\partial r_{s,1}}{\partial e_{1}}\\
\frac{\partial R_{1}}{\partial s_{1}} & \frac{\partial R_{1}}{\partial e_{1}}\end{array}\right]\]
has negative eigenvalues. One can check that thi submatrix always
has negative eigenvalues, and\[
\frac{\partial R_{2}}{\partial e_{2}}=\frac{V_{e,2}\sigma_{f,2}}{\bar{K}_{e,2}}-D-k_{e,2}=D_{t,2}-D,\]
so that $E_{101}$ is stable if and only if\begin{equation}
D>D_{t,2}.\label{eq:StabilityE10_2}\end{equation}
In the particular case of single-substrate growth on $S_{1}$, this
condition is always satisfied, so that $E_{101}$ is stable whenever
it exists.

\subsubsection{$E_{011}$ ($e_{1}=0$, $e_{2}>0$, $c>0$) }

The existence and stability conditions for $E_{011}$ can be derived
by methods analogous to those shown above for $E_{101}$.

\subsubsection{$E_{111}$ ($e_{1}>0$, $e_{2}>0$, $c>0$)}

This steady state exists if and only if the equations\begin{align}
0= & D(s_{f,1}-s_{1})-V_{s,1}e_{1}\sigma_{1}c\label{eq:E111s1}\\
0= & D(s_{f,2}-s_{2})-V_{s,2}e_{2}\sigma_{2}c\label{eq:E111s2}\\
0= & V_{e,1}\frac{\sigma_{1}}{\bar{K}_{e,1}+e_{1}\sigma_{1}}-\left(Y_{1}V_{s,1}e_{1}\sigma_{1}+Y_{2}V_{s,2}e_{2}\sigma_{2}+k_{e,1}\right)\label{eq:E111e1}\\
0= & V_{e,2}\frac{\sigma_{2}}{\bar{K}_{e,2}+e_{2}\sigma_{2}}-\left(Y_{1}V_{s,1}e_{1}\sigma_{1}+Y_{2}V_{s,2}e_{2}\sigma_{2}+k_{e,2}\right)\label{eq:E111e2}\\
0= & Y_{1}V_{s,1}e_{1}\sigma_{1}+Y_{2}V_{s,2}e_{2}\sigma_{2}-D\label{eq:E111c}\end{align}
have positive solutions, $s_{i},e_{i},c>0$. It follows from (\ref{eq:E111s1})--(\ref{eq:E111s2})
that $c>0$ if and only if $s_{i}<s_{f,i}$.

To prove the existence conditions, it is useful to simplify the above
equations. To this end, observe that $c$ can be eliminated from eqs.~(\ref{eq:E111s1})--(\ref{eq:E111s2})
to obtain the relation\begin{align}
V_{s,2}e_{2}\sigma_{2}\left(s_{f,1}-s_{1}\right) & =V_{s,1}e_{1}\sigma_{1}\left(s_{f,2}-s_{2}\right),\label{eq:EliminateC}\end{align}
and eqs.~(\ref{eq:E111e1})--(\ref{eq:E111c}) immediately yield
\begin{equation}
e_{i}=\frac{V_{e,i}}{D+k_{e,i}}-\frac{\bar{K}_{e,i}}{\sigma_{i}},\; i=i,2,\label{eq:E111eSS}\end{equation}
We conclude that $E_{111}$ exists if and only there exist $0<s_{i}<s_{f,i}$,
$e_{i}>0$ satisfying (\ref{eq:E111c})--(\ref{eq:E111eSS}). We show
below that such $s_{i}$ and $e_{i}$ exist if and only if $0<D<D_{t,1},D_{t,2},D_{c}$.

To prove the sufficiency of the condition, $0<D<D_{t,1},D_{t,2},D_{c}$,
substitute (\ref{eq:E111eSS}) in (\ref{eq:E111c}) and (\ref{eq:EliminateC})
to obtain the equations\begin{align}
0 & =f_{1}(\sigma_{1},\sigma_{2})\equiv\frac{\sigma_{1}}{\overline{\sigma}_{1}}+\frac{\sigma_{2}}{\overline{\sigma}_{2}}=1,\label{eq:GrowthIsocline}\\
0 & =f_{2}(\sigma_{1},\sigma_{2})\equiv\frac{V_{s,1}V_{e,1}}{D+k_{e,1}}\left(\sigma_{1}-\underline{\sigma}_{1}\right)\left(s_{f,2}-s_{2}\right)-\frac{V_{s,2}V_{e,2}}{D+k_{e,2}}\left(\sigma_{2}-\underline{\sigma}_{2}\right)\left(s_{f,1}-s_{1}\right),\label{eq:ConsumptionCurve}\end{align}
where \begin{align*}
\overline{\sigma}_{i} & \equiv\frac{1}{Y_{i}V_{s,i}\bar{K}_{e,i}}\left(D+k_{e,i}\right)\left(D+Y_{1}V_{s,1}\bar{K}_{e,1}+Y_{2}V_{s,2}\bar{K}_{e,2}\right),\\
\underline{\sigma}_{i} & \equiv\left(D+k_{e,i}\right)\frac{\bar{K}_{e,i}}{V_{e,i}}.\end{align*}
Now, (\ref{eq:GrowthIsocline}) defines a line on the $\sigma_{1},\sigma_{2}$-plane
with the intercepts, $\overline{\sigma}_{1}$ and $\overline{\sigma}_{2}$
(full line in Fig.~\ref{fig:StabilityProofs}b). As $D$ increases,
it moves further from the origin, and passes through the point, $\left(\sigma_{f,1},\sigma_{f,2}\right)$,
precisely when $D=D_{c}$. On the other hand, (\ref{eq:ConsumptionCurve})
defines a curve which has the geometry of the dashed curve in Fig.~\ref{fig:StabilityProofs}b.
To see this, observe that $\left(\underline{\sigma}_{1},\underline{\sigma}_{2}\right)$
and $\left(\sigma_{f,1},\sigma_{f,2}\right)$ lie on the curve. Moreover,
$\left(\underline{\sigma}_{1},\underline{\sigma}_{2}\right)$ lies
to the left and below $\left(\sigma_{f,1},\sigma_{f,2}\right)$ because
\[
D<D_{t,i}=\frac{V_{e,i}\sigma_{f,i}}{\bar{K}_{e,i}}-k_{e,i}\Leftrightarrow\underline{\sigma}_{i}<\sigma_{f,i},\]
and the slope of the curve is positive since\[
-\frac{\partial f_{1}/\partial\sigma_{1}}{\partial f_{1}/\partial\sigma_{2}}=\frac{\frac{V_{s,1}V_{e,2}}{D+k_{e,2}}\frac{ds_{1}}{d\sigma_{1}}\left(\sigma_{2}-\underline{\sigma}_{2}\right)+\frac{V_{s,2}V_{e,1}}{D+k_{e,1}}\left(s_{f,2}-s_{2}\right)}{\frac{V_{s,2}V_{e,1}}{D+k_{e,1}}\frac{ds_{2}}{d\sigma_{2}}\left(\sigma_{1}-\underline{\sigma}_{1}\right)+\frac{V_{s,1}V_{e,2}}{D+k_{e,2}}\left(s_{f,1}-s_{1}\right)}>0\]
for all $\underline{\sigma}_{i}<\sigma_{i}<\sigma_{f,i}$. Finally,
$\left(\underline{\sigma}_{1},\underline{\sigma}_{2}\right)$ lies
below the growth isocline because\[
\frac{\underline{\sigma}_{1}}{\overline{\sigma}_{1}}+\frac{\underline{\sigma}_{2}}{\overline{\sigma}_{2}}=\frac{Y_{1}V_{s,1}\bar{K}_{e,1}+Y_{2}V_{s,2}\bar{K}_{e,2}}{D+Y_{1}V_{s,1}\bar{K}_{e,1}+Y_{2}V_{s,2}\bar{K}_{e,2}}<1\]
for all $D>0$. It follows that the curves defined by (\ref{eq:GrowthIsocline})
and (\ref{eq:ConsumptionCurve}) intersect at a unique point, say,
$\left(\widetilde{\sigma}_{1},\widetilde{\sigma}_{2}\right)$, such
that $\underline{\sigma}_{i}<\widetilde{\sigma}_{1}<\sigma_{f,1}.$
Since\[
\underline{\sigma}_{i}<\widetilde{\sigma}_{i}\Leftrightarrow e_{i}=\frac{V_{e,i}}{D+k_{e,i}}-\frac{\bar{K}_{e,i}}{\widetilde{\sigma}_{i}}>0\]
$E_{111}$ exists and is unique.

The necessity of the existence conditions follows from the fact that
for all $0<\sigma_{i}<\sigma_{f,i}$,\[
e_{i}<\frac{V_{e,i}}{D+k_{e,i}}-\frac{\bar{K}_{e,i}}{\sigma_{f,i}}=\frac{\bar{K}_{e,i}}{\sigma_{i}\left(D+k_{e,i}\right)}\left(D_{t,i}-D\right).\]
Hence, if $D_{t,i}\le0$ or $D\ge D_{t,i}$, $E_{111}$ does not exist
because $e_{i}<0$. Likewise, if $D\ge D_{c}$, $E_{111}$ does not
exist because there are no $0<s_{i}<s_{f,i}$ satisfying (\ref{eq:GrowthIsocline}).

We conclude that $E_{111}$ exists if and only if\[
0<D<D_{t,1},D_{t,2},D_{c}.\]
We were unable to prove the stability condition for $E_{111}$. However,
extensive numerical simulations suggest that it is stable whenever
it exists.
\end{document}